\documentclass[A4]{amsart}
\usepackage{amssymb,stmaryrd}
\usepackage{amsfonts}
\usepackage{amstext}
\usepackage{algorithmic}
\usepackage{algorithm}
\usepackage{graphicx}
\usepackage{epstopdf}
\usepackage[all]{xy}
\usepackage{MnSymbol}
\numberwithin{equation}{section}
\newtheorem{theorem}{Theorem}[section]
\newtheorem{lemma}[theorem]{Lemma}
\newtheorem{proposition}[theorem]{Proposition}

\theoremstyle{definition}

\theoremstyle{remark}

\renewcommand{\S}{{\mathcal{S}}}
\newcommand{\R}{{\mathbb{R}}}

\newcommand{\CR}{{\mathcal{R}}}

\newcommand{\wedgeq}{{\wedge\kern-5pt\cdot}}

\newcommand{\cg}{{g}}

\newcommand{\tens}{\otimes}

\newcommand{\id}{{\rm id}}

\newcommand{\extd}{{\rm d}}
\newcommand{\del}{{\partial}}
\newcommand{\eps}{\epsilon}

\newcommand{\la}{{\triangleright}}

\begin{document}

\title[Noncommutative spherically symmetric spacetimes]{Noncommutative spherically symmetric spacetimes at semiclassical order}


\author{Christopher Fritz \& Shahn Majid }
\address{Department of Physics and Astronomy, University of Sussex, Brighton BN1 9RH, UK\\ + \\ Queen Mary, University of London\\
School of Mathematics, Mile End Rd, London E1 4NS, UK}
\thanks{This work is supported in part by the Science and Technology Facilities Council (grant number ST/J000477/1)}

\email{c.fritz@sussex.ac.uk, s.majid@qmul.ac.uk}


\begin{abstract} Working within the recent formalism of Poisson-Riemannian geometry, we completely solve the case of generic spherically symmetric metric and spherically symmetric Poisson-bracket to find a unique answer for the quantum differential calculus, quantum metric and quantum Levi-Civita connection at semiclassical order  $O(\lambda)$. Here $\lambda$ is the deformation parameter, plausibly the Planck scale. We find that $r,t,\extd r,\extd t$ are all forced to be central, i.e. undeformed at order $\lambda$, while for each value of $r,t$ we are forced to have a fuzzy sphere of radius $r$ with a unique  differential calculus which is necessarily nonassociative at order $\lambda^2$. We give the spherically symmetric quantisation of the FLRW cosmology in detail and also recover a previous analysis for the Schwarzschild black hole, now showing that the quantum Ricci tensor for the latter vanishes at order $\lambda$. The quantum Laplace-Beltrami operator for spherically symmetric models turns out to be undeformed at order $\lambda$ while more generally in Poisson-Riemannian geometry we show that it deforms to
\[ \square f+{\lambda\over 2}\omega^{\alpha\beta}({\rm Ric}^\gamma{}_\alpha-S^\gamma{}_{;\alpha})(\widehat\nabla_\beta\extd f)_\gamma + O(\lambda^2)\]
in terms of the classical Levi-Civita connection $\widehat\nabla$, the contorsion tensor $S$, the Poisson-bivector $\omega$ and the Ricci curvature of the Poisson-connection that controls the quantum differential structure. The Majid-Ruegg spacetime $[x,t]=\lambda x$ with its standard calculus and unique quantum metric provides an example with nontrivial correction to the Laplacian at order $\lambda$.
\end{abstract}

\maketitle 

\section{Introduction}

In recent years it has come to be fairly widely accepted that quantum gravity effects could render spacetime better modelled as a noncommutative or `quantum' geometry than a classical one\cite{Ma:pla}. The remarkable discovery here is that such a {\em quantum spacetime hypothesis} is highly restrictive in that not every classical Riemannian or pseudo-Riemannian geometry $(M,\cg)$ can be quantised while also respecting symmetries\cite{BegMa5,MaTao}, starting with the quantum anomaly for differential calculus or no-go theorems introduced in \cite{BegMa1,BegMa2}. More recently a theory of `Poisson-Riemannian geometry' in \cite{BegMa6} provided a systematic analysis of the constraints on the classical geometry for the quantisation to exist at least at lowest deformation order. This emergence of a well-defined order $\lambda$ deformation theory in \cite{BegMa6} means that a specific paradigm of physics, namely of lowest order quantum gravity effects, emerges out of the  quantum spacetime hypothesis in much the same way as classical mechanics emerges from quantum mechanics at first order in $\hbar$. In our case $\lambda$ is plausibly the Planck scale so although this is a Poisson-Level theory, it includes quantum gravity effects and could also be called `semi-quantum gravity'\cite{BegMa6}. 

Our primary aim in the present paper is to analyse the content of these constraints in the spherically symmetric case. This includes the example of the Schwarzschild black hole already covered in \cite{BegMa6} but now taken further and also, which is new, the FLRW or big-bang cosmological models. In our class of quantisations we assume that both the metric and the Poisson tensor are spherically symmetric and find generically that $t$ must be central.  The radius variable $r$ and the differentials $\extd t,\extd r$ are then also central as an outcome of our analysis. This means that the only quantisation that can take place is on the spheres at each fixed $r,t$ and we find that this is necessarily the `nonassociative fuzzy sphere' quantisation of $S^2$ proposed in \cite{BegMa2} as a cochain twist and later in \cite{BegMa6} within Poisson-Riemannian geometry. This result is both positive and negative. It is positive because our analysis says that this simple form of quantisation is unique under our assumptions, it is negative because it is hard to extract physical predictions in this model and we show in particular that more obvious sources such as corrections to the quantum scaler curvature and quantum Laplace-Beltrami operator vanish at order $\lambda$, in line with the twist as a kind of `gauge transformation'. We do still have changes to the form of the quantum metric (and quantum Ricci tensor) and more subtle effects such as nonassociativity of the differential calculus at order $\lambda^2$. Indeed, quantisation by cochain twisting remains the main source of quasi-associative examples in the physics literature\cite{BegMa2,BegMa3,Schupp, Szabo,AscSch} so it is significant that our general analysis shows that we are generically {\em forced} into this setting by spherical symmetry. 

Our second goal is to the theoretical development of the quantum Laplace operator and quantum Ricci tensor in Poisson-Riemannian geometry, which continues on from the development in \cite{BegMa6}. We need of course a Poisson bracket $\{\ ,\ \}$ on our Riemannian manifold as the semiclassical data for the quantum spacetime coordinate algebra. Next, as in \cite{BegMa1,BegMa6}, we need a `preconnection' (or Lie-Rinehart a.k.a. contravariant connection\cite{Haw,Heu,Burs,Fern}) compatible with the Poisson tensor and which controls the quantisation of the differential structure. See also \cite{Wat} using contravariant connections in a different context from noncommutative geometry. However, to keep things accessible within the tools of ordinary differential geometry, it is observed in \cite{BegMa6} that every ordinary covariant derivative $\nabla$ pulls back to a preconnection $\gamma(f,\ )=\nabla_{\widehat f}$ along Hamiltonian vector fields $\widehat f=\{f,\ \}$ (where $f$ is a function on the manifold) and we focus attention on this class of preconnections given by actual covariant derivatives. This is not strictly necessary but is physically reasonable in the first instance given that  covariant derivatives already arise extensively in general relativity. The downside is that the covariant derivatives that we obtain will have directions that do not affect the noncommutative geometry, so there may for example be a unique preconnection or contravariant connection but a bigger moduli of covariant derivatives with extra directions that do not play an immediate role for the quantisation (but which could couple to physical fields later on). Of this covariant derivative we require\cite{BegMa6}
\begin{enumerate}
\item Poisson compatibility  $\nabla_{\widehat f}\extd h-\nabla_{\widehat h}\extd f=\extd\{f,h\}$
\item Metric compatibility $\nabla_{\widehat f}\cg=0$ where $\cg$ is the metric 1-1 form.
\item A condition on the curvature and   torsion of $\nabla$ (see Section~2) which provides for a unique quantum Levi-Civitia connection.\end{enumerate}
In fact we will assume stronger versions of the first two, namely requiring them not only along hamiltonian vector fields, so $\nabla g=0$ for the second item\cite{BegMa6}. Again, this is not strictly needed for the noncommutative geometry but makes sense in practice as a natural simplification. In this setting it is shown in \cite{BegMa6} that the entire exterior algebra $\Omega(M)$ quantises at lowest order, i.e. with a new non-super-commutative wedge product $\wedge_1$, the metric similarly quantises to a quantum metric $\cg_1$ and the Levi-Civita connection $\widehat\nabla$ similarly quantises to a `best possible' quantum Levi-Civita $\nabla_1$ within the recent systematic framework of `bimodule noncommutative geometry'\cite{DVM1,Mou,BegMa4,BegMa5,MaTao}. This in general could have $\nabla_1 \cg_1=O(\lambda)$ as an antisymmetric correction to metric compatibility at order $\lambda$. In classical geometry $\nabla \cg$ is of course necessarily symmetric so there is no antisymmetric part to be concerned about, but in Poisson-Riemannian geometry\cite{BegMa6} the vanishing of the symmetric part alone determines $\nabla_1$ and the antisymmetric part may or may not then vanish. The optional `quantum Levi-Civita condition' (3) asserts that this effect vanishes and we have full metric compatibility.  

We only require all this to order $\lambda$ so that this is a semiclassical (or semi-quantum) theory. However, if $\lambda$ is of order the Planck scale then this is already so small that, away from singularities that blow up the effective value, the next order in $\lambda$ is hardly relevant and the semiclassical one is the main one of interest. In particular, if the preconnection $\nabla$ has curvature then the differential calculus will be non-associative at order $\lambda^2$, so we do not need care about $\nabla$ being flat in our order $\lambda$ paradigm. This already takes us beyond conventional (associative) noncommutative geometry, i.e. the curvature of $\nabla$ is perfectly allowed in our paradigm but may lead to subtle effects at order $\lambda^2$. This is one of the mentioned outcomes of our analysis,  that spherical symmetry generically forces nonassociativity at order $\lambda^2$. 

Another possible effect we will consider concerns the form of the   quantum metric. If we write this as $g_1=\extd x^\mu \bullet\tilde g_{\mu\nu}\tens_1 \extd x^\nu$ (where $\bullet$ is the quantum product) then we find in the general analysis that
\[ \tilde g_{\mu\nu}=g_{\mu\nu}+{\lambda\over 2}h_{\mu\nu}\]
at order $\lambda$ where $h_{\mu\nu}$ is a certain antisymmetric tensor (or 2-form) built from the classical data. The physical interpretation of this is not clear but {\em if} we suppose that it is $\tilde g$ whose expectation values are the observed `effective metric' then we see that, aside from its values being in a noncommutative algebra, it acquires a non-symmetric component. This makes contact with other scenarios where non-symmetric metrics have been studied and means that the metric acquires a spin 1 component. On the other hand $h_{\mu\nu}$ is not tensorial i.e. transforms in a more complicated way if we change coordinates, albeit in such a way that when proper account is taken of the quantum tensor product $\tens_1$, our constructions themselves are coordinate invariant. We look at this closely on one of the models in Section~4.3.  The same applies for the quantum Ricci tensor. 

Finally, there could be effects that modify the Laplace Beltrami operator as for the black hole in \cite{Ma:alm} in an earlier approach. We show that $\square_1=(\ ,\ )_1\nabla_1\extd$, where $(\ ,\ )_1$ is the inverse to $\cg_1$,  gets generically an order $\lambda$ correction given by the Ricci curvature of $\nabla$ and the covariant derivative of the contorsion vector (Theorem~\ref{square1}). We include the Majid-Ruegg spacetime $[x,t]=\lambda x$ as an example, with its unique form of noncommutative geometry previously established in \cite{BegMa5}.  Plane waves for the quantum Laplacian at first order are provided by Kummer $M$ and $U$ functions. One of the surprising outcomes of the paper is that such examples are relatively rare and for many others including all central time spherically symmetric models there is no quantum correction to the Laplacian at first order. We also find no order $\lambda$ correction to the quantum Laplacian for the Bertotti-Robinson metric quantisation of \cite{MaTao} though in this case one can easily compute the quantum geometry to all orders (this was done in \cite{MaTao} and one see that there are order $\lambda^2$ corrections to the Laplacian. 

A third goal of the paper is the in-depth study of examples both to explore effects such as the above and to put the methods of Poisson-Riemannian geometry\cite{BegMa6} into practice.  After some general formalism and the quantum Laplacian in Theorem~\ref{square1} and quantum Ricci tensor in Section~\ref{riccisec}, we study the  bicrossproduct model in Section~\ref{bicross} (both for new results and to check the formalism), the non-time central Bertotti-Robinson model in Section~\ref{BRsec} and the further treatment of the fuzzy sphere in Section~\ref{fuzzyS} (including the result that its quantum Ricci tensor is proportional to the quantum metric). These are all 2D models as part of the theoretical development. Next we in look in Section~3 for any rotationally invariant quantisation of the classical spatially-flat FLRW metric
\[  \cg=-\extd t\tens\extd t  +a^2(t)(\extd r\tens\extd r+\delta_{ij}r^2\extd z^i\tens\extd z^i)\]
where $z^i=x^i/r$ are equivalent to the angular variables in the polar decompostion of $\R^3$ (here $z^i z^j\delta_{ij}=1$ so that only two of the three are independent). We find that everything works in the sense that, as for the black hole, there does exist $\nabla$ meeting our requirements. It is not the Levi-Civita connection, which we will rather denote $\widehat\nabla$, and it pulls back to a unique preconnection, so there is a unique noncommutative geometry at order $\lambda$ in our context. We also apply the machinery of \cite{BegMa6} to find the unique quantum metric $\cg_1$ at order $\lambda$ (it looks the same if we choose the correct quantum coordinates and ordering) and a unique quantum Levi-Civita bimodule connection $\nabla_1$ which is torsion free and obeys $\nabla_1 \cg_1=0$ both at order $\lambda$. On the other hand, the quantum wave operator $\square_1$ is undeformed at this order. Section~4 then proceeds to our general results for spherically symmetric metrics including a revisit to the Schwarzschild black hole with a stronger statement than in \cite{BegMa6} and full details of the resulting noncommutative Riemannian geometry (we find that it is quantum Ricci flat) at order $\lambda$. Section~5 contains a non-spherically symmetric example, namely pp gravitational waves, but with similar non-deformation features for the quantum scaler curvature and Laplacian. The paper concludes with some discussion in Section~6.

\section{Formalism}

Throughout this paper by `quantum' we mean extended to noncommutative geometry to order $\lambda$. There is a physical assumption that quantities will extend further to all orders according to axioms yet to be determined  but we do not consider the details of that yet (the idea is to proceed order by order strictly as necessary). This is for convenience and one could more precisely say `semiquantum' as in \cite{BegMa6}.  We use $\bullet$ for the deformed and $;$ for the covariant derivative with respect to the Poisson compatible `quantising' connection $\nabla$. 

\subsection{Poisson-Riemannian geometry and the quantum Laplacian}

We start with a short recap of the formulae we need from \cite{BegMa6} along with some modest new results in the same spirit. We let $M$ be a smooth manifold with exterior algebra $\Omega$, equipped with a metric tensor $\cg$ and Levi-Civita connection $\widehat\nabla$ on $\Omega^1$. We let $\nabla:\Omega^1\to \Omega^1\tens\Omega^1$ be another connection on $\Omega^1$ and let $\widehat\Gamma,\Gamma$ be the respective Christoffel symbols so that 
\[ \nabla_{\beta}\extd x^\alpha=-\Gamma^\alpha{}_{\beta \gamma} \extd x^\gamma\]
and similarly for $\widehat\nabla$. The tensor product here is over $C^\infty(M)$ and we view the covariant derivative abstractly as a map or in practice with its first output against $\del/\del x^\beta$ to define $\nabla_\beta$. Its action on the component tensor of a 1-form $\omega=\omega_\alpha\extd x^\alpha$ is $\nabla_\beta\omega_\alpha=(\nabla_\beta\omega)\alpha=\del_\beta\omega_\alpha-\omega_\gamma\Gamma^\gamma{}_{\beta\alpha}$, which fixes its extension to other tensors. The torsion and curvature tensors are
\begin{equation*} T^\alpha{}_{\beta\gamma}=\Gamma^\alpha{}_{\beta \gamma}- \Gamma^\alpha{}_{\gamma \beta},\quad R^{\alpha}{}_{\beta \gamma \delta}  = \Gamma^{\alpha}{}_{\delta \beta,\gamma}-\Gamma^{\alpha}{}_{\gamma \beta,\delta}  + \Gamma^{\kappa}{}_{\delta \beta}\Gamma^{\alpha}{}_{\gamma \kappa}- \Gamma^{\kappa}{}_{\gamma \beta}\Gamma^{\alpha}{}_{\delta \kappa} . \end{equation*}
in the conventions of \cite{BegMa6}. In the presence of a metric we have a contorsion tensor $S$ defined by
\[ \Gamma^\alpha{}_{\beta \gamma}=\widehat\Gamma^\alpha{}_{\beta \gamma}+ S^\alpha{}_{\beta \gamma}\]
where metric compatibility $\nabla g=0$ is equivalent to the first of
\[ S^ \alpha{}_{\beta \gamma}={1\over 2}\cg^{\alpha \delta}(T_{\delta \beta \gamma}+T_{\beta \delta \gamma}+T_{\gamma \delta \beta}),\quad T^ \alpha{}_{\beta \gamma}=S^ \alpha{}_{\beta \gamma}-S^ \alpha{}_{\gamma\beta}\]
and also implies that the lowered $R_{\alpha\beta\gamma\delta}$ is antisymmetric in the first pair of indices (as well as the second). Also note that the lowered index contorsion here obeys $S_{\alpha \beta \gamma}=-S_{\gamma \beta \alpha}$. In this setup, torsion becomes the relevant field which determines metric compatible $\nabla$ via $S$. This set up is slightly more than we strictly needed for the noncommutative geometry itself as explained in the introduction.

We let $\omega^{\alpha \beta}$ define the Poisson bracket $\{f,h\}=\omega^{\alpha \beta}(\del_\alpha f)\del_\beta h$. This is a bivector field and it is shown in \cite{BegMa6} that $\nabla$ in our full sense is Poisson compatible if and only if
\begin{equation}\label{nablaT} \nabla_\gamma \omega^{\alpha \beta}+T^\alpha{}_{\delta \gamma}\omega^{\delta \beta}+T^\beta{}_{\delta \gamma}\omega^{\alpha \delta}=0\end{equation}
or equivalently in the Riemannian case that
\begin{equation}\label{nablaS} \widehat{\nabla}_\gamma \omega^{\alpha \beta}+S^\alpha{}_{\delta \gamma}\omega^{\delta \beta}+S^\beta{}_{\delta \gamma}\omega^{\alpha \delta}=0 \end{equation}
We also want $\omega$ to be a Poisson tensor even though this is not strictly needed at order $\lambda$, 
\begin{equation}\label{omegapoisson} \sum_{{\rm cyclic}(\alpha,\beta,\gamma)} \omega^{\alpha \mu}\omega^{\beta \gamma}{}_{,\mu} = 0 \end{equation}
which given Poisson-compatibility is equivalent  \cite{BegMa6} to 
\begin{equation}\label{omegapoissonT}\sum_{{\rm cyclic}(\alpha,\beta,\gamma)}\omega^{\alpha \mu}\omega^{\beta \nu} T^\gamma{}_{\mu\nu}=0.\end{equation}

Given the Poisson tensor and  (\ref{nablaT}) respectively we quantize the product of  functions with each other and 1-forms by functions, 
\[ f\bullet h=fh+{\lambda\over 2}\{f,h\},\quad   f\bullet \eta=f\eta+{\lambda\over 2}\omega^{ab}f_{,a}\nabla_b\eta,\quad \eta\bullet f=\eta f-{\lambda\over 2}\omega^{ab}f_{,a}\nabla_b\eta\]
to order $\lambda$, so that
\begin{equation}\label{fetacomm} [x^\alpha,\eta]=\lambda \omega^{\alpha \beta}\nabla_\beta\eta\end{equation}
to order $\lambda$ in the quantum algebra. It is shown in \cite{BegMa6} that we also can quantize the wedge product of 1-forms and higher,
\begin{equation}\label{wedge1} \extd x^\alpha\wedge_1 \extd x^\beta=\extd x^\alpha\wedge\extd x^\beta+{\lambda\over 2} \omega^{\gamma\delta}\nabla_\gamma\extd x^\alpha\wedge\nabla_\delta\extd x^\beta+\lambda H^{\alpha \beta} \end{equation}
to order $\lambda$. This gives anticommutation relations
\begin{equation}\label{anticomm} \{\extd x^\alpha,\extd x^\beta\}_1=\lambda\omega^{\gamma\delta}\Gamma^\alpha{}_{\gamma\mu}\Gamma^\beta{}_{\delta\nu}\extd x^\mu\wedge \extd x^\nu+ 2\lambda H^{\alpha\beta}.\end{equation}
Here the extra `non-functorial' term needed is given by a family of 2-forms
\[ H^{\alpha\beta}={1\over 4} \omega^{\alpha \gamma}(\nabla_\gamma  T^\beta{}_{\nu\mu}- 2R^\beta{}_{\nu \mu \gamma})\extd x^\mu \wedge \extd x^\nu.\]
The exterior derivative $\extd$ is taken as undeformed in the underlying vector spaces. Note that because the products by functions is modified, $\tens_1$, by which mean the tensor product over the quantum algebra, taken to order $\lambda$, is not the usual tensor product. It is characterised by
\[ \eta\tens_1 f\bullet \zeta=\eta\bullet f\tens_1\zeta\]
for all functions $f$ and any $\eta,\zeta$. The quantum and classical tensor products are identified by a natural transformation $q$ to order $\lambda$, as explained further in \cite{BegMa6}. If we denote by $A$ the vector space $C^\infty(M)$ with this modified product, which can always be taken to be associative, then $\tens_1$ means $\tens_A$, i.e. over this quantum coordinate algebra.   

Next, although not strictly necessary, it is useful to optionally require that the $\nabla_Q$ is geometrically well behaved in noncommutative geometry in that its associated quantum torsion commutes with the quantum product by functions (a `bimodule map'). This is a quantum tensoriality property similar to that which requires centrality of the quantum metric. We say that such a $\nabla$ is {\em regular}, which amounts to
\begin{equation}\label{Tcompat} \omega^{\alpha \beta}\nabla_\beta T^\gamma{}_{\mu\nu}=0\end{equation}
With or without this simplifying assumption, there is a quantum metric $\cg_1\in \Omega^1\tens_1\Omega^1$ to order $\lambda$ given by
\begin{equation}\label{g1}
\cg_1=\cg_{\mu \nu}\extd x^\mu\tens_1\extd x^\nu +{\lambda\over 2}\omega^{\alpha \beta}\Gamma_{\mu\alpha \kappa}\Gamma^\kappa{}_{\beta \nu}\extd x^\mu\tens_1\extd x^\nu + {\lambda\over 2}\CR_{\mu \nu}\extd x^\mu \tens_1\extd x^\nu \end{equation}
and obeying $\wedge_1(\cg_1)=0$ as well as a `reality' property ${\rm flip}(*\tens *)g_1=g_1$.  In our case $x^\mu{}^*=x^\mu$ since our classical manifold has real coordinates and also acts trivially on all classical (real) tensor components, while $\lambda^*=-\lambda$. It's action on $\bullet$ is more nontrivial namely to reverse order while on $\wedge_1$ to reverse order with sign according to the degrees. For the most part this $*$-operation takes care of itself given that our classical tensors are real, so we will not emphasise it. The first two terms in (\ref{g1}) are the functorial part $g_Q$ and the last term is a correction. Here (correcting an overall sign error in the preprint version of \cite[(5.5)]{BegMa6})
\[ \CR_{\mu\nu}={1\over 2}\cg_{\alpha\beta}\omega^{\alpha \gamma}(\nabla_\gamma T^\beta{}_{\mu\nu}-R^\beta{}_{\mu\nu\gamma}+ R^\beta{}_{\nu\mu\gamma})\]
is antisymmetric and can be viewed as the {\em generalised Ricci 2-form} 
\[ \CR={1\over 2}\CR_{\mu\nu}\extd x^\nu \wedge\extd x^\mu=\cg_{\mu\nu}H^{\mu\nu}\]
(note the sign and factor in our conventions for 2-form components). Next we let $(\ ,\ )_1:\Omega^1\tens_1\Omega^1\to A$ be the inverse metric as a bimodule map. We define $A$-valued coefficients $g_{1\mu\nu},\tilde g_{\mu\nu}$ by $\cg_1=\cg_1{}_{\mu\nu}\extd x^\mu \tens_1 \extd x^\nu=\extd x^\mu\bullet \tilde\cg_{\mu\nu}\tens_1 \extd x^\nu$ so that
\[ \tilde g_{\mu\nu}=g_1{}_{\mu\nu}+{\lambda\over 2}\omega^{\alpha\beta}\Gamma^\gamma{}_{\alpha\mu} g_{\gamma\nu,\beta}=g_{\mu\nu}+{\lambda\over 2} h_{\mu\nu}\]
to order $\lambda$, where we also write $h_{\mu\nu}$ for the leading order correction to $g_{\mu\nu}$. Explicitly, from (\ref{g1}), this gives us
\begin{equation}\label{h} h_{\mu\nu}=\CR_{\mu\nu}+\omega^{\alpha\beta}(\Gamma_{\mu\alpha\kappa}\Gamma^\kappa{}_{\beta\nu}  +\Gamma^\gamma{}_{\alpha\mu}g_{\gamma\nu,\beta})= -h_{\nu\mu} \end{equation}
where we use  metric compatibility of $\nabla$ in the form $g_{\gamma\nu,\beta}=\Gamma_{\gamma\beta\nu}+\Gamma_{\nu\beta\gamma}$ to replace the second term to more easily verify antisymmetry.  We let $\tilde\cg^{\mu\nu}$ be the $A$-valued matrix inverse so that $\tilde\cg_{\mu\nu}\tilde\cg^{\nu\gamma}=\delta_\mu^\gamma=\tilde\cg^{\gamma\nu}\tilde\cg_{\nu\mu}$ then it follows from the bimodule properties of $(\ ,\ )_1$ that 
\begin{equation}\label{tildegupper} (\extd x^\mu, \extd x^\nu)_1=\tilde\cg^{\mu\nu}=g^{\mu\nu}-{\lambda\over 2}\tilde h^{\mu\nu}\end{equation}
where
\begin{eqnarray*} \tilde h^{\mu\nu}&=&g^{\mu\alpha}g^{\nu\beta} h_{\alpha\beta}+g^{\mu\alpha} \{g_{\alpha\beta},g^{\beta\nu}\}=\CR^{\mu\nu}+\omega^{\alpha\beta}(\Gamma^\mu{}_{\alpha \kappa}\Gamma^\kappa{}_{\beta\gamma}g^{\nu\gamma}+ \Gamma^\mu{}_{\alpha\gamma}g^{\nu\gamma}{}_{,\beta})\\
&=&\CR^{\mu\nu}-\omega^{\alpha\beta} g^{\eta\zeta} \Gamma^\mu{}_{\alpha \eta}\Gamma^\nu{}_{\beta\zeta}= -\tilde h^{\nu\mu}\end{eqnarray*}
and $\CR$ has indices raised by $g$. As an application, in bimodule noncommutative geometry there is a quantum dimension\cite{BegMa5} which we can now compute.

\begin{proposition}\label{dim1} In the setting above, the  `quantum dimension' to order $\lambda$ is
\[ \dim_1:=(\ ,\ )_1(g_1)=\dim(M)+{ \lambda\over 2}\{g_{\mu\nu},g^{\mu\nu}\}.\]
\end{proposition}
\proof  Given the above results, we have 
\begin{eqnarray*}\dim_1&=&(\extd x^\mu\bullet \tilde g_{\mu\nu},\extd x^\nu)_1=\tilde g_{\mu\nu}\bullet\tilde g^{\mu\nu}+([\extd x^\mu,g_{\mu\nu}],\extd x^\nu)\\
&=&\dim(M)+{\lambda\over 2} (h_{\mu\nu}-h_{\nu\mu})g^{\mu\nu}+\lambda\omega^{\alpha\beta}g_{\mu\nu,\alpha}\Gamma^\mu{}_{\beta\gamma} g^{\gamma\nu}\\
&=&\dim(M)-\lambda\omega^{\alpha\beta}g^{\mu\nu}{}_{,\alpha}\Gamma_{\nu\beta\mu} 
\end{eqnarray*} where the middle term vanishes as $g^{\mu\nu}$ is symmetric and we transferred to the derivative to the inverse metric. We can now use metric compatibility in the form $\Gamma_{\mu\beta\nu}+\Gamma_{\nu\beta\mu}=g_{\mu\nu,\beta}$ to obtain the answer. \endproof

Finally, the theory in \cite{BegMa6} says that  there is a quantum torsion free quantum metric compatible (or quantum Levi-Civita) connection $\nabla_1:\Omega^1\to \Omega^1\tens_1\Omega^1$ to order $\lambda$ if and only if 
\begin{equation}\label{nabla1g1} \widehat\nabla \CR+ \omega^{\alpha\beta}\,\cg_{\rho\sigma}\,S^\sigma{}_{\beta\nu}(R^\rho{}_{\mu\gamma\alpha}+\nabla_\alpha S^\rho{}_{\gamma\mu})\,\extd x^\gamma\tens\extd x^\mu \wedge \extd x^\nu=0\end{equation}
In fact the theory always gives a unique `best possible' $\nabla_1$ at this order for which the symmetric part of $\nabla_1 g_1$ vanishes. This leaves open that $\nabla_1g_1=O(\lambda)$, namely proportional to the left hand side  of (\ref{nabla1g1}). The construction of $\nabla_1$ takes the form
\[ \nabla_1=\nabla_Q+ q^{-1}Q(S) +\lambda K\]
where the first two terms are functorial and the last term is a further correction. Translating the formulae in \cite{BegMa6} into indices and combining, one has
\begin{lemma}cf\cite{BegMa6}\label{Gamma1} Writing $\nabla_1\extd x^\iota=-\Gamma_1{}^\iota{}_{\mu\nu}\extd x^\mu\tens_1\extd x^\nu$, the construction of \cite{BegMa6} can be written explicitly as 
\[ \Gamma_1{}^\iota{}_{\mu\nu}= \widehat\Gamma^{\iota}{}_{\mu\nu} + \frac{\lambda}{2} \omega^{\alpha\beta} \left(\widehat\Gamma^{\iota}{}_{\mu\kappa,\alpha}\Gamma^{\kappa}{}_{\beta\nu} -\widehat\Gamma^{\iota}{}_{\kappa\tau}\Gamma^{\kappa}{}_{\alpha\mu}\Gamma^{\tau}{}_{\beta\nu} +   \widehat\Gamma^{\iota}{}_{\alpha\kappa}(R^{\kappa}{}_{\nu\mu\beta}  + \nabla_{\beta} S^{\kappa}{}_{\mu\nu})\right).\]
\end{lemma}
\proof It is already stated in \cite{BegMa6} that 
\[  \nabla_{Q}(\extd x^{\iota})  = -\left( \Gamma^{\iota}{}_{\mu\nu} + \frac{\lambda}{2} \omega^{\alpha\beta} (\Gamma^{\iota}{}_{\mu\kappa,\alpha}\Gamma^{\kappa}{}_{\beta\nu} - \Gamma^{\iota}{}_{\kappa\tau}\Gamma^{\kappa}{}_{\alpha\mu}\Gamma^{\tau}{}_{\beta\nu} -  \Gamma^{\iota}{}_{\beta\kappa} R^{\kappa}{}_{\nu\mu\alpha} )\right) \extd x^{\mu} \otimes_{1} \extd x^{\nu}\]
Next, we carefully we write the term $\omega^{ij}\nabla_i \circ\nabla_j(S)$ in $Q(S)$ in \cite[Lemma~3.2]{BegMa6} as curvature plus an extra term involving $\nabla S$ and $\Gamma$, to give
\begin{eqnarray*} q^{-1}Q(S)(\extd x^{\iota}) &=& \big( S^{\iota}{}_{\mu\nu} + \frac{\lambda}{2} \omega^{\alpha\beta} (S^{\iota}{}_{\mu\kappa,\alpha}\Gamma^{\kappa}{}_{\beta\nu}- S^{\iota}{}_{\kappa\tau}\Gamma^{\kappa}{}_{\alpha\mu}\Gamma^{\tau}{}_{\beta\nu})+{\lambda\over 4} R_\omega(S)^\iota{}_{\mu\nu}\\
&&\quad -{\lambda\over 2}\omega^{\alpha\beta}\Gamma^\iota{}_{\alpha\kappa}\nabla_\beta S^\kappa{}_{\mu\nu}
 \big) \extd x^{\mu} \otimes_{1} \extd x^{\nu}\end{eqnarray*}
where 
\[ R_\omega(S)^\iota{}_{\mu\nu}=\omega^{\alpha\beta}\left(R^{\iota}{}_{\kappa \alpha\beta}S^\kappa {}_{\mu\nu} - R^{\kappa}{}_{\mu \alpha\beta}S^{\iota}{}_{\kappa\nu} - R^{\kappa}{}_{\nu \alpha\beta}S^{\iota}{}_{\mu \kappa}\right)\]
is the curvature of $\nabla$ evaluated on the Poisson bivector and acting on the contorsion tensor $S$. 
Finally, we take $K$ given explicitly in  \cite[Corollary~5.9]{BegMa6},
\[ 
K(\extd x^\iota) = \left( \frac{1}{2} \omega^{\alpha\beta} ( S^{\iota}{}_{\alpha\kappa} \nabla_{\beta} S^{\kappa}{}_{\mu\nu} - S^{\iota}{}_{\beta\kappa} R^{\kappa}{}_{\nu\mu \alpha})-{1\over 4}R_\omega(S)^\iota{}_{\mu\nu}  \right)\extd x^\mu\tens_1\extd x^\nu
\]
and combine all the terms to give the compact formula stated. \endproof
 
 As a bimodule connection there is also a generalised braiding $\sigma_1:\Omega^1\tens_1\Omega^1\to \Omega^1\tens_1\Omega^1$ that expresses the right-handed Leibniz rule for a bimodule left connection, 
\begin{equation}
\sigma_{1}(\extd x^{\alpha} \otimes_{1} \extd x^\beta) = \sigma_{Q}(\extd x^{\alpha} \otimes_{1} \extd x^\beta) + \lambda \omega^{\beta\mu} (\nabla_\mu S)(\extd x^\alpha)\end{equation}
which comes out as
\begin{equation}
\label{sigma1}
\sigma_1(\extd x^\alpha\tens_1\extd x^\beta)=\extd x^{\beta} \otimes_{1} \extd x^{\alpha} + \lambda  \left(\omega^{\mu\nu}\Gamma^{\alpha}{}_{\mu\gamma}\Gamma^{\beta}{}_{\nu\delta} - \omega^{\mu\beta}(R^{\alpha}{}_{\gamma\delta\mu}+S^\alpha{}_{\delta\gamma;\mu}) \right) \extd x^{\delta} \otimes_{1} \extd x^{\gamma} 
\end{equation} 

The bimodule noncommutative geometry  also has a natural definition of quantum Laplacian \cite{BegMa5} and we can now compute this

\begin{theorem}\label{square1} In Poisson-Riemannian geometry the quantum Laplacian to order $\lambda$ is 
\[ \square_1 f:=(\ ,\ )_1\nabla_1\extd f=\square f+\tfrac{\lambda}{2} \omega^{\alpha\beta}({\rm Ric}^\gamma{}_\alpha-S^\gamma{}_{;\alpha})(\widehat\nabla_\beta \extd f)_\gamma\]
\end{theorem}
\proof Here  ${\rm Ric}^\gamma{}_\alpha=g^{\gamma\nu}R^\beta{}_{\nu\beta\alpha}=-R^\gamma{}_{\nu\beta\alpha}g^{\nu\beta}$ and $(\widehat\nabla_\alpha\extd f)_\gamma=f_{,\alpha \gamma}-\widehat\Gamma^\iota{}_{\alpha\gamma}f_{,\iota}$ as usual. Let us also note that $\extd$ is not deformed but can look different, namely write $\extd f=  (\tilde\del_\alpha f)\bullet\extd x^\alpha$ so that
\[ \tilde\del_\mu= \del_\mu +{\lambda\over 2}\omega^{\alpha\beta}\Gamma^\nu{}_{\beta\mu}\del_\alpha\del_\nu\]
and we similarly write $\nabla_1\extd x^\iota=-\tilde\Gamma_1{}^\iota{}_{\mu\nu}\bullet\extd x^\mu\tens_1\extd x^\nu$ so that
\[ \tilde\Gamma_1{}^\iota{}_{\mu\nu}=\Gamma_1{}^\iota{}_{\mu\nu}+{\lambda\over 2}\omega^{\alpha\beta}\widehat\Gamma^\iota{}_{\nu\kappa,\alpha}\Gamma^\kappa{}_{\beta\mu}=\widehat\Gamma^\iota{}_{\mu\nu}+{\lambda\over 2}\gamma^{\iota}{}_{\mu\nu},\]
say, using symmetry of the last two indices of $\widehat\Gamma$. Then by the bimodule and derivation properties at the quantum level, we deduce
\[ \square_1 f=(\ ,\ )_1(\extd \tilde\del_\nu f\tens_1\extd x^\nu+ (\tilde\del_\alpha f)\bullet\nabla_1\extd x^\alpha)=\left(\tilde\del_\mu\tilde\del_\nu f -(\tilde\del_\alpha f)\bullet\tilde\Gamma_1{}^\alpha{}_{\mu\nu}\right)\bullet \tilde g^{\mu\nu}\]
We then expand this out to obtain as the classical $\square f$ and five corrections time $\lambda/2$ as follows:

(i) From the deformed product with $\tilde g^{\mu\nu}$ we obtain
\[ \{\del_\mu\del_\nu f-(\del_\alpha f)\widehat\Gamma^\alpha{}_{\mu\nu},g^{\mu\nu}\}\]

(ii) From the deformation in $\tilde g^{\mu\nu}$ we obtain
\[ -(\del_\mu\del_\nu f-(\del_\alpha f)\widehat\Gamma^\alpha{}_{\mu\nu})\tilde h^{\mu\nu}=0\]
by the antisymmetry of $\tilde h^{\mu\nu}$ compared to symmetry of $\widehat\Gamma^\alpha{}_{\mu\nu}$ and of $\del_\mu\del_\nu f$. So there is no contribution from this aspect at order $\lambda$. 

(iii) From the deformation in $\tilde\del_\mu\tilde\del_\nu f$ we obtain
\[  \omega^{\alpha\beta}\Gamma^\gamma{}_{\beta\mu}g^{\mu\nu}\del_\alpha\del_\gamma\del_\nu f + g^{\mu\nu}\del_\mu(\omega^{\alpha\beta}\Gamma^\gamma{}_{\beta\nu}\del_\alpha\del_\gamma f)=2 \omega^{\alpha\beta}\Gamma^\gamma{}_{\beta\mu}g^{\mu\nu}\del_\alpha\del_\gamma\del_\nu f + g^{\mu\nu}(\del_\alpha\del_\gamma f)\del_\mu(\omega^{\alpha\beta}\Gamma^\gamma{}_{\beta\nu})\]

(iv) From the deformation in $-\tilde\del_\alpha f\bullet\widehat\Gamma^\alpha{}_{\mu\nu}$ we obtain
\[ -\omega^{\alpha\beta}\Gamma^\gamma{}_{\beta\kappa}g^{\mu\nu}\widehat\Gamma^\kappa{}_{\mu\nu}\del_\alpha\del_\gamma f - \{\del_\alpha f,\widehat\Gamma^\alpha{}_{\mu\nu}\}g^{\mu\nu}\]

(v) From the deformation in $\tilde\Gamma^1$ and our above formulae for that, we obtain 
\[ -\gamma^\iota{}_{\mu\nu}g^{\mu\nu}\del_\iota f = -(\del_\iota f)\omega^{\alpha\beta}\left(2\widehat\Gamma^\iota{}_{\mu\kappa,\alpha}\Gamma^\kappa{}_{\beta\nu} g^{\mu\nu}+\widehat\Gamma^\iota{}_{\beta\kappa}{\rm Ric}^\kappa{}_\alpha+ \widehat\Gamma^\iota{}_{\alpha\kappa} S^\kappa{}_{;\beta}\right)\]
where $S^\kappa=S^\kappa{}_{\mu\nu}g^{\mu\nu}$ is the `contorsion vector field' and $;$ is with respect to $\nabla$. 

Now, comparing we see that the cubic derivatives of $f$ in (i) and (iii) cancel using metric compatibility to write a derivative of the metric in terms of $\Gamma$. Similarly the 1-derivative term from (i) is $-\del_\iota f$ times
\[ \{\widehat\Gamma^\iota{}_{\mu\nu},g^{\mu\nu}\}=\omega^{\alpha\beta}\widehat\Gamma^\iota{}_{\mu\kappa,\alpha}g^{\mu\nu}{}_{,\beta}g_{\eta\nu}g^{\kappa\eta}=-\omega^{\alpha\beta}\widehat\Gamma^{\iota}_{\mu\kappa,\alpha}g^{\mu\nu}(\Gamma_{\eta\beta\nu}+\Gamma_{\nu\beta\eta})g^{\kappa\eta}=-2\omega^{\alpha\beta}\widehat\Gamma^\iota{}_{\mu\kappa,\alpha}g^{\mu\nu}\Gamma^\kappa{}_{\beta\nu}\]
where we inserted $g_{\eta\nu}$, turned $\del_\beta$ onto this and used metric compatibility of $\nabla$. In the last step we used that $\widehat\Gamma$ is torsion free so symmetric in the last two indices. The result exactly cancels with a term in (v) giving 
\[\square_1 f=\square f +{\lambda\over 2}(\del_\iota f)\omega^{\alpha\beta}\widehat\Gamma^\iota{}_{\alpha\gamma}\left({\rm Ric}^\gamma{}_\beta-\S^\gamma{}_{;\beta} \right)+O(\del^2)f\]
where we have not yet analysed corrections with quadratic derivatives of $f$. Turning to these, the remainder of (i) and (iv)  contribute
\begin{eqnarray*} &&\kern-15pt -\{\del_\gamma f,\widehat\Gamma^\gamma\}-(\del_\alpha\del_\gamma f)\omega^{\alpha\beta}\Gamma^\gamma{}_{\beta\kappa}\widehat\Gamma^\kappa=-\omega^{\alpha\beta}(\del_\alpha\del_\gamma f)\widehat\Gamma^\gamma{}_{;\beta}\\
&=&\omega^{\alpha\beta}(\del_\alpha\del_\gamma f)S^\gamma{}_{;\beta}-(\del_\alpha\del_\gamma f)g^{\mu\nu}\omega^{\alpha\beta}\Gamma^\gamma{}_{\mu\nu;\beta}\\
&=&\omega^{\alpha\beta}(\del_\alpha\del_\gamma f)S^\gamma{}_{;\beta}-(\del_\alpha\del_\gamma f)g^{\mu\nu}\omega^{\alpha\beta}\left(\Gamma^\gamma{}_{\mu\nu,\beta}+\Gamma^\gamma{}_{\beta\kappa}\Gamma^\kappa{}_{\mu\nu}-\Gamma^\gamma{}_{\kappa\nu}\Gamma^\kappa{}_{\beta\mu}-\Gamma^\gamma{}_{\mu\kappa}\Gamma^\kappa{}_{\beta\nu}\right).
\end{eqnarray*}
Meanwhile in (iii), we use poisson-compatibility in the direct form\cite{BegMa6}
\[ \omega^{\alpha\beta}{}_{,\mu}=\omega^{\beta\eta}\Gamma^\alpha{}_{\eta\mu}+\omega^{\eta\alpha}\Gamma^\beta{}_{\eta\mu}\]
to obtain
\[  (\del_\alpha\del_\gamma f)g^{\mu\nu}\left(\omega^{\alpha\beta}\Gamma^\gamma{}_{\beta\nu,\mu}+\omega^{\beta\eta}\Gamma^\alpha{}_{\eta\mu}\Gamma^\gamma{}_{\beta\nu}-\omega^{\alpha\beta}\Gamma^\kappa{}_{\beta\nu}\Gamma^\gamma{}_{\kappa\mu}\right)\]
using $g^{\mu\nu}$ symmetric to massage the last term. The middle term vanishes as it is antisymmetric in $\alpha,\gamma$ and the remaining two terms together with the above terms from $\Gamma^\gamma{}_{\mu\nu;\beta}$ combine to give $(\del_\alpha\del_\gamma f)g^{\mu\nu}\omega^{\alpha\beta}R^\gamma{}_{\nu\mu\beta}$. This gives our 2-derivative corrections at order $\lambda$ as 
\[ {\lambda\over 2}(\del_\alpha\del_\gamma f)\omega^{\alpha\beta}(S^\gamma{}_{;\beta}-{\rm Ric}^\gamma{}_\beta).\]
We then combine our results to the expression stated. 
\endproof

\subsection{Quantum Riemann and Ricci curvatures}\label{QRiemRicci} \label{riccisec}

The quantum Riemann curvature in noncommutative geometry is defined by 
\begin{equation}\label{Rnabla1} {\rm Riem}_1=(\extd\tens_1\id-(\wedge_1\tens_1\id)(\id\tens_1\nabla_1))\nabla_1\end{equation}
and we start by obtaining an expression for it to semiclassical order in terms of tensors. It will be convenient to define components by 
\[ {\rm Riem}_1(\extd x^\alpha):= -{1\over 2} \widehat{R}_1{}^\alpha{}_{\beta\mu\nu}\extd x^\mu\wedge\extd x^\nu\tens_1\extd x^\beta:=-\frac{1}{2}\widetilde{R_1}{}^\alpha{}_{\beta\mu\nu}\bullet(\extd x^\mu\wedge\extd x^\nu)\tens_1\extd x^\beta\]
\[ \widetilde{R_1}{}^\alpha{}_{\beta\mu\nu}= \widehat{R}_1{}^\alpha{}_{\beta\mu\nu}+{\lambda\over 2}\omega^{\delta\gamma}\left(\widehat{R}{}^\alpha{}_{\beta\eta\nu,\delta}\Gamma^\eta{}_{\gamma\mu}
+\widehat{R}{}^\alpha{}_{\beta\mu\eta,\delta}\Gamma^\eta{}_{\gamma\nu}\right)\]
depending on how the coefficients enter.  If we write $\Gamma_1=\widehat\Gamma+ {\lambda\over 2}\widehat\gamma$ then
\begin{eqnarray}\label{Riem1}
{\rm Riem}_1 (\extd x^\alpha) &=& (\extd\tens_1\id-(\wedge_1\tens_1\id)(\id\tens_1\nabla_1))\nabla_1(\extd x^\alpha)\nonumber
\\
&=& - (\extd\tens_1\id-(\wedge_1\tens_1\id)(\id\tens_1\nabla_1))\Gamma_1{}^\alpha{}_{\mu\beta} \extd x^\mu \tens_1 \extd x^\beta \nonumber
\\
&=& - \left( \Gamma_1{}^\alpha{}_{\mu\beta,\nu} \extd x^\nu \wedge \extd x^\mu \tens_1 \extd x^\beta + (\Gamma_1{}^\alpha{}_{\mu\gamma} \extd x^\mu) \wedge_1( \Gamma_1{}^\gamma{}_{\nu\beta} \extd x^\nu) \tens_1 \extd x^\beta \right) \nonumber
\\ \nonumber
&=& - \left( \widehat\Gamma^\alpha{}_{\nu\beta,\mu} \extd x^\mu \wedge \extd x^\nu \tens_1 \extd x^\beta + (\widehat\Gamma^\alpha{}_{\mu\gamma} \extd x^\mu) \wedge_1 (\widehat\Gamma^\gamma{}_{\nu\beta} \extd x^\nu) \tens_1 \extd x^\beta \right) 
\\ \nonumber
&& + \frac{\lambda}{2} \left( \widehat\gamma^\alpha{}_{\mu\beta,\nu} + \widehat\Gamma^\alpha{}_{\nu\gamma} \widehat\gamma^\gamma{}_{\mu\beta} - \widehat\gamma^\alpha{}_{\mu\gamma} \widehat\Gamma^\gamma{}_{\nu\beta} \right) \extd x^\mu \wedge \extd x^\nu \tens_1 \extd x^\beta \nonumber
\\
&=& - \frac{1}{2} \widehat{R}^\alpha{}_{\beta\mu\nu} \extd x^\mu \wedge \extd x^\nu \tens_1 \extd x^\beta  + \frac{\lambda}{2} \widehat\gamma^\alpha{}_{\mu\beta\hat{;}\nu} \extd x^\mu \wedge \extd x^\nu \tens_1 \extd x^\beta \nonumber
\\
&& -\frac{\lambda}{2} \omega^{\eta\zeta} \nabla_\eta \left( \widehat\Gamma^\alpha{}_{\mu\gamma} \extd x^\mu \right) \wedge \nabla_\zeta \left(\widehat\Gamma^\gamma{}_{\nu\beta} \extd x^\nu \right) \tens_1 \extd x^\beta  \nonumber
\\
&& - \lambda  \widehat\Gamma^\alpha{}_{\mu\gamma} \widehat\Gamma^\gamma{}_{\nu\beta} H^{\mu\nu} \tens_1 \extd x^\beta
\end{eqnarray}
to $O(\lambda^2)$, where $\hat{;}$ is with respect to the Levi-Civita connection. The term $\widehat\gamma^\alpha{}_{\gamma\beta}\widehat\Gamma^\gamma{}_{\nu\mu}$ does not contribute due to the antisymmetry of the wedge product. This implies for the components 
\begin{eqnarray*}\widehat{R}_1{}^\alpha{}_{\beta\mu\nu}&=&\widehat{R}^\alpha{}_{\beta\mu\nu}+\lambda\left({1\over 2}(\widehat\gamma^\alpha{}_{\nu\beta\hat{;}\mu}-\widehat\gamma^\alpha{}_{\mu\beta\hat{;}\nu})+\omega^{\eta\zeta}(\widehat\Gamma^\alpha_{\mu\gamma,\eta}-\widehat\Gamma^\alpha{}_{\kappa\gamma}\Gamma^\kappa{}_{\eta\mu})(\widehat\Gamma^\gamma_{\nu\beta,\zeta}-\widehat\Gamma^\gamma{}_{\kappa\beta}\Gamma^\kappa{}_{\zeta\nu})\right.\\
&&\kern60pt \left. +{\omega^{\eta\kappa}\over 2}\widehat{\Gamma}^\alpha{}_{\eta\gamma}\widehat{\Gamma}^\gamma{}_{\zeta\beta}( R^\zeta{}_{\mu\nu\kappa}-R^\zeta{}_{\nu\mu\kappa}-T^\zeta{}_{\mu\nu;\kappa})\right)
\end{eqnarray*}
where we inserted a previous formula for $H$ in terms of the curvature and torsion of $\nabla$. One can similarly read off $\widehat{\gamma}$ from the quantum Levi-Civita connection in Lemma~\ref{Gamma1}.

Next, following \cite{BegMa5}, we consider the classical map $i:\Omega^2\to \Omega^1\tens\Omega^1$ that sends a 2-form to an antisymmetric 1-1 form in the obvious way.

\begin{proposition}\label{i1} The map $i$ quantises to a bimodule map such that $\wedge_1 i_1=\id$ to $O(\lambda^2)$ by
\[ i_1(\extd x^\mu\wedge\extd x^\nu)={1\over 2}(\extd x^\mu\tens_1\extd x^\nu- \extd x^\nu\tens_1\extd x^\mu) +\lambda I(\extd x^\mu\wedge\extd x^\nu)\]
for any tensor map $I(\extd x^\mu\wedge\extd x^\nu)=I^{\mu\nu}{}_{\alpha\beta}\extd x^\alpha\tens \extd x^\beta$ where the tensor $I$ is antisymmetric in $\mu,\nu$ and symmetric in $\alpha,\beta$. The functorial choice here is
\[ I^{\mu\nu}{}_{\alpha\beta}=-{1\over 4}\omega^{\kappa\tau}(\Gamma^\mu{}_{\kappa\alpha}\Gamma^\nu{}_{\tau\beta}+\Gamma^\mu{}_{\tau\alpha}\Gamma^\nu{}_{\kappa\beta}).\]
 \end{proposition}
\proof The functorial construction in \cite{BegMa6} gives $i_Q:\Omega^2\to \Omega^1\tens_1\Omega^1$ necessarily obeying $\wedge_Qi_Q=\id$. Here $\nabla(i)=0$ since the connection on $\Omega^2$ is descended from the connection on $\Omega^1\tens\Omega^1$ so that $i_Q=q^{-1}i$ on identifying the vector spaces. This gives the expression stated for the canonical $I$ and this also works for $\wedge_1$ since this on ${1\over 2}(\extd x^\mu\tens_1\extd x^\nu-\extd x^\nu\tens_1\extd x^\mu)$ differs from $\wedge_Q$ by ${\lambda\over 2}(H^{\mu\nu}-H^{\nu\mu})=0$. Finally,  if we change the canonical $I$  to any other tensor with the same symmetries then its wedge is not changed and we preserve all required properties.  Note that canonical choice can also be written as 
\begin{equation}\label{i1H} i_1(\extd x^\mu\wedge_1 \extd x^\nu)={1\over 2}(\extd x^\mu\tens_1\extd x^\nu - \extd x^\nu\tens_1\extd x^\mu)-{\lambda\over 2}\omega^{\alpha\beta}\Gamma^\mu{}_{\alpha\kappa}\Gamma^\nu{}_{\beta\tau}\extd x^\tau\tens_1\extd x^\kappa+\lambda i(H^{\mu\nu})\end{equation}
when we allow for the relations of $\wedge_1$. \endproof

Now we can follow \cite{BegMa5} and use $i_1$ to lift the first output of ${\rm Riem}_1$ and take a trace of this to compute the quantum Ricci tensor. To take the trace it is convenient, but not necessary, to use the quantum metric and its inverse, so
\begin{equation}\label{Ricnabla1} {\rm Ricci_1}=((\ ,\ )_1\tens_1\id\tens_1\tens\id)(\id\tens_1 i_1\tens_1\id)(\id\tens_1 {\rm Riem}_1)(\cg_1)\end{equation}

We now calculate ${\rm Ricci_1}$ from \eqref{Ricnabla1} taking first the `classical' antisymmetric lift 
\[ i_1(\extd x^\mu\wedge\extd x^\nu)={1\over 2}( \extd x^\mu\tens_1 \extd x^\nu- \extd x^\nu\tens_1\extd x^\mu)\]
corresponding to $I=0$. Then using the second form of the components of ${\rm Riem}_1$,
\begin{eqnarray*}
{\rm Ricci}_1 &=&((\ ,\ )_1\tens_1\id\tens_1\tens\id)(\id\tens_1 i_1\tens_1\id)(\id\tens_1 {\rm Riem_1})(\cg_1)
\\
&=& -\frac{1}{2}((\ ,\ )_1\tens_1\id\tens_1\tens\id)(\id\tens_1 i_1\tens_1\id) (\extd x^\alpha \bullet \tilde g_{\alpha\beta} \tens_1 \widetilde{R_1}^\beta{}_{\gamma\mu\nu} \bullet( \extd x^\mu \wedge \extd x^\nu) \tens_1 \extd x^\gamma)
\\
&=& -\frac{1}{2} ((\ ,\ )_1\tens_1\id\tens_1\tens\id) (\extd x^\alpha \tens_1 \tilde g_{\alpha\beta} \bullet \widetilde{R_1}^\beta{}_{\gamma\mu\nu} \bullet i_1 (\extd x^\mu \wedge \extd x^\nu) \tens_1 \extd x^\gamma)
\\
&=& -\frac{1}{2} (\extd x^\alpha, \tilde g_{\alpha\beta} \bullet \widetilde{R_1}^\beta{}_{\gamma\mu\nu} \bullet \extd x^\mu)_1 \bullet \extd x^\nu \tens_1 \extd x^\gamma
\\
&=& -\frac{1}{2} \tilde g^{\alpha\mu} \bullet \tilde g_{\alpha\beta} \bullet \widetilde{R_1}^\beta{}_{\gamma\mu\nu} \bullet \extd x^\nu \tens_1 \extd x^\gamma -\frac{1}{2} (\extd x^\alpha, [\widehat{R}_{\alpha\gamma\mu\nu}, \extd x^\mu]) \extd x^\nu \tens_1 \extd x^\gamma
\end{eqnarray*}
In the fourth line we used the fact that the Riemann tensor is already antisymmetric in $\mu$ and $\nu$. Note that $\tilde g^{\alpha\mu} \bullet \tilde g_{\alpha\beta} =(\tilde g^{\mu\alpha}+\lambda \tilde h^{\mu\alpha})\bullet \tilde g_{\alpha\beta}= \delta^\mu{}_\beta+\lambda\tilde h^\mu{}_\beta$ to $O(\lambda^2)$ where we lower an index using the classical metric. Meanwhile, putting in  general $I$ adds a term
\[ -{\lambda\over 2} \widehat{R}_{\alpha\gamma\mu\nu} ((\ ,\ )\tens\id\tens\tens\id) (\extd x^\alpha \tens_1 I ( \extd x^\mu \wedge \extd x^\nu ) \tens_1 \extd x^\gamma)\]
and we therefore obtain
\begin{eqnarray} \label{Ricci1}
{\rm Ricci}_1 &=& -\frac{1}{2} \widetilde{R_1}^\mu{}_{\gamma\mu\nu} \bullet \extd x^\nu \tens_1 \extd x^\gamma - \frac{\lambda}{2}\left( \tilde h^{\mu\beta} \widehat{R}_\beta{}_{\gamma\mu\nu} - \omega^{\eta\zeta} g^{\alpha\beta} \widehat{R}_{\alpha\gamma\mu\nu,\eta} \Gamma^\mu{}_{\zeta\beta} \right) \extd x^\nu \tens_1 \extd x^\gamma \nonumber \\
&&-{\lambda\over 2}\widehat{R}^\alpha{}_{\gamma\eta\zeta}I^{\eta\zeta}{}_{\alpha\nu}\extd x^\nu\tens_1 \extd x^\gamma
\end{eqnarray}
The idea of \cite{BegMa5} is then to use the freedom in $I$ to arrange that  
\[ \wedge_1({\rm Ricci}_1)=0,\quad {\rm flip}(*\tens *){\rm Ricci}_1={\rm Ricci_1}\]
 to order $\lambda$ so that ${\rm Ricci}_1$ enjoys the same quantum symmetry and `reality' properties (to order $\lambda$) as $\cg_1$. (A further `reality' condition on the map $i_1$ in \cite{BegMa5} just amounts in our case to the entries of the tensor $I$ being real.) If we write components
\[ {\rm Ricci}_1:=  -\frac{1}{2} \widetilde{R_1}_{\mu\nu}\bullet\extd x^\nu\tens_1\extd x^\mu= -\frac{1}{2}\extd x^\nu\bullet {\widetilde{\widetilde{R_1}}}_{\nu\mu}\tens_1\extd x^\mu\]
then (\ref{Ricci1}) is equivalent to
\begin{equation}\label{Ricci1tens}  \widetilde{R_1}_{\mu\nu}=\widetilde{R_1}^\alpha{}_{\mu\alpha\nu} + \lambda\left( \tilde h^{\alpha\beta} \widehat{R}_\beta{}_{\mu\alpha\nu} - \omega^{\eta\zeta} g^{\alpha\beta} \widehat{R}_{\alpha\mu\kappa\nu,\eta} \Gamma^\kappa{}_{\zeta\beta}+\widehat{R}^\alpha{}_{\mu\eta\zeta}I^{\eta\zeta}{}_{\alpha\nu} \right)\end{equation}
and
\[ {\widetilde{\widetilde{R_1}}}_{\nu\mu}=\widetilde{R_1}_{\mu\nu}-\lambda\omega^{\alpha\beta}\widehat{R}_{\mu\delta,\alpha}\Gamma^\delta{}_{\beta\nu}\]
respectively, where ${\widehat R}_{\mu\nu}$ is the classical Ricci tensor. This second version is useful for the quantum reality condition, which says that if  we write ${\widetilde{\widetilde{\rm R_1}}}_{\mu\nu}=\widehat R_{\mu\nu}+\lambda \rho_{\mu\nu}$ then the quantum correction $\rho_{\mu\nu}$  is required to be antisymmetric. Remember that this will have contributions from $\widetilde{R_1}$ as  well as the terms directly visible in (\ref{Ricci1tens}). 

We then define the quantum Ricci scaler as
\begin{equation}\label{S1} S_1=(\ ,\ )_1{\rm Ricci}_1=-{1\over 2}\widetilde{R_1}_{\mu\nu}\bullet \tilde g^{\nu\mu}\end{equation}
which does not depend on the lifting tensor $I$ due to the antisymmetry of the first two indices of $\widehat{R}$. There does not appear to be a completely canonical choice of Ricci in noncommutative geometry as it depends on the choice of lifting for which we have not done a general analysis, but this constructive approach allows us to begin to explore it. The reader should note that the natural conventions in our context reduce in the classical limit to $-{1\over 2}$ of the usual Riemann and Ricci curvatures, which we have handled by putting this factor into the definition of the tensor components so that these all have limits that match standard conventions.

\subsection{Laplacian in the bicrossproduct model} \label{bicross}

As a check of all our formulae above we will compare our analysis with the bicrossproduct model \cite{BegMa5} where the noncommutative geometry has been fully computed. The 2D version has coordinates $t,r$ with $r$ invertible and Poisson bracket $\{r,t\}=r$ or $\omega_{10}=r$ in the coordinate basis. The work \cite{BegMa5} used $r$ rather than $x$ as this is also the radial geometry of a higher-dimensional model. The Poisson-compatible `quantising' connection is given by $\Gamma^0{}_{01}=-r^{-1}, \Gamma^0{}_{10}=r^{-1}$ or in abstract terms $\nabla\extd r=0$ and $\nabla\extd t=r^{-1}(\extd t\tens\extd r-\extd r\tens\extd t)$. Letting $v=r\extd t-t\extd r$, we have $\nabla \extd r=\nabla v=0$ so a pair of central 1-forms $v,\extd r$ at least at first order. We take classical metric $g=\extd r\tens\extd r+b v\tens v$ where $b$ is a nonzero real parameter which clearly has inverse $(\extd r,\extd r)=1$, $(v,v)=b^{-1}$, $(\extd r,v)=(v,\extd r)=0$. The classical Riemannian geometry is that of either a strongly gravitating particle or an expanding universe as explained in \cite{BegMa5} according to the sign of $b$. The Levi-Civita connection is 
\[\widehat \nabla v = -{2 v\over r}\tens \extd r,\quad \widehat\nabla\extd r={2 b v\over r}\tens v\]
This model has trivial curvature of $\nabla$ but in other respects is a good test of our formulae with nontrivial torsion and contorsion and curvature of the Levi-Civita connection. The contorsion tensor can be written\cite{BegMa5}
\[ S^\kappa{}_{\alpha\beta}=2b\eps_{\alpha\mu}x^\mu\eps_{\beta\nu}g^{\kappa\nu},\quad S^\mu = 2 {x^\mu \over r^2}\]
where $x^0=t$ and $x^1=r$ and $\eps_{01}=1$ is antisymmetric. We write 
\[ \extd f= f_{,r}\extd r+ f_{,t}\extd t= (\del_r f)\extd r+(\del_v f)v;\quad \del_v f= {1\over r}f_{,t},\quad \del_r f=f_{,r}+{t\over r}f_{,t}\]
Then 
\begin{eqnarray*}\square f&=&(\ ,\ )\widehat\nabla \extd f=(\ ,\ )((\del_r f)\widehat\nabla \extd r+ (\del_v f)\widehat\nabla v+\extd\del_r f\tens\extd r+\extd \del_v f\tens v)\\
&=&{2\over r}(\del_rf) +\del_r^2 f+ b^{-1}\del_v^2 f
\end{eqnarray*}
is the classical Laplacian. 

Now we repeat the same computation working in the quantum algebra. We let $\nu=r\bullet\extd t-t\bullet \extd r=v+{\lambda\over 2}\extd r$ and the (full) quantum metric, inverse quantum metric and quantum Levi-Civita connection in \cite{BegMa5} are
\[ g_1=(1+b\lambda^2)\extd r\tens_1\extd r+b\nu\tens_1\nu-b\lambda \nu\tens_1\extd r\]
\[ (\nu,\nu)_1=b^{-1},\quad (\extd r,\nu)_1=0,\quad (\nu,\extd r)_1={\lambda\over 1+ b\lambda^2},\quad (\extd r,\extd r)_1={1\over 1+b\lambda^2}\]
\[ \nabla_1\extd r={8 b\over r(4+7b\lambda^2)}v\tens_1\nu - {12 b\lambda \over r(4+7b\lambda^2)}v\tens_1\extd r\]
\[ \nabla_1\nu=-{4 b \lambda \over r(4+7b\lambda^2)}v\tens_1\nu - {8 (1+b\lambda^2)) \over r(4+7b\lambda^2)}v\tens_1\extd r\]
(there is a typo in the coefficient of $\beta'$ in \cite{BegMa5}). Next note that because $\nabla v=\nabla\extd r=0$, we do not have any corrections to the products of anything with these basic 1-forms and this allows us to equally well write
\[ \extd f= (\del_r f)\bullet \extd r+ (\del_v f)\bullet v\]
with the classical derivatives if we use this basis. Then working to $O(\lambda)$ for the quantum connection (which agrees with the one found in \cite[Prop 7.1]{BegMa6}), 
\begin{eqnarray*}\square_1 f&=&(\ ,\ )_1\nabla_1 \extd f=(\ ,\ )_1( (\del_r f)\bullet\nabla_1 \extd r+ (\del_v f)\bullet\nabla_1 v+\extd \del_r f\tens_1 \extd r+\extd\del_v f \tens_1 v)\\
&=&(\ ,\ )_1((\del_r f)\bullet{2 b\over r} v\tens_1 v-(\del_r f)\bullet{2 b\lambda\over r}v\tens_1\extd r-(\del_v f)\bullet{2\over r}v\tens_1\extd r-(\del_v f)\bullet{2 b\lambda\over r}v\tens_1 v  \\ &&\quad +(\del^2_r f)\extd r\tens_1 \extd r+(\del_v\del_r f) v\tens_1\extd r+(\del_r\del_v f)\extd r\tens_1 v+(\del_v^2 f )v\tens_1 v)\\
&=&(\del_r f){2 b\over r} b^{-1}+ {\lambda\over 2}\del_v\del_r f 2b b^{-1}-(\del_v f){2\over r}{\lambda\over 2}-(\del_v f){2 b\lambda\over r} b^{-1} \\
&&\quad  +(\del^2_r f)+(\del_v\del_r f) {\lambda\over 2}-(\del_r\del_v f){\lambda\over 2}+(\del_v^2 f ) b^{-1}  \\ 
&=&\square f+\lambda(-{3\over r}\del_v f +{1\over 2}[\del_v,\del_r]f+ \del_v\del_r f )\\
&=&\square f+\lambda\left({1\over r}{\del\over\del t}{\del\over\del r} f+ {t\over r^2}{\del^2\over\del t^2}f -{1\over r^2}{\del\over\del t}f\right)
\end{eqnarray*}
where we used $h\bullet r^{-1}=hr^{-1}+{\lambda\over 2}\del_v h$ for any function $h$ and, to $O(\lambda)$,
\[ (v,v)_1=b^{-1},\quad (\extd r,v)_1=-{\lambda\over 2},\quad (v,\extd r)_1={\lambda\over 2},\quad (\extd r,\extd r)_1=1. \]
Now we compare these results with the general theory of Section~2.1. Working now in the $t,r$ coordinate basis, the metric tensor is 
\[ g_{\mu\nu}=\begin{pmatrix}b r^2  &- b r t \\ -b r t & 1+b t^2\end{pmatrix},\quad g^{\mu\nu}=\begin{pmatrix}{1+b t^2\over b r^2}  & {t\over r}\\ {t\over r} & 1\end{pmatrix}\]
which gives $\CR_{10}=br$ and hence $\CR=\CR_{10}\extd t\wedge\extd r=bv\wedge\extd r$ as in \cite[Sec 7.1]{BegMa6}. Meanwhile expanding $g_1$ to $O(\lambda)$ we have
\begin{eqnarray*}g_1&=&\extd r\tens_1\extd r+bv\tens_1 v+b{\lambda\over 2}(\extd r\tens_1 v-v\tens_1\extd r)\\
&=&g_{\mu\nu}\extd x^\mu\tens_1\extd x^\nu+ b{\lambda\over 2}\extd r\tens_1 v- \lambda b v\tens_1\extd r\\
&=&g_{\mu\nu}\extd x^\mu\tens_1\extd x^\nu+{\lambda\over 2}bt\extd r\tens_1\extd r+{\lambda\over 2}br\extd r\tens_1\extd t-\lambda br \extd t\tens_1\extd r\\
&=&g_{\mu\nu}\extd x^\mu\tens_1\extd x^\nu+{\lambda\over 2}\CR_{01}(\extd t\tens_1\extd r-\extd r\tens_1\extd t)+{\lambda\over 2}\omega^{10}g_{\mu0}\Gamma^0{}_{10}\Gamma^0{}_{01}\extd x^\mu\tens_1\extd r
\end{eqnarray*}
which is exactly the content of (\ref{g1}). We used $r\extd t =r\bullet\extd t-{\lambda\over 2}\extd r$ to move all coefficients to the left in order to compare. We can do a similar trick with $r\extd t=(\extd t)\bullet r+{\lambda\over 2}\extd r$ to put the coefficients in the middle, giving
\[ g_1=\extd r\tens_1\extd r+b(\extd t)\bullet r^2\tens_1\extd t+b(\extd r)\bullet t^2\tens_1\extd r-b(\extd r)\bullet t\bullet r\tens_1 \extd t-b(\extd t)\bullet r\bullet t\tens_1\extd r+b\lambda(\extd r\tens_1 v-v\tens_1\extd r)\]
giving
\[ \tilde g_{\mu\nu}=g_{\mu\nu}+{\lambda\over 2} \begin{pmatrix} 0 &-3br  \\  3br& 0 \end{pmatrix}.\]
Comparing with the formula (\ref{h}), we have
\[ h_{01}=\CR_{01}+ g_{00}\omega^{10}\Gamma^0{}_{10}\Gamma^0{}_{01}+\omega^{10}\Gamma^0{}_{10}g_{01,0}=-3br\]
and similarly $h_{10}=3br$, exactly as required. 

We now similarly check the formula (\ref{Gamma1}) for $\Gamma_1$. The classical Levi-Civita here is\cite[Appendix]{BegMa5}
\[ \widehat\Gamma^0_{\mu\nu}=\begin{pmatrix}-2b t & r^{-1}(1+2bt^2)\\ r^{-1}(1+2bt^2) & -2r^{-2}t(1+bt^2)\end{pmatrix},\quad 
\widehat\Gamma^1_{\mu\nu}=\begin{pmatrix}-2b r & 2bt \\ 2bt & -2b r^{-1}t^2\end{pmatrix}\]
Working  to $O(\lambda)$, the quantum connection from \cite[Prop 7.1]{BegMa6} is
\begin{eqnarray*}\nabla_1\extd r&=&{2 b\over r}(v\tens_1 v-\lambda v\tens_1\extd r)\\
&=& {2 b\over r}v\tens_1( r\bullet \extd t-t\bullet\extd r-{\lambda\over 2}\extd r) -\lambda {2 b\over r}v\tens_1\extd r \\
&=& 2 b v\tens_1 \extd t-2b v(r^{-1}\bullet t)\tens_1\extd r -\lambda {3 b\over r}v\tens_1\extd r \\
&=&-\widehat\Gamma^1_{\mu\nu}\extd x^\mu\tens_1\extd x^\nu-2\lambda b(\extd t-r^{-1}t\extd r)\tens_1\extd r 
\end{eqnarray*}
similarly to the computation of $v\tens_1 v$ as above, which agrees with the correction in $\Gamma_1{}^1{}_{\mu\nu}$ of
\begin{eqnarray*} &&  \frac{\lambda}{2} \omega^{\alpha\beta} \widehat\Gamma^1{}_{\mu0,\alpha}\Gamma^0{}_{\beta\nu} -{\lambda\over 2}\omega^{\alpha\beta}\widehat\Gamma^1{}_{00}\Gamma^0{}_{\alpha\mu}\Gamma^0{}_{\beta\nu}  -{\lambda\over 2}\omega^{01}S^1{}_{0\kappa} \nabla_1 S^{\kappa}{}_{\mu\nu}\\
&&\quad =-{\lambda\over 2}\widehat\Gamma^1{}_{\mu0,\nu} -\lambda b \eps_{\mu\nu}-{\lambda\over 2}r \widehat\Gamma^1{}_{0\kappa} \nabla_1 S^{\kappa}{}_{\mu\nu}=2\lambda b\begin{pmatrix}0 & 1 \\ 0 & -r^{-1}t\end{pmatrix}
\end{eqnarray*}
in Lemma~\ref{Gamma1}. Here \[ \nabla_1 S^0{}_{\mu\nu}= \left(
\begin{array}{cc}
 -\frac{2 b t}{r} & \frac{2 (1+b t^2)}{r^2} \\
 \frac{2 b t^2}{r^2} & -\frac{2 t \left(1+b t^2\right)}{r^3} \\
\end{array}
\right) ,\quad \nabla_1 S^1{}_{\mu\nu}=\left(
\begin{array}{cc}
 -2 b & \frac{2 b t}{r} \\
 \frac{2 b t}{r} & -\frac{2 b t^2}{r^2} \\
\end{array}
\right)\]
where $\nabla_1$ in this context means with respect to $r$. Similarly, the semiquantum connection $\nabla_1 v=-{2\over r}(v\tens_1\extd r+b\lambda v\tens_1 v)$ implies
\begin{eqnarray*}\nabla_1\extd t&=&\nabla_1(r^{-1}\bullet v+(r^{-1}t)\bullet\extd r)=\extd r^{-1}\tens_1 v+\extd(r^{-1}t)\tens_1\extd r+r^{-1}\bullet\nabla_1 v+(r^{-1}t)\bullet\nabla_1\extd r\\
&=& -r^{-2}\extd r \tens_1 (r\bullet \extd t-t\bullet\extd r-{\lambda\over 2}\extd r)-r^{-2}t\extd r\tens_1\extd r+r^{-1}\extd t\tens_1\extd r\\
&&-2 r^{-2}(v\tens_1\extd r+b\lambda v\tens_1 v)+2b(r^{-1}t)\bullet r^{-1}(v\tens_1v-\lambda v\tens_1\extd r) \\
&=& -r^{-1}\extd r \tens_1\extd t + (r^{-2}\bullet t)\extd r\tens_1\extd r+{\lambda\over 2}r^{-2}\extd r\tens_1 \extd r-r^{-2}t\extd r\tens_1\extd r+r^{-1}\extd t\tens_1\extd r\\
&&-2 r^{-2}v\tens_1\extd r-\lambda b r^{-2} v\tens_1 v+2br^{-2}tv\tens_1(r\bullet \extd t-t\bullet\extd r-{\lambda\over 2}\extd r)-\lambda 2br^{-2}t v\tens_1\extd r \\
&=& -r^{-1}\extd r \tens_1\extd t +r^{-2}t \extd r \tens_1\extd r-{\lambda\over 2}r^{-2}\extd r\tens_1 \extd r-r^{-2}t\extd r\tens_1\extd r+r^{-1}\extd t\tens_1\extd r\\
&&-2 r^{-2}v\tens_1\extd r-\lambda b r^{-2} v\tens_1 v+2b(r^{-2}t)\bullet rv\tens_1\extd t-2b(r^{-2}t)\bullet t v\tens_1 \extd r-\lambda 3br^{-2}t v\tens_1\extd r \\
&=&-\widehat\Gamma^0{}_{\mu\nu}\extd x^\mu\tens_1\extd x^\nu-{\lambda\over 2}r^{-2}\extd r \tens_1 \extd r -\lambda b r^{-2}v\tens_1 v- \lambda  br^{-1}v\tens_1\extd t-\lambda br^{-2}t v\tens_1\extd r \\
&=&-\widehat\Gamma^0{}_{\mu\nu}\extd x^\mu\tens_1\extd x^\nu-{\lambda\over 2}r^{-2}\extd r \tens_1 \extd r -2 \lambda b (\extd t-r^{-1}t\extd r)\tens_1\extd t 
\end{eqnarray*}
which agrees with the correction to $\Gamma_1{}^0{}_{\mu\nu}$ of 
\begin{eqnarray*}&& \frac{\lambda}{2} \omega^{\alpha\beta} \widehat\Gamma^0{}_{\mu0,\alpha}\Gamma^0{}_{\beta\nu}   -{\lambda\over 2}\omega^{\alpha\beta}\widehat\Gamma^0{}_{00}\Gamma^0{}_{\alpha\mu}\Gamma^0{}_{\beta\nu}-{\lambda\over 2}\omega^{01}S^0{}_{0\kappa} \nabla_1 S^{\kappa}{}_{\mu\nu}\\
&&\quad =-{\lambda\over 2}\widehat\Gamma^0{}_{\mu0,\nu} -\lambda b r^{-1}t\eps_{\mu\nu}-{\lambda\over 2}r \widehat\Gamma^0{}_{0\kappa} \nabla_1 S^{\kappa}{}_{\mu\nu}=\lambda\left(
\begin{array}{cc}
 2 b & 0 \\
 -\frac{2 b t}{r} & \frac{1}{2 r^2} \\
\end{array}
\right)\end{eqnarray*}
in Lemma~\ref{Gamma1}.

Finally, the contracted contorsion tensor obeys
\[ S^\mu{}_{;0}=0,\quad S^\mu{}_{;1}=- 2{x^\mu\over r^3}\]
while the curvature of $\nabla$ vanishes. Hence the Laplacian in Theorem~\ref{square1} is 
\begin{eqnarray*}\square_1 f&=&\square f+{ \lambda\over 2}\omega^{10}(-S^\mu{}_{;1})(\widehat\nabla_0\extd f)_\mu =\square f+{\lambda\over r^2}x^\mu(\widehat\nabla_0 \extd f)_\mu\end{eqnarray*}
which when expanded out using the values of $\Gamma^0_{\mu\nu}$ agrees with our previous algebraic computation from within the noncommutative geometry in \cite{BegMa5}. This both checks all of our tensor formulae in a coordinate basis and shows how to change quantum variables given that the noncommutative algebra was all done with $v,\extd r$. 

When $b<0$ the interpretation is that of a strong gravitational source and curvature singularity at $r=0$. Being conformaly flat after a change of variables to $r'=1/r, t'=t/r$ the massless waves or zero eigenfunctions of the classical Laplacian are plane waves in $t',r'$ space of the form $e^{\imath \omega t'}e^{\pm{\imath \omega r'\over\sqrt{-b}}}$ or 
\[ \psi^\pm_\omega(t,r)=e^{\imath {\omega t\over r}} e^{\pm \imath{ \omega\over r \sqrt{-b}}}\]
while the massive modes are harder to describe due to the conformal factor. One can similarly solve the expanding universe case where $b>0$ and the interpretation of the $r,t$ variables is swapped. In the deformed case the same change of variables and separating off $\psi=e^{\imath\omega t'}f$ gives an equation for null modes 
\[ \left({\del^2 \over \del r'{}^2} + \lambda_P\omega \left({2\over r'}+ {\del \over\del r'}\right) - {\omega^2 \over b}\right)f=0\]
where $\lambda=\imath\lambda_P$, which is solved by 
\begin{eqnarray*} 
f &=& r' e^{-\frac{1}{2}\omega r'\left( \lambda_P+\frac{\imath \sqrt{b\lambda_P^2+4}}{\sqrt{-b}} \right)} \left( A M(1-\frac{\imath\sqrt{-b}\lambda_P}{\sqrt{b\lambda_P^2+4}},2,\imath\sqrt{b\lambda_P^2+4} \frac{\omega r'}{\sqrt{-b}}) \right.
\\
&& \left. + B U( 1-\frac{\imath\sqrt{-b}\lambda_P}{\sqrt{b\lambda_P^2+4}},2,\imath\sqrt{b\lambda_P^2+4} \frac{\omega r'}{\sqrt{-b}}) \right)  
\end{eqnarray*}
for constants $A$ and $B$. $M(a,b,z)$ and $U(a,b,z)$ denote the Kummer $M$ and $U$ functions  (or  hypergeometric ${}_1F_1$, $U$ respectively in Mathematica). In the limit $\lambda_P\to 0$, this becomes 
\[ f = \frac{\imath}{2}\frac{\sqrt{-b}A}{\omega} e^{\imath\frac{\omega r'}{\sqrt{-b}}}+\frac{\imath}{2}\frac{\sqrt{-b}(B-A)}{\omega} e^{-\imath\frac{\omega r'}{\sqrt{-b}}}  \]
which means we recover our two independent solutions $\psi^\pm_\omega$ as a check. Bearing in mind that our equations are only justified to order $\lambda$, we can equally well write
\[ \psi_\omega(t,r)={e^{\imath{\omega t\over r}}\over r} e^{-\imath\frac{\omega }{r \sqrt{-b}}\left( 1-{\imath\sqrt{-b}\lambda_P\over 2} \right)} \left( A M( 1-\frac{\imath\sqrt{-b}\lambda_P}{2},2,\imath \frac{2\omega}{r\sqrt{-b}})  + B U( 1-\frac{\imath\sqrt{-b}\lambda_P}{2},2,\imath \frac{2\omega }{r\sqrt{-b}}) \right)  \]
and proceed to analyse the behaviour for small $\lambda_P$ in terms of integral formulae. Thus
\[ M(1-a,2,z)={1\over\Gamma(1-a)\Gamma(1+a)}\int^1_0e^{zu}({1-u\over u})^a\extd u= \int^1_0e^{zu}(1+a\ln({1-u\over u}))\extd u +O(a^2) \]
which we evaluate for $z=2 \imath s$ and $s$ real in terms of the function
\[ \tau_M(s)=\imath {M^{(1,0,0)}(1,2,2\imath s)\over M(1,2,2 \imath s)}={\int_0^1e^{2 \imath s u}\ln({1-u\over u})\extd u\over  \imath \int_0^1e^{2 \imath s u}\extd u}\]
shown in Figure~1. This function in the principal region (containing $s=0$) is qualitatively identical to the trig function $-2\tan(s/2)$ but blows up slightly more slowly as $s\to \pm {\pi}$. This gives us  $M(1-a,2,2\imath s)={1 \over 2\imath s}(e^{2 i s}-1)(1+ a\imath\tau_M(s))+O(a^2)$ and hence with $a=\imath {\sqrt{-b}\lambda_P\over 2}$ and $s={\omega\over  r \sqrt{-b}}$, we have up to normalisation
\[ \psi_\omega^M(t,r)=e^{\imath{\omega t\over r}}\sin({\omega\over r\sqrt{-b}})e^{-{\omega\over 2r}\lambda_P\left(1+{r \over \omega}\sqrt{-b}\tau_M({\omega\over  r \sqrt{-b}})\right)}+O(\lambda_P^2),\quad |r|> {|\omega|\over\pi \sqrt{-b}}\]
as one of our independent solutions. Notice that for $\lambda_P\ne 0$ our solution blows up and our approximations break down as $r$ approaches a certain minimum distance as shown to the classical Ricci singularity at $r=0$,  depending on the frequency. This is a geometric `horizon' of some sort (with scale controlled by $\sqrt{-b}$) but frequency dependent, and very different effect from the usual Planck scale bound $|r|>>|\omega|\lambda_P$ needed in any case for our general analysis. Meanwhile for large $|r|$, the effective $\lambda_P$ is suppressed as $\tau_M'(0)=-1$. 

For the other mode, the similar integral
\[ U(1-a,2,z)={1\over \Gamma(1-a)}\int_0^\infty e^{-z u} ({1+u\over u})^a \extd u\]
is not directly applicable as it is not valid on the imaginary axis but we can still proceed in a similar way for the other mode by defining
\[ \tau_U(s)=\imath {U^{(1,0,0)}(1,2,2\imath s)\over U(1,2,2 \imath s)}=T(s)+\imath S(s)\]
where the real function $T(s)$ resembles ${\pi\over 2}\tanh(s)$ (but is vertical at the origin) and $S(s)$ resembles $-\ln(e^{-\gamma}+2|s|)$ as also shown in Figure~1, where $\gamma\approx 0.577$ is the Euler constant. Then $U(1-a,2,2\imath s)={1 \over 2\imath s}(1+ a\imath\tau_U(s))+O(a^2)$ giving up to normalisation
\[ \psi_\omega^U(t,r)=e^{\imath{ \omega t\over r}}e^{-\imath\frac{\omega }{r \sqrt{-b}}}e^{-{\omega\over 2r}\lambda_P\left(1+{r \over \omega}\sqrt{-b}\tau_U({\omega\over  r \sqrt{-b}})\right)}+O(\lambda_P^2)\]
as a second solution. This still has our general Planck scale lower bound needed for the general analysis but no specific geometric bound at finite radius as $\tau_U$ does not blow up and moreover has only a mild log divergence as $s\to \infty$ or $r\to 0$. There is no particular suppression of $\lambda_P$ as $s\to 0$ or $r\to\infty$ and indeed $\tau_U$ tends to a constant nonzero imaginary value (the meaning of which is unclear as it can be absorbed in a normalisation).

\begin{figure}
\[ \includegraphics[scale=.7]{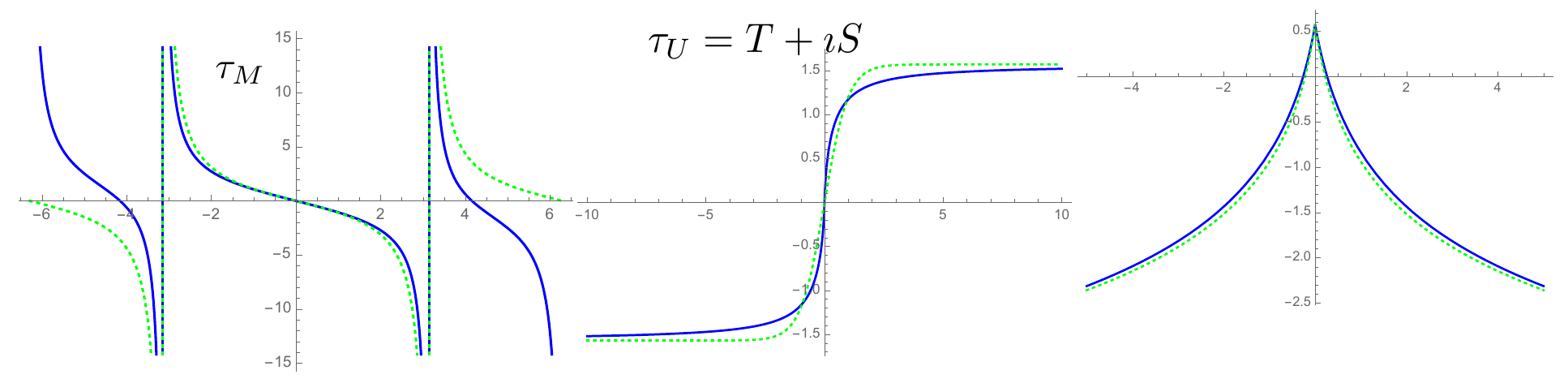}\]
\caption{Functions $\tau_M$ and $\tau_U$ related to differentials of Kummer $M(1-a,2,2\imath s)$ and $U(1-a,2,2\imath s)$ at $a=0$ and similar to $\tan$, $\tanh$ and a shifted $\ln$ (shown dashed for reference)}
\end{figure}

Both of our solutions have been exhibited as deviations from the classical solutions and consequently there can reasonably be expected to be physical predictions in the form of a likely change to the group velocity along the lines of \cite{AmeMa} and of gravitational redshift along the lines of \cite{Ma:alm}. The details of such effects remain to be computed for our curved metric but we have indicated some relevant qualitative features of the modes.

In this model we can in fact write down the full quantum Laplacian in noncommutative geometry in the setting of \cite{BegMa5}. This is not our topic here so we will be brief. Writing \[\extd f= (\del_r f)\bullet \extd r+\del_v f\bullet \nu\]
where $\del_r,\del_v$ are now quantised versions of the ones before and are derivations of the noncommutative algebra since $\extd r, v$ are central. They obey
\[ \del_v f(r)=0,\quad \del_v f(t)=r^{-1}\bullet \del_0 f,\quad \del_rf(r)=f',\quad \del_rf(t)=-r^{-1}\bullet t\del_0 f\]
as easily found using the derivation rule, the values on $r,t$ and the relations $t\bullet r^{-1}=r^{-1}\bullet (t+\lambda)$ in the algebra. Here $\del_0 f(t)= \lambda^{-1}(f(t+\lambda)-f(t))$ is a finite difference. Then for any $f$ in the algebra, one can compute $(\ ,\ )_1\nabla_1\extd f$ using the full expressions above to obtain
\[ \square_1 f= {(8-6b\lambda^2)\del_r f-\lambda(8+4b\lambda^2)\del_v f\over 4+ 7b\lambda^2}\bullet r^{-1}+ {\del_r^2f+\lambda\del_v\del_r f\over 1+b \lambda^2}+b^{-1}\del^2_vf\]
for $\nabla_1$ the quantum Levi-Civita connection stated above for this model. 

The work \cite{BegMa5} already contained the full quantum Ricci as proportional to the quantum metric. The first ingredient for this is the quantum Riemann tensor in \cite{BegMa5} and expanding this gives 
\begin{eqnarray*} {\rm Riem}_1(\extd r)&=&-2 b {1\over r^2}\bullet  \nu\wedge_1\extd r\tens_1 (r\bullet \extd t-t\bullet\extd r)+ {7 b\lambda\over r}\extd t\wedge\extd r\tens\extd r\\
&=&-2 b \extd t\wedge_1\extd r\tens_1\extd t+ {2b}{1\over r^2}\bullet t \bullet r \bullet (\extd t\wedge_1\extd r)\tens_1\extd r+ {7 b\lambda\over r}\extd t\wedge\extd r\tens\extd r\\
&=&-2 b \extd t\wedge\extd r\tens_1\extd t+ {2b}\left({t\over r}+{\lambda\over 2 r^2}\{t,r\}+{\lambda\over 2}\{{1\over r^2},t\}r\right) \bullet (\extd t\wedge\extd r)\tens_1\extd r+ {7 b\lambda\over r}\extd t\wedge\extd r\tens\extd r\\
&=&-2 b \extd t\wedge\extd r\tens_1\extd t+ {2b}({t\over r}) \bullet (\extd t\wedge\extd r)\tens_1\extd r+ {4 b\lambda\over r}\extd t\wedge\extd r\tens\extd r\\
&=&-\frac{1}{2}\widehat{R}^1{}_{\beta\mu\nu} \extd x^\mu \wedge \extd x^\nu \tens_1 \extd x^\beta+{5\lambda} {b\over r} \extd t\wedge\extd r\tens_1 \extd r+O(\lambda^2).
\end{eqnarray*}
where we use $\nu,\extd r$ central for the second equality, then $\extd t\wedge_1\extd r=\extd t\wedge\extd r$ and $\nabla_1(\extd t\wedge \extd r)=-r^{-1}\extd t\wedge\extd r$. There is a similar formula for ${\rm Riem}_1(\extd t)$ obtained from ${\rm Riem}_1(\nu)=r\bullet{\rm Riem}_1(\extd t)- t\bullet{\rm Riem}_1(\extd r)$ given in \cite{BegMa5} and expanding. The result is
\[ {\rm Riem}_1(\extd x^\alpha)=-\frac{1}{2}\widehat{R}^\alpha{}_{\beta\mu\nu} \extd x^\mu \wedge \extd x^\nu \tens_1 \extd x^\beta +{5 b \lambda\over r}\extd t\wedge\extd r\tens_1\extd x^\alpha\]
and agrees with direct computation (using Maple) from the tensorial formula (\ref{Riem1}), which provides a rather non-trivial check of that. 

Next the lifting map $i_1$ was given by the method in \cite{BegMa5} uniquely (by the time the reality property is included) as,
\[ i_1(\nu\wedge_1\extd r)={1\over 2}(\nu\tens_1\extd r-\extd r\tens_1\nu)+{7\lambda\over 4}g_1+O(\lambda^2)\]
according to the order $\lambda$ part of the full calculation in \cite[Sec~6.2.1]{BegMa5} (the 9/4 in \cite[eqn (5.21)]{BegMa5} was an error and should be 7/4).  This is equivalent at semiclassical order to
\[ i_1(\extd t \wedge\extd r)={1\over 2}(\extd t\tens_1\extd r-\extd r\tens_1\extd t)+{7\lambda\over  4 r}g\]
where $\extd t\wedge_1\extd r=\extd t\wedge\extd r$ since $\nabla\extd r=0$ and only $H^{00}$ is non-zero. The first term is the functorial term and the second term is $\lambda I(\extd t\wedge \extd r)$. 
It was then shown in \cite{BegMa5} that ${\rm Ricci}_1=g_1/r^2$. Expanding the quantum metric from \cite{BegMa5} as recalled above, the quantum Ricci is to $O(\lambda^2)$,
\begin{eqnarray*} {\rm Ricci}_1&=&{1\over r^2}\bullet \left(\extd r\tens_1\extd r+br\bullet \nu\tens_1\extd t-b t\bullet\nu\tens_1\extd r-b\lambda \nu\tens_1\extd r \right)\\  
&=&{1\over r^2}\extd r\tens_1\extd r+{b\over  r}(v+{\lambda\over 2}\extd r)\tens_1\extd t- ({b\over r^2}\bullet t)(v+{\lambda\over 2}\extd r)\tens_1\extd r-{b\over r^2}\lambda v\tens_1\extd r \\
&=&-\frac{1}{2}\widehat{R}_{\mu\nu} \extd x^\mu \tens_1 \extd x^\nu-{\lambda\over 2} {b t\over r^2}\extd r\tens_1\extd r+{\lambda\over 2}{b\over r}\extd r\tens_1\extd t=-\frac{1}{2}\widehat{R}_{\mu\nu} \extd x^\mu \tens_1 \extd x^\nu+{\lambda\over 2}{b\over r^2}\extd r\tens_1 v+O(\lambda^2).\end{eqnarray*}
where for the second equality uses $\nabla\extd r=\nabla v=0$ so that bullet products with these are
classical products. For the third equality we used $\{{1\over r^2},t\}=-{2\over r^2}$ from the $\bullet$ which cancels the last term. We obtain exactly this answer by direct calculation from (\ref{Ricci1}) as a nontrivial check on our new general  formulae.

\subsection{Laplacian in the 2D Bertotti-Robinson model}\label{BRsec} By way of contrast we note that the bicrossproduct spacetime algebra has an alternative differential structure for which the full quantum geometry was also already solved, in \cite{MaTao}. We have the same Poisson bracket as above but this time the zero curvature `quantising' connection
$ \nabla\extd r= {1\over r} \extd r\tens\extd r,\quad \nabla\extd t=-{\alpha\over r}\extd t$ or non-zero Christoffel sysmbols $ \Gamma^1{}_{11}=-r^{-1}$ and $\Gamma^0{}_{10}=\alpha r^{-1}$ and the de Sitter metric in the form $g=a r^{-2} \extd r\tens\extd r + b r^{\alpha-1}(\extd r\tens \extd t+\extd r\tens\extd r)+c r^{2\alpha}\extd t\tens\extd t$ where only the nonzero combination $\bar\delta=c \alpha^2/(b^2-ac)$ of parameters is relevant up coordinate transformations. One can easily compute the classical Levi-Civita connection in these coordinates as
\[ \widehat{\Gamma}^0{}_{00} =- \frac{bc\alpha r^\alpha}{b^2-ac}, \quad \widehat{\Gamma}^0{}_{10} =- \frac{ac\alpha}{r(b^2-ac)} , \quad  \widehat{\Gamma}^0{}_{11} =- \frac{ b a \alpha r^{-\alpha - 2}}{b^2-ac} \]
\[ \widehat{\Gamma}^1{}_{00} = \frac{c^2 \alpha r^{1+2\alpha}}{b^2-ac} ,\quad  \widehat{\Gamma}^1{}_{10} = \frac{bc\alpha r^\alpha}{b^2-ac} ,\quad \widehat{\Gamma}^1{}_{11} = \frac{-b^2(1 - \alpha) + ac}{r(b^2-ac)} \]
Combing this with the `quantising' connection yields the contorsion tensor
\[ S^0{}_{00}=-S^1_{01}=-S^1{}_{10}={bc\alpha r^\alpha\over b^2-ac},\quad S^0{}_{10}=-S^1{}_{11}=S^0{}_{01}+{\alpha\over r}={b^2\alpha\over b^2-ac}\]
\[ S^0{}_{11}={ba \alpha r^{-\alpha-2}\over b^2-ac},\quad S^1{}_{00}=-{c^2\alpha r^{1+2\alpha}\over b^2-ac}\]
From here we compute $S^\mu={\alpha\over b^2-ac}(br^{-\alpha},-c r)$ for the $t,r$ components giving $\nabla_{\nu} S^\mu = 0$ so that in conjunction with  flatness of $\nabla$, Theorem~\ref{square1} shows that there is no order $\lambda$ correction to the Laplace operator.

We can also find the geometric quantum Laplacian to all orders directly from the full quantum geometry at least after a convenient but non-algebraic coordinate  transformation in \cite{MaTao}. If we allow this then the model has generators $R,T$ with the only non-zero commutation relations $[T,R]=\lambda'$, $[R,\extd R]=\lambda'\sqrt{\bar\delta}\extd R$ where $\lambda'=\lambda\sqrt{b^2-ac}$ and the quantum metric and quantum Levi-Civita connection\cite{MaTao}
\[ g_1=\extd R \bullet e^{2 T\sqrt{\bar\delta}}\tens_1\extd R-\extd T\tens_1\extd T,\]
\[  \nabla_1\extd T=-\sqrt{\bar\delta}e^{2 T\sqrt{\bar\delta}}\bullet\extd R\tens_1\extd R,\quad \nabla_1\extd R=-\sqrt{\bar\delta}(\extd R\tens_1\extd T+\extd T\tens_1\extd R),\]
which immediately gives us 
\[ (\extd R,\extd R)_1=e^{-2 T\sqrt{\bar\delta}},\quad (\extd T,\extd T)_1=-1,\quad \square_1 T=(\ ,\ )\nabla_1\extd T=-\sqrt{\bar\delta},\quad \square_1 R=0.\] Finally, for a general normal-ordered function $f(T,R)$ with $T's$ to the left, we have 
\[ \extd f={\del f\over\del T}\bullet \extd T+ \del^1 f\bullet \extd R;\quad \del^1 f(R)={f(R)-f(R-\lambda'\sqrt{\bar\delta})\over\lambda'\sqrt{\bar\delta}}\]
due to the standard form of the commutation relations. With these ingredients and following exactly the same method as above, we have
\begin{eqnarray*}\square_1 f&=&(\ , )_1\nabla_1(\extd f)=-\sqrt{\bar\delta}{\del f\over\del T}-{\del^2 f\over\del T^2} +(\del^1)^2 f\bullet e^{-2T{\sqrt{\bar\delta}}}= \square f+O(\lambda^2)   \end{eqnarray*}
when we expand $\del^1={\del \over \del R}-{\lambda'\sqrt{\bar\delta}\over 2}{\del^2\over\del R^2}+O(\lambda^2)$ and write the bullet as classical plus Poisson bracket. This confirms what we found from Theorem~\ref{square1}. We can also use identities from quantum mechanics applied to $R,T$ in our case to further write
\[ \square_1f=-\sqrt{\bar\delta}{\del f\over\del T}-{\del^2 f\over\del T^2}+e^{-2 T\sqrt{\bar\delta}}\bullet\Delta_1 f \]
where
\[ \Delta_1 f(R)={f(R+2\lambda'\sqrt{\bar\delta})-2 f(R-\lambda'\sqrt{\bar\delta})+f(R)\over (\lambda'\sqrt{\bar\delta})^2}.\]
We see that the quantum Laplacian working in the quantum algebra with normal-ordered quantum wave functions has exactly the classical form except that the derivative in the $R$ direction is a finite difference one.  It is also clear that we have eigenfunctions $\psi(T,R)=e^{\imath\omega T}e^{\imath k R}$.  All if this is an identical situation to the standard Minkowski spacetime bicrossproduct model in \cite{AmeMa} except that there time became a finite difference and there was no actual quantum geometry. Like there, one {\em could} claim that there is an order $\lambda$ correction provided classical fields are identified with normal ordered ones, but from the point of view of Poisson-Riemannian geometry this is an artefact of such an assumption (the Poisson geometry being closer to Weyl ordering). We have focussed on the 2D case but the same conclusion holds for the Bertotti-Robinson quantum metric on $S^{n-1}\times dS_2$ in \cite{MaTao} keeping the angular coordinates to the left along with $T$; then only the double $R$-derivative deforms namely to $\Delta_1$ on normal-ordered functions. The work \cite{MaTao} were already obtained the quantum Ricci and scaler curvatures which have the same form as classically (normal ordered in the former case). 

\subsection{Fuzzy nonassociative sphere revisited}\label{fuzzyS}

The case of the sphere in Poisson-Riemannian geometry is covered in \cite{BegMa6} mainly in very explicit cartesian coordinates where we broke the rotational symmetry. However, the results are fully rotationally invariant as is more evident if  we work with $z^i$, $i=1,2,3$ and the relation $\sum_i z^i{}^2=1$. We took $\nabla=\widehat\nabla$ (the Levi-Civita connection) so $S=0$, and $\omega$ the inverse of the canonical volume 2-form on the unit sphere. Then  the results of \cite{BegMa6} give us a particular `fuzzy sphere' differential calculus
\[ [z^i,z^j]_\bullet=\lambda\eps^{ij}{}_k z^k,\quad  [z^i,\extd z^j]_\bullet=\lambda z^j \eps^i{}_{mn} z^m\extd z^n.\]
to order $\lambda$. These are initially valid for $i=1,2$ but must hold in this form for $i=1,2,3$ by rotational symmetry of both the Poisson bracket and the Levi-Civita connection. One also finds from the algebra that $z^m\bullet \extd z^m=0$ (sum over $m=1,2,3$) at order $\lambda$ on differentiating the radius 1 relation. Here $\Omega^1$ is a projective module with $\extd z^i$ as a redundant set of generators and a relation. We also have degree 1 relations
\[ \{\extd z^i,\extd z^j\}_\bullet=\lambda (3 z^iz^j-\delta_{ij}){\rm Vol}\]
to order $\lambda$ as derived in \cite{BegMa6} for $i=1,2$ and which then holds for $i=1,2,3$. This can also be derived by applying $\extd$ to the bimodule relations and using $\extd z^i\wedge\extd z^j=\eps^{ij}{}_kz^k {\rm Vol}$ at the classical level on the unit sphere. We will also use the antisymmetric lift $\widetilde{\rm Vol}={1\over 2}(z^3)^{-1}(\extd z^1\tens\extd z^2-\extd z^2\tens\extd z^1)$ at the classical level. The classical sphere metric $g_{\mu\nu}$ is given in \cite{BegMa6} in the $z^1,z^2$ coordinates but we can also write it as
\[ g=\sum_{i=1}^3\extd z^i\tens \extd z^i\]
Similarly, the inverse metric and metric inner product are
\[ g^{\mu\nu}=\delta_{\mu\nu}-z^\mu z^\nu,\quad (\extd z^i,\extd z^j)=\delta_{ij}-z^iz^j \]
for $\mu,\nu=1,2$, which extends as the second equality for $i,j=1,2,3$.  The sphere is a 2-dimensional manifold so only two of the $z^i$ are independent in any coordinate patch but the expressions themselves are rotationally invariant when expressed in terms of all three. 

The work \cite{BegMa6} also computes the quantum metric and quantum Levi-Civita connection at order $\lambda$. We have 
\begin{align*} g_1=&g_{\mu\nu}\extd z^\mu\tens_1\extd z^\nu-{\lambda\over 2 (z^3)^2}\extd z^3\tens_1\eps_{3ij}z^i\extd z^j+\lambda \widetilde{\rm Vol}\\
=&g_{\mu\nu}\extd z^\mu\tens_1\extd z^\nu +\frac{\lambda}{2(z^3)^2}\eps_{3ij}\left(z^3\extd z^i\tens_1 \extd z^j-z^i \extd z^3\tens_1 \extd z^j\right)\end{align*}
\begin{align*} \nabla_1\extd z^\mu&=\nabla_Q\extd z^\mu=-z^\mu\bullet g_1=-\widehat\Gamma^\mu{}_{\alpha\beta}\extd z^\alpha\tens_1\extd z^\beta-\lambda z^\mu\widetilde{\rm Vol}+{\lambda\over 2 }\left(\extd z^3\tens_1 (\eps^{\mu\beta}g_{\beta\gamma}+{z^\mu z^\beta\over (z^3)^2}\eps_{\beta\gamma})\extd z^\gamma\right)\\
&= - \widehat{\Gamma}^\mu{}_{\alpha\beta} \extd z^\alpha \tens_1 \extd z^\beta- \frac{\lambda}{2 (z^3)^2} \left( \epsilon_{3ij} z^\mu z^3 \extd z^i \tens_1 \extd z^j - \epsilon^\mu{}_{\nu 3}  \extd z^3 \tens_1 \extd z^\nu \right)\end{align*}
where we massaged the formulae in \cite{BegMa6}. The classical Christoffel symbols are $\widehat\Gamma^\mu{}_{\alpha\beta}=z^\mu g_{\alpha\beta}$. 

If we work with coefficients $\tilde g_{ij}$ in the middle for the metric then the given quantum metric corresponds to the correction term
\[ h = \frac{(2-(z^3)^2)}{(z^3)^3}\epsilon_{3ij} \extd z^i \tens_1 \extd z^j=  \frac{2(2-(z^3)^2)}{(z^3)^2}\widetilde{\rm Vol}\]
which we see is antisymmetric. For the corresponding inverse metric we get from (\ref{tildegupper}) that
\[ (\extd z^i,\extd z^j)_1= g^{ij}+{\lambda\over 2}\eps_{ijk}z^k\]
to order $\lambda$ when $i,j=1,2$ but which extends to $i,j=1,2,3$ with $g^{ij}=\delta_{ij}-z^iz^j$. 
For the connection it is a nice check that the formula in Lemma~\ref{Gamma1} gives the same answer for $\nabla_1$. Then we can calculate the quantum Riemann tensor from \eqref{Rnabla1} or directly from the above formulae for $\nabla_1$. 
\begin{eqnarray*}
{\rm Riem}_1 (\extd z^\alpha) &=& (\extd\tens_1\id-(\wedge_1\tens_1\id)(\id\tens_1\nabla_1))\nabla_1(\extd z^\alpha)
\\
&=& - \left( \extd (\Gamma_1{}^\alpha{}_{\mu\beta} ) \wedge \extd z^\mu \tens_1 \extd z^\beta + \Gamma_1{}^\alpha{}_{\mu\gamma} \extd z^\mu \wedge_1 \Gamma_1{}^\gamma{}_{\nu\beta} \extd z^\nu \tens_1 \extd z^\beta \right)
\end{eqnarray*}
which can be broken down into three terms as follows

(i) The first term gives
\begin{eqnarray*}
\extd (\Gamma_1{}^\alpha{}_{\mu\beta} ) \wedge \extd z^\mu \tens_1 \extd z^\beta &=& - \widehat{\Gamma}^\alpha{}_{\mu\beta,\nu} \extd z^\nu \wedge \extd z^\mu \tens_1 \extd z^\beta- \frac{\lambda}{2} \del_\mu \left( \frac{z^\alpha}{z^3}  \right) \epsilon_{3\nu\beta} \extd z^\mu \wedge \extd z^\nu \tens_1 \extd z^\beta 
\\
&=& - \widehat{\Gamma}^\alpha{}_{\mu\beta,\nu} \extd z^\nu \wedge \extd z^\mu \tens_1 \extd z^\beta- \frac{\lambda}{2 (z^3)^2} \epsilon_{3\nu\beta} \left( z^3 \extd z^\alpha  - z^\alpha \extd z^3 \right) \wedge \extd z^\nu \tens_1 \extd z^\beta 
\\
&=& - \widehat{\Gamma}^\alpha{}_{\mu\beta,\nu} \extd z^\nu \wedge \extd z^\mu \tens_1 \extd z^\beta - \frac{\lambda}{2 z^3} \left( z^3 {\rm Vol} \tens_1 \extd z^\alpha - z^\alpha {\rm Vol} \tens_1 \extd z^3 \right)
\end{eqnarray*}
The last step comes from expanding the expression in the previous line and simplifying, this will prove useful in comparing to the other terms.

(ii) Expanding $\wedge_1$ gives a further two terms at $\mathcal{O}(\lambda)$. But first, using the formula for the classical Christoffel symbols and metric compatibility note that 
\begin{equation*}
\nabla_\alpha \widehat\Gamma^\iota{}_{\mu\nu} \extd z^\nu = \nabla_\alpha z^\iota{} g_{\mu\nu} \extd z^\nu = (\delta^\iota{}_{\alpha} g_{\mu\nu} + z^\iota z_\nu g_{\alpha\mu} ) \extd z^\nu
\end{equation*}
Now consider 
\begin{eqnarray*}
\omega^{\eta\zeta} \nabla_\eta \widehat\Gamma{}^\alpha{}_{\mu\gamma} \extd z^\mu \wedge \nabla_\zeta \widehat\Gamma{}^\gamma{}_{\nu\beta} \extd z^\nu \tens_1 \extd z^\beta &=& \omega^{\eta\zeta} (\delta^\alpha{}_{\eta} g_{\mu\gamma} + z^\alpha z_\gamma g_{\eta\mu} ) \extd z^\mu \wedge (\delta^\gamma{}_{\zeta} g_{\nu\beta} + z^\gamma z_\beta g_{\zeta\nu} ) \extd z^\nu \tens_1 \extd z^\beta
\\
&=& \omega^{\alpha\gamma} g_{\mu\gamma} g_{\nu\beta} \extd z^\mu \wedge \extd z^\nu \tens_1 \extd z^\beta 
\\
&=& \frac{1}{z^3} (\epsilon^\alpha{}_{\mu3} + \epsilon_{3\mu\gamma} z^\alpha z^\gamma)  g_{\nu\beta} \extd z^\mu \wedge z^\nu \tens_1 \extd z^\beta 
\\
&=& {\rm Vol} \tens_1 \extd z^\alpha
\end{eqnarray*}
where the cancellations in the second line result from the antisymmetry of $\mu,\nu$ and $\eta,\zeta$. For the second term use 
\[H^{\mu\nu} = \frac{1}{2}( z^\mu z^\nu - \delta^{\mu\nu}) {\rm Vol} \]
giving
\begin{eqnarray*}
\widehat\Gamma{}^\alpha{}_{\mu\gamma} \widehat\Gamma{}^\gamma{}_{\nu\beta} H^{\mu\nu} \tens_1 \extd z^\beta  &=& \frac{1}{2} z^\alpha z^\gamma g_{\mu\gamma} g_{\nu\beta}( z^\mu z^\nu - \delta^{\mu\nu}) {\rm Vol} \tens_1 \extd z^\beta
\\
&=& \frac{1}{2} z^\alpha z^\nu \left( \frac{(z^1)^2 + (z^2)^2}{(z^3)^4} - \frac{1}{(z^3)^4} \right) \delta_{\nu\beta} {\rm Vol} \tens_1 \extd z^\beta
\\
&=& -\frac{z^\alpha}{2z^3} {\rm Vol} \tens_1 \extd z^3
\end{eqnarray*}
Combining these two (remembering to add an overall 1/2 to the first) results in
\begin{eqnarray*}
\widehat\Gamma{}^\alpha{}_{\mu\gamma} \extd z^\mu \wedge_1 \widehat\Gamma{}^\gamma{}_{\nu\beta} \extd z^\nu \tens_1 \extd z^\beta &=&\widehat\Gamma{}^\alpha{}_{\mu\gamma} \widehat\Gamma{}^\gamma{}_{\nu\beta} \extd z^\mu \wedge \extd z^\nu \tens_1 \extd z^\beta + \frac{\lambda}{2 z^3} \left( z^3 {\rm Vol} \tens_1 \extd z^\alpha - z^\alpha {\rm Vol} \tens_1 \extd z^3 \right)
\end{eqnarray*}

(iii) The last term involves the $\mathcal{O}(\lambda)$ of $\Gamma_1{}^\alpha{}_{\mu\gamma} \Gamma_1{}^\gamma{}_{\nu\beta} \extd z^\mu \wedge  \extd z^\nu \tens_1 \extd z^\beta$ and is
\[ z^\gamma g_{\nu\beta} (z^\alpha z^3 \epsilon_{3\mu\gamma} \extd z^\mu - \epsilon^\alpha{}_{\gamma3} \extd z^3) \wedge \extd z^\nu \tens_1 \extd z^\beta + z^\alpha g_{\mu\gamma} \extd z^\mu \wedge (z^\gamma z^3 \epsilon_{3\nu\beta} \extd z^\nu - \epsilon^\gamma{}_{\beta3} \extd z^3) \tens_1 \extd z^\beta \]
The second term, which given in components is
\[z^\alpha g_{\mu\gamma} (z^\gamma z^3 \epsilon_{3\nu\beta} + \frac{1}{z^3}\epsilon^\gamma{}_{\beta} \delta_{\delta\nu} z^\delta) \]
can be shown to be symmetric in $\mu,\nu$ and therefore vanishes, whereas the first can be expanded and simplified to give
\[ z^\gamma g_{\nu\beta} (z^\alpha z^3 \epsilon_{3\mu\gamma} \extd z^\mu - \epsilon^\alpha{}_{\gamma3} \extd z^3) \wedge \extd z^\nu \tens_1 \extd z^\beta = -\frac{1}{(z^3)^2} (1-(z^3)^2) {\rm Vol} \tens_1 \extd z^\alpha -2\frac{z^\alpha}{z^3} {\rm Vol} \tens_1 \extd z^3 \]
Now, taking together the above terms gives the semiclassical Riemann tensor as  
\begin{equation*} {\rm Riem}_1 (\extd z^\alpha ) =-\frac{1}{2}\widehat{R}^\alpha{}_{\beta\mu\nu} \extd x^\mu \wedge \extd x^\nu \tens_1 \extd x^\beta + \frac{\lambda}{2 (z^3)^2} (1+(z^3)^2) {\rm Vol} \tens_1 \extd z^\alpha  \end{equation*}
Where the classical Riemann tensor is $R^\iota{}_{\gamma\mu\nu} \extd z^\mu \wedge \extd z^\nu \tens_1 \extd z^\gamma = \extd z^\iota \wedge g $. This is the same result as the general tensorial calculation  using (\ref{Riem1}), as a useful check of those formulae. 

For the Ricci tensor, the form of the quantum lift from Proposition~\ref{i1} is
\[ i_1 (\extd z^\mu \wedge \extd z^\nu )  = \frac{1}{2} \left( \extd z^\mu \tens_1 \extd z^\nu - \extd z^\nu \tens_1 z^\mu \right) + \lambda I(\extd z^\mu \wedge \extd z^\nu)\]
The functorial choice here comes out as $I(\extd z^\mu \wedge \extd z^\nu)=0$, but we leave this general. In 2D the lift map has three independent components which, in tensor notation, we parametrize as $\alpha:=I^{12}{}_{11}$, $\beta:=I^{12}{}_{22}$ and $\gamma:=I^{12}{}_{12}$, with the remaining components being related by symmetry. Then the tensorial formula (\ref{Ricci1})  gives us
\begin{eqnarray*}
{\rm Ricci}_1 &= &-\frac{1}{2}g_1 - \frac{3\lambda}{2} \widetilde{\rm Vol}\\
&& - \frac{\lambda}{(z^3)^2} \left((\alpha z^1 z^2 + \gamma((z^2)^2 - 1)) \extd z^1 \tens_1 \extd z^1 - (\beta z^1 z^2 + \gamma((z^1)^2 - 1)) \extd z^2 \tens_1 \extd z^2 \right.
\\
&& \left. + (\gamma z^1 z^2 + \alpha((z^1)^2 - 1)) \extd z^1 \tens_1 \extd z^2 - (\gamma z^1 z^2 + \beta((z^2)^2-1)) \extd z^2 \tens_1 \extd z^1 \right)
\end{eqnarray*}
Next, following our general method, we require $I$ to be such that $\wedge_1{\rm Ricci}_1=0$, i.e.  quantum symmetric. This results in the constraint
\[\gamma = -\frac{1}{4z^1 z^2} \left( 3z^3 + 2\alpha ((z^1)^2 - 1) + 2 \beta ((z^2)^2 - 1) \right) \]
with $\alpha$ and $\beta$ undetermined. We also want ${\rm Ricci}_1$ to be hermitian or `real' in the sense ${\rm flip}(*\tens *){\rm Ricci}_1={\rm Ricci_1}$  which already holds for $-{1\over 2}g_1$. Since $\lambda$ is imaginary this requires the matrix of coefficents in the order $\lambda$ terms displayed above to be antisymmetric as all tensors are real. This imposes three more constraints which are fortunately not independent and give us a unique suitable lift, namely with 
\[ \alpha={3\over 4 z^3} (1-(z^2)^2 )
,\quad \beta={3\over 4 z^3} (1-(z^1)^2 )
,\quad \gamma={3\over 4} {z^1 z^2 \over z^3} .\]
The result (and similarly in any rotated coordinate chart) is
 \begin{eqnarray*}
i_1 (\extd z^1 \wedge \extd z^2 )  = \frac{1}{2} \left( \extd z^1 \tens_1 \extd z^2 - \extd z^2 \tens_1 z^1 \right) - \frac{3\lambda}{4z^3} g 
\end{eqnarray*}
\[ {\rm Ricci}_1 = -\frac{1}{2}g_1\]
where the latter in our conventions is analogous to the classical case.  And from this or from \eqref{S1} we get the quantum scalar curvature 
\[ {S}_1 = -\frac{1}{2}\widehat{S},\quad \widehat{S}= \widehat{R}_{\mu\nu} g^{\mu\nu}=2\]
the same as classically in our conventions, so this has no corrections at order $\lambda$. As remarked in the general theory, the quantum Ricci scalar is independent of the choice of lift $I$.  

We also find no correction to the Laplacian at order $\lambda$ since the classical Ricci tensor is proportional to the metric hence the contraction in Theorem~\ref{square1} gives $\omega^{\alpha\beta}(\widehat\nabla_\beta\extd f)_\alpha$ which factors through $\widehat\nabla\wedge\extd f=0$ due to zero torsion of the Levi-Civita connection. 

We close with some other comments about the model. In fact the parameter $\lambda$ in this model is dimensionless and if we want to have the usual  finite-dimensional `spin $j$' representations of our algebra then we need 
\[ \lambda=\imath/\sqrt{ j(j+1)}\]
 for some natural number $j$ as a quantisation condition on the deformation parameter. Here our reality conventions require $\lambda$ imaginary. It is also known from \cite{BegMa2} that this differential algebra arises from twisting by a cochain at least to order $\lambda^2$ {\em but} in such a way that the twisting
also induces the correct differential structure at order $\lambda$, i.e. as given by the Levi-Civita connection. We let $U(so_{1,3})$ have generators and relations
\[ [M_i,M_j]=\eps_{ijk}M_k,\quad [M_i,N_j]=\eps_{ijk}N_k,\quad [N_i,N_j]=-\eps_{ijk}M_k  \]
acting on the classical $z^i$ (i.e. converting \cite{BegMa2} to the coordinate algebra) as,
\[  M_i\la z^j= \eps_{ijk}z^k,\quad N_i\la z^j= z^iz^j-\delta_{ij}.\]
This is the action of $so_{1,3}$ on the `sphere at infinity'. The cochain we need is then\cite{BegMa2}
\[ F^{-1}=1+\lambda f+{\lambda^2\over 2}f^2+\cdots,\quad f={1\over 2}M_i\tens N_i\]
to be constructed order by order in such a way that the algebra remains associative (and gives the quantisation of $S^2$ as a quotient of $U(su_2)$). On the other hand cochain twisting extends the differential calculus to all orders as a graded quasi-algebra in the sense of \cite{BegMa3}. Specifically, if we start with the classical algebra and exterior algebra on the sphere, the deformed products at order $\lambda$ are
\[ z^i\bullet z^j=(F^{-1}\la z^i)(F^{-2}\la z^j)=z^iz^j+{\lambda\over 2}\eps_{ijk}z^k \]
\[ z^i\bullet\extd z^j=( F^{-1}\la z^i )\extd F^{-2}\la z^j=z^i\extd z^j+{\lambda\over 2}z^j\eps_{imn}z^m\extd z^n\]
\[ \extd z^j\bullet z^i=( F^{-1}\la \extd z^j )\extd F^{-2}\la z^i=(\extd z^j)z^i-{\lambda\over 2}z^i\eps_{jmn}z^m\extd z^n-{\lambda\over 2}\eps_{ijm}\extd z^m\]
giving relations 
\[ [z^i,\extd z^j]_\bullet={\lambda\over 2}((z^i\eps_{jmn}+z^j\eps_{imn})z^m\extd z^n+\eps_{ijm}\extd z^m)=\lambda z^j\eps_{jmn}z^m\extd z^n\]
in agreement with the quantisation of the calculus by the Levi-Civita connecton. For the last step we let 
\[ w^i=\eps^i{}_{jk}z^j\extd z^k.\]
and note that classically $z^i w^j\eps_{ijk}=-\extd z^k$ using the differential of the sphere relation and hence $z^iw^j-z^jw^i=-\eps_{ijk}\extd z^k$, which we use. This twisting result in \cite{BegMa2} is in contrast to other cochain twist or deformation theory quantisations such as in \cite{Schupp}, which consider only the coordinate algebra. It means that although the differential calculus is not associative at order $\lambda^2$, corresponding to the curvature of the sphere, different brackets are related via an associator and hence strictly controlled. One can then twist other aspects of the noncommutative geometry using the formalism of \cite{BegMa3}, see also more recently \cite{AscSch}. 

To get a sense of how these equations fit together even though nonassociative, we work now in the quantum algebra so {\em from now till the end of the section all products are deformed ones}. We have the commutation relations
\[ [z^i,z^j]=\lambda\eps^{ij}{}_k z^k,\quad  [z^i,\extd z^j]=\lambda w^i z^j\]
to order $\lambda$. Then, if we apply $\extd$ to the first relation we have
\begin{align*}\lambda\eps_{ijk}\extd z^k&= [\extd z^i,z^j]+[z^i,\extd z^j]=\lambda(w^iz^j-w^jz^i)=\lambda\eps_{jik}\eps_{kmn}z^mw^n=-\lambda\eps_{ijk}\eps_{kmn} z^m\eps_{nab}z^a\extd z^b\\
&=-\lambda\eps_{ijk}(\delta^k_a\delta^m_b-\delta^k_b\delta^m_a)z^m z^a\extd z^b=\lambda\eps_{ijk}\extd z^k - \lambda z^m\eps_{ijk}z^k\extd z^m\end{align*}
which confirms that $\sum z^m\extd z^m=O(\lambda)$ (which is to be expected since it is zero classically). In fact we only need the commutation relations for $i,j=1,2$ to arrive at this deduction. Moreover, 
\[ 0=\extd(\sum_m z^m z^m)=2z^m \extd z^m-\lambda  w^m z^m+[\extd z^3 ,z^3]+\lambda w^3 z^3\]
and $w^mz^m =O(\lambda)$ since zero classically, which tells us that 
\[ [z^3,\extd z^3]=\lambda w^3 z^3+ 2 z^m\extd z^m.\]
Hence $z^m\extd z^m=0$ at order $\lambda$ if the $z^3$ commutation relations hold as claimed. In fact assuming only the $i,j=1,2$ commutation relations one can deduce (so long as $z^3$ is invertible) that $[z^3,\extd z^j]=\lambda w^3 z^j$ for $j=1,2$ by looking at $[(z^3)^2,\extd z^i]=2 z^3[z^3,\extd z^i]$ on the one hand and using the radius relation on the other hand. From this and $\lambda\eps_{i3k}\extd z^k=[\extd z^i,z^3]+[z^i,\extd z^3]$ we deduce that $[z^i,\extd z^3]=\lambda w^iz^3$ as claimed. Then by the same calculation as for the $[z^3,\extd z^i]$ relation we can deduce $[z^3,\extd z^3]=\lambda w^3 z^3$ as well. Thus, we have internal consistency of the quantum algebra relations even if we do not have associativity of the relations involving the $\extd z^i$ and can use the algebra to deduce the rotationally invariant form of the commutation relations as claimed earlier.

\section{Semiquantum FLRW model}\label{flrw}

We work with coordinates $t,x^i$, $i=1,2,3$, possibly with singularities. We also use polars $r,\theta,\phi$ for the spatial coordinates so that $\extd^2\Omega=\sin^2(\theta)\extd\phi\tens\extd\phi$ is the unit sphere metric.  It is already known from \cite{BegMa6} that for  bivector $\omega$ to be rotationally invariant leads in polars to
\begin{equation}\label{FLRWPoissonC} \omega^{23}={f(t,r)\over \sin \theta}=-\omega^{32},\quad \omega^{01}=g(t,r)=-\omega^{10} \end{equation}
for some functions $f,g$ and other components zero. Our approach is to solve (\ref{nablaS}) for $S$ using the above form of $\omega$ and $\widehat\nabla$ for the chosen metric, which in the present section is the spatial flat FLRW one
\[ \cg=-\extd t\tens\extd t+ a(t)^2(\extd r\tens\extd r+ r^2 \extd^2\Omega)\]
with
\begin{equation*}
\label{FLRWflatChr}
\begin{array}{llll}
\displaystyle \widehat{\Gamma}^{0}_{11} = \dot{a}a,\quad  & \displaystyle \widehat{\Gamma}^{0}_{22} = \dot{a}ar^{2} ,\quad & \displaystyle \widehat{\Gamma}^{0}_{33} = \dot{a}ar^{2} \sin^{2}(\theta),\quad & \\
\displaystyle \widehat{\Gamma}^{1}_{01} = \frac{\dot{a}}{a},\quad  & \displaystyle \widehat{\Gamma}^{1}_{22} = -r  ,\quad & \widehat{\Gamma}^{1}_{33} = - r \sin^{2}(\theta)\\
\displaystyle \widehat{\Gamma}^{2}_{02} = \frac{\dot{a}}{a},\quad & \widehat{\displaystyle \Gamma}^{2}_{21} = \frac{1}{r},\quad & \displaystyle \widehat{\Gamma}^{2}_{33} = -\sin(\theta)\cos(\theta) \\
\displaystyle \widehat{\Gamma}^{3}_{03} = \frac{\dot{a}}{a} ,\quad & \widehat{\displaystyle \Gamma}^{3}_{31} = \frac{1}{r},\quad & \displaystyle \widehat{\Gamma}^{3}_{23} = \cot(\theta)
\end{array}
\end{equation*}
 Remarkably, if $\omega$ is generic in the sense that the functions $a,f,g$ are algebraically independent and invertible then it turns out that one can next solve the Poisson-compatibility condition \eqref{nablaS} for $S$ {\em uniquely} using computer algebra. This is relevant if we drop the requirement \eqref{omegapoisson} that $\omega$ obeys the Jacobi identity which is to say if we allow the coordinate algebra to be nonassociative at order $\lambda^2$ and if we drop \eqref{nabla1g1} which is to say we allow a possible quantum effect where $\nabla_1g_1=O(\lambda)$ in its antisymmetric part. Such a theory appears quite natural for this reason, but for the present purposes we do want to go further and impose \eqref{omegapoisson} as well as the condition \eqref{nabla1g1} for the existence of a fully quantum Levi-Civita connection. 

\begin{proposition}\label{FLRWPoisson}  In a FLRW spacetime with spherically symmetric Poisson tensor, to have a Poisson-compatible connection obeying \eqref{nabla1g1} and \eqref{omegapoisson} requires up to normalisation that $g(r,t) = 0$ and $f(r,t) = 1$. The contorsion tensor in this case is 
\[S_{022} = a\dot{a} r^2 ,\quad S_{122} = a^2 r,\quad S_{033} = a\dot{a} r^2 \sin^2(\theta) ,\quad S_{133} = a^2 r \sin^2(\theta) \]
\[ S_{120} = S_{123} = S_{223} = S_{320} = S_{130} = S_{132} = S_{230} = S_{233}=0\]
up to the outer antisymmetry of $S_{\mu\nu\gamma}$. The remaining components $S_{\mu 0 \nu}$, $S_{\mu 1 \nu}$ are  undetermined but are irrelevant to the combination $\omega^{\alpha\beta}\nabla_\beta$ (the contravariant connection), which is  uniquely determined. 
\end{proposition}
\proof 
As already noted in \cite{BegMa6} for $\omega$ of the rotationally invariant form \eqref{FLRWPoissonC} to obey \eqref{omegapoisson} comes down to
\begin{equation}\label{FLRWomegajac} g\del_t f= g\del_r f=0\end{equation}
which tells us that either $f=k$ a constant or $g=0$. We examine the former case, then the Poisson compatibility condition \eqref{nablaS} becomes
\[ S_{201} = 0,\quad S_{301} = 0  , \quad S_{001}g - \del_r g = 0,\quad S_{314} = 0,\quad S_{233} = 0,\quad S_{322} =0 \]
\[ S_{031}k r^2a^2 + S_{112} g \sin(\theta) = 0, \quad r^2 a^2 k \sin(\theta)S_{021} + S_{311} g = 0,\quad a\dot a \sin(\theta) g +k S_{231} = 0 \]
\[ S_{203}g - r^2 \sin(\theta) (r a^2 - S_{231}) = 0 , \quad S_{002} a^2 g + S_{112} g = 0 , \quad \sin(\theta) g + S_{230} k r = 0\]
\[ kr^4 a^2 \sin(\theta) a \dot a + S_{312} g + kS_{022}=0,\quad a^2 r^3 k \sin^2(\theta) + S_{203} g + kS_{331} = 0 \]
\[ r^4 a^3 \sin^2(\theta) S_{033} - S_{312} = 0,\quad \sin(\theta) g - r k S_{320} = 0,\quad k r^2 S_{031} - S_{002} g \sin(\theta) = 0 \] 
\[a^2 S_{003} - S_{322} = 0 ,\quad S_{021} r^2 \sin(\theta) + S_{003} = 0,\quad 2k a\dot{a} r^2 \sin(\theta) - S_{022}\sin(\theta) - kS_{033} = 0 \]
\[ g a\dot a \sin(\theta) - k S_{321} = 0 , \quad r^2 a^2 \del_t g + r^2 a \dot a g + S_{012} = 0 ,\quad (2ra^2 - S_{122})k\sin^2(\theta) + S_{331} = 0.\]
This is not enough to determine all the components of $S$ and hence $\nabla$ but it determines enough components for the  Ricci two-form to be uniquely determined when $k\ne 0$ as 
\[ \CR_{01} = \frac{1}{r^{2} }  \left( 5 r^2 a \dot a \del_t g - \dot a^2  + g r^2 a \dot a + \del_t^2 g r^2 a \dot a - \del_r^2 g r^2 - 2 r \del_r g +6g \right) \]
\[\CR_{23} = \frac{1}{k r^{2} } \sin(\theta) \left( k^2 r^4 a^2 + g a^2 r^2 - g^2 \right). \] 
We can now impose the Levi-Civita condition, using the \textit{Physics} package in \textit{Maple}, to expand \eqref{nabla1g1} and solve for $g$ simultaneously with the above requirements of \eqref{nablaS} (details omitted). This results in $g=0$ as the only unique solution permitting Poisson compatibility and a quantum Levi-Civita connection for $f$ a (nonzero) constant.  The case $f=0$ also has this conclusion and we exclude  this so as to exclude the unquantized case $\omega=0$ in our analysis. We now go back and examine the second case, setting $g=0$ and leaving $f$ arbitrary. Now \eqref{nablaS} results in a number of equations but they include 
\[ \del_r f = 0, \quad \del_t f = 0 \] 
independently of the contorsion tensor. Hence this takes us back to $g=0$ and $f$ constant again. We can absorb the latter constant in $\lambda$, i.e. we take $k=1$ up to the overall normalisation of $\omega$. Then the above-listed content of (\ref{nablaS}) setting $k=1$ and $g=0$ gives us the values of $S$ and 12 undetermined components as stated. 

In the process above we also solved \eqref{nabla1g1} so this holds for the stated $S$ with $f=1$ and $g=0$. As this depended on a Maple solution, we also check it analytically as follows. We set 
\begin{equation}\label{Q} Q_{\gamma\mu\nu}=\omega^{\alpha\beta}\,\cg_{\rho\sigma}\,S^\sigma{}_{\beta\nu}(R^\rho{}_{\mu\gamma\alpha}+\nabla_\alpha S^\rho{}_{\gamma\mu}) - \omega^{\alpha\beta}\,\cg_{\rho\sigma}\,S^\sigma{}_{\beta\mu}(R^\rho{}_{\nu\gamma\alpha}+\nabla_\alpha S^\rho{}_{\gamma\nu})\end{equation}
while from the above with $k=1,g=0$, the Ricci two-form for our solution is
\begin{equation}
\label{FLRWRicci2Form}
\CR = - \frac{1}{2} k a^2 r^2 \sin(\theta) \extd \theta \wedge \extd \phi
\end{equation} 
and is independent of the undetermined components. This allows us to compute $\widehat\nabla_0 \CR_{\mu\nu} = \widehat\nabla_1 \CR_{\mu\nu}=0$ as well as 
\[ \widehat\nabla_2 \CR_{\mu\nu} = a r k \sin(\theta) \begin{pmatrix} 
0 & 0 & -\dot{a}r & 0 
\\ 
0 & 0 & -a & 0
\\
\dot{a}r & a & 0 & 0
\\
0 & 0 & 0 & 0
\end{pmatrix},\quad 
\widehat\nabla_3 \CR_{\mu\nu} = a r k \sin(\theta) \begin{pmatrix} 
0 & 0 & \dot{a}r & 0 
\\ 
0 & 0 & a & 0
\\
-\dot{a}r & -a & 0 & 0
\\
0 & 0 & 0 & 0
\end{pmatrix}
\]
Further calculation yields $Q_{0\mu\nu} =Q_{1\mu\nu}=0$ and 
\[ Q_{2\mu\nu} = a r k \sin(\theta) \begin{pmatrix} 
0 & 0 & -\dot{a}r & 0 
\\ 
0 & 0 & -a & 0
\\
\dot{a}r & a & 0 & 0
\\
0 & 0 & 0 & 0
\end{pmatrix},\quad 
Q_{3\mu\nu} =a r k \sin(\theta) \begin{pmatrix} 
0 & 0 & \dot{a}r & 0 
\\ 
0 & 0 & a & 0
\\
-\dot{a}r & -a & 0 & 0
\\
0 & 0 & 0 & 0
\end{pmatrix}
\]
Substituting into \eqref{nabla1g1} we see that this holds in the form $\frac{1}{2}Q_{\gamma\alpha\beta} - \frac{1}{2}\widehat\nabla_\gamma \CR_{\mu\nu} = 0$.   
\endproof

Thus we see that if we want the Poisson bracket to obey the Jacobi identity so as to keep an associative coordinate algebra and if we want a full quantum Levi-Civita connection without on $O(\lambda)$ correction to the antisymmetric part of the quantum metric compatibility tensor $\nabla_1g_1$, then rotational invariance forces us to a model in which time is central and in which the other commutation relations are also determined uniquely from $\omega^{\alpha\beta}\nabla_\beta$. To work these out it is convenient (though not essential) to work with the angular variables in terms of  $z^i=x^i/r$ as redundant unit sphere variables at each $r,t$, with $\extd z^i={1\over r}\extd x^i-{x^i\over r^2}\extd r$. Now, using the contorsion tensor above Christoffel symbols of the `quantising' connection come out as
\begin{equation}
\label{CartSphChr}
\begin{array}{cc}
\displaystyle {\Gamma}^{0}_{\mu\nu} = \begin{pmatrix} 
0 & -S^0{}_{01} & -S^0{}_{02} & -S^0{}_{03}
\\ 
0 & a \dot a - S^0{}_{11} & -S^0{}_{12} & -S^0{}_{13}
\\
0 & 0 & 0 & 0
\\
0 & 0 & 0 & 0
\end{pmatrix}
,& 
\displaystyle {\Gamma}^{1}_{\mu\nu} = \begin{pmatrix} 
S^1{}_{00} & \frac{\dot a}{a} & S^1{}_{02} & S^1{}_{03} 
\\ 
\frac{\dot a}{a} + S^1{}_{10}  & 0 & S^1{}_{12} & S^1{}_{13}
\\
0 & 0 & 0 & 0
\\
0 & 0 & 0 & 0
\end{pmatrix}
\\
\displaystyle {\Gamma}^{2}_{\mu\nu} = \begin{pmatrix} 
S^2{}_{00} & S^2{}_{01} & \frac{\dot a}{a} & S^2{}_{03}  
\\ 
S^2{}_{10} & S^2{}_{11} & r^{-1} & S^2{}_{12}
\\
0 & 0 & \frac{z^1(1-(z^2)^2)}{(z^3)^2} & \frac{(z^1)^2 z^2}{(z^3)^2}
\\
0 & 0 & \frac{(z^1)^2 z^2}{(z^3)^2} & \frac{z^1(1-(z^1)^2)}{(z^3)^2}
\end{pmatrix}
,&
\displaystyle {\Gamma}^{3}_{\mu\nu} = \begin{pmatrix} 
S^3{}_{00} & S^3{}_{01}  & S^3{}_{03} & \frac{\dot a}{a}
\\ 
S^3{}_{10} & S^3{}_{11}  & S^3{}_{12} & r^{-1}
\\
0 & 0 & \frac{z^2(1-(z^2)^2)}{(z^3)^2} & \frac{z^1 (z^2)^2}{(z^3)^2}
\\
0 & 0 & \frac{z^1 (z^2)^2}{(z^3)^2} & \frac{z^2(1-(z^1)^2)}{(z^3)^2}
\end{pmatrix}
\end{array}
\end{equation}
while the bimodule relations are independent of the undetermined components of $S$ and come out as
\[ [z^i,z^j]=\lambda \eps^{ij}{}_k z^k,\quad [z^i,\extd z^j]=\lambda z^j\eps^i{}_{mn}z^m\extd z^n\]
which is the fuzzy unit sphere of Section~\ref{fuzzyS}. Our quantum algebra at least at order $\lambda$ is thus classical in the $r,t$ directions and a standard fuzzy sphere in the angular ones. We also have
\[ [r,x^i]=0,\quad [r,\extd x^i]=0,\quad [x^i,\extd r]=0\]
so that $r,t,\extd t,\extd r$ are all central.  The undetermined contorsion components do not enter these relations from
\eqref{fetacomm} because only $\omega^{23}$ is nonzero so contraction with the Christoffel symbols selects only the $\Gamma^\mu{}_{2\nu}$ and $\Gamma^\mu{}_{3\nu}$ components which depend on only the corresponding $S$ components. 

For the rest of this section for the sake of brevity, we shall concentrate on the case where the undetermined and irrelevant $S$ components  are all set to zero, returning later when analyzing general spherically symmetric metrics to see what happens when these are included. For the record, changing to Cartesians, the nonzero bimodule relations are
\[ [x^i,x^j]=\lambda  r \eps^{ij}{}_k x^k,\quad [x^i,\Omega^j]={\lambda\over r}x^j\eps^i{}_{mn} x^m\Omega^n\]
by letting $\extd z^i=\Omega^i/r$ while our choice of the undetermined contorsion tensor components allows us to write down a nice expression for the `quantising' connection
\[ \Gamma^i{}_{jk}=-{x^m\over r^2}\eps^i{}_{kn}\eps^n{}_{jm},\quad \Gamma^i{}_{0j}={\dot a\over a}\delta^i{}_j\,\quad \Gamma^i{}_{j0}=0\]
The torsion comes out as
\[ T^i{}_{jk}= \frac{x^m}{r^{2}}\epsilon^{i}{}_{mn}\epsilon^{n}{}_{jk},\quad T^i{}_{0j}={\dot a\over a}\delta^i{}_j \]
and the Riemann, Ricci and scaler curvatures are
\begin{equation}
\label{Rnablaex}
R^{i}{}_{jkl}   =  \displaystyle \frac{1}{r^{2}} \epsilon^{i}{}_{jm}\epsilon^{m}{}_{kl} + \frac{1}{r^{4}} \left( x_{j}x^{m} \epsilon^{i}{}_{mn}\epsilon^{n}{}_{kl} + x^{i}x^{m} \epsilon_{jmn}\epsilon^{n}{}_{kl} \right)
\end{equation}
\begin{equation}\label{Ricnablaex} R_{ij}={1\over r^4}(\delta_{ij}r^2-x_i x_j),\quad S={2\over a^2 r^2}\end{equation}
and it should be noted that $R^{0}{}_{jkl}=R^{i}{}_{0kl}=R^{i}{}_{j0l} = 0$ and $R_{i0}=R_{0i}=0$. 

\subsection{Construction of quantum metric and quantum Levi-Civita connection}

Having solved for a Poisson bracket and Poisson compatible metric-compatible connection  we are in a position to read off, according to the theory in \cite{BegMa6}, the full exterior algebra and the quantum metric to lowest order. First compute 
\begin{equation}
H^{ij}  = \displaystyle  -\frac{1}{2r^3} \left( \epsilon^{i}{}_{nk} x_{m}x^{j}x^{k} - r^{2} \epsilon^{i}{}_{nk} \delta^{j}{}_{m} x^{k} \right) \extd x^{n} \wedge \extd x^{m}
\end{equation}
from which we get
\begin{equation}\label{Rmn}
\mathcal{R}_{mn}=\frac{a^2}{r}\eps_{mnk}x^k,\quad  \mathcal{R} = \frac{a^{2}}{2r} \epsilon_{mnk} x^{k} \extd x^{n} \wedge \extd x^{m}
\end{equation}
As with the curvature, all time components are equal to zero. From $\Gamma$ and $H^{ij}$ we have
\begin{equation*}
\label{wedge1ex}
\extd x^{i} \wedge_{1} \extd x^{j}  =\extd x^{i} \wedge \extd x^{j} + \frac{\lambda}{2r^3} \left(r^2 \epsilon^i{}_{nm} x^j + r^2 \epsilon^{ij}{}_m x_n + r^2 \epsilon^i{}_{nk} \delta^j{}_m x^{k} + \epsilon^i{}_{nk} x^k x^j x_m \right) \extd x^m \wedge \extd x^n
\end{equation*}
\begin{equation}
\label{wedge1com1}
\extd r \wedge_{1} \extd x^{i}  = \extd r \wedge \extd x^{i},\quad \extd t \wedge_{1} \extd x^{i}  = \extd t \wedge \extd x^{i}, \quad \extd x^{i} \wedge_{1} \extd t  = \extd x^{i} \wedge \extd t
\end{equation}
\begin{equation*}
\{\extd x^i,\extd x^j\}_1=\frac{\lambda}{r^3} \left(r^2 \epsilon^i{}_{nm} x^j + r^2 \epsilon^{ij}{}_m x_n + r^2 \epsilon^i{}_{nk} \delta^j{}_m x^{k} + \epsilon^i{}_{nk} x^k x^j x_m \right) \extd x^m \wedge \extd x^n
\end{equation*}
Similary, from $\Gamma$ and $\CR$ we compute $g_{1}$ from (\ref{g1}). Remarkably, the correction term  $\frac{\lambda}{2} \omega^{\alpha\beta} g_{\mu\rho} \Gamma^{\rho}_{\,\,\alpha\kappa} \Gamma^{\kappa}_{\,\,\beta\nu}$ exactly cancels the $\lambda \CR_{\mu\nu}$ so that $g_1{}_{\mu\nu}=g_{\mu\nu}$ and 
\begin{equation}
\label{g1ex}
g_{1}  = g_{\mu\nu} \extd x^{\mu} \otimes_{1} \extd x^{\nu}
\end{equation}
Moreover since the components $g_{\mu\nu}$ depend only on time, we also have that $g_{1\mu\nu} = \tilde g_{1\mu\nu}$. It is a nice check to verify that  $\wedge_{1}(g_{1}) = 0$ is satisfied as it must from our general theory. The second version of the metric is subtly different and equality depends on the form of the FLRW metric. One can also compute 
\begin{equation*}
\omega^{\mu\nu}{[\nabla_{\mu},\nabla_{\nu}]}  T^{\alpha}{}_{\beta\gamma} = 0 ,\quad \omega^{\mu\nu}{[\nabla_{\mu},\nabla_{\nu}]}  S^{\alpha}{}_{\beta\gamma} = 0
\end{equation*}
\begin{equation*}
\widehat{\nabla}_{i} \mathcal{R}_{mn} =- \frac{a^{2}}{r^{3}} \left( \epsilon_{mnk}x^{k}x_{i} - r^{2}\epsilon_{mni}  \right) - \frac{a\dot{a}}{r} \left( \epsilon_{ink}x^{k} - \epsilon_{imk}x^{k} \right)
\end{equation*}
and see once again that (\ref{nabla1g1}) holds  as it must by construction in Proposition~\ref{FLRWPoisson}. 

Hence a quantum Levi-Civita connection for $g_1$ exists by the theory from \cite{BegMa6} and from Lemma~\ref{Gamma1} we find it to be
\begin{equation*}
\label{QuantumLC}
\nabla_{1}(\extd x^{\iota}) =   \widehat{\Gamma}^{\iota}_{\mu\nu}  \extd x^{\mu} \otimes_1 \extd x^{\nu} 
\end{equation*}
which, like the quantum metric earlier, keeps its undeformed coefficients in the coordinate basis {\em if} we keep all coefficients to the left and use $\tens_1$. The theory in \cite{BegMa6} ensures that this is quantum torsion free and quantum metric compatible as a bimodule connection with generalised braiding $\sigma_1$ from (\ref{sigma1}) which computes as
\begin{equation}
\begin{array}{ll}
\label{sigma1ex}
\sigma_1(\extd x^{a} \otimes_{1} \extd x^{b})   &=  \extd x^{b} \otimes_{1} \extd x^{a} + {\lambda \over r^3} \left( \epsilon_k{}^{bk} x^k x_m x_n + \epsilon_{kn}{}^{b} x^k x^a x_m + 2r^2 \epsilon_k^{ab} x_k \delta_{mn} \right.
\\
&+ \left.  2r^2 \epsilon^{b}{}_{mk} x_k \delta^{a}{}_{n} - r^2 \epsilon^a{}_{mk} x^k \delta^{b}{}_{m} + r^2 \epsilon_{bck} x^k \delta^{ab} \right) \extd x^{n} \otimes_1 \extd x^{m}
\\
\sigma_1(\extd t \otimes_{1} \extd x^{a}) & = \extd x^{a} \otimes_{1} \extd t
\\
\sigma_1(\extd x^{a} \otimes_{1} \extd t) & = \extd t \otimes_{1} \extd x^{a}
\end{array}
\end{equation}
It is a reassuring but rather nontrivial check to verify directly from our results for $\nabla_1,\sigma_1,g_1$ that $\nabla_1 g_1=0$ as implied by the general theory in \cite{BegMa6}. Lastly, we compute the quantum lift map from Proposition~\ref{i1} as
\begin{eqnarray}
\label{FLRWi1}
i_1(\extd x^a \wedge \extd x^b) &=& \frac{1}{2} \left( \extd x^a \tens_1 \extd x^b - \extd x^b \tens_1 \extd x^a \right)
\nonumber \\
&-& \frac{\lambda}{4r} \epsilon^{ab}{}_{m} x_{n} \left( \extd x^m \tens_1 \extd x^n + \extd x^n \tens_1 \extd x^m \right)
\nonumber \\
i_1(\extd t \wedge \extd x^\alpha) &=& \frac{1}{2} \left( \extd t \tens_1 \extd x^\alpha - \extd x^\alpha \tens_1 \extd t \right)  
\\
i_1(\extd x^\alpha \wedge \extd t) &=& \frac{1}{2} \left( \extd x^\alpha \tens_1 \extd t - \extd t \tens_1 \extd x^\alpha \right) \nonumber
\end{eqnarray} 
with we have taken the functorial choice $I=0$.
\subsection{Laplace operator and Curvature Tensors}

We first observe that $[\extd x^m,g_{mn}]=0$ for the FLRW metric since either the coefficients $g_{mn}$ depend only on $t$ or are constant in our basis. Hence  the inverse metric is simply $(\extd x^a,\extd x^b)_1=g^{ab}$ undeformed similarly to  the coefficients of $g_1$, since then \begin{eqnarray*}
(f \bullet \extd x^{\alpha}, g_{\mu\nu}\bullet \extd x^{\mu})_{1} \bullet\extd x^{\nu} 
 & =&  (f\bullet \extd x^{a}, \extd x^{\mu}\bullet g_{\mu\nu})_{1}\bullet \extd x^{\nu} - (f\bullet \extd x^{\alpha}, [\extd x^{\mu},g_{\mu\nu}])_{1}\bullet\extd x^\nu
\\
 & =& \ f \bullet(\extd x^{\alpha}, \extd x^{\mu})_{1} \bullet g_{\mu\nu}\bullet \extd x^{\nu}= f \bullet \extd x^\alpha
\end{eqnarray*}
as required, where we also need that $\{g^{am},g_{mn}\}=0$ which holds for the FLRW metric. Similarly on the other side.  It follows that the quantum dimension is the same as the classical dimension, namely 4, in our model. Similarly, because $\nabla_1, g_1, (\ ,\ )_1$ also have their classical form, from Theorem~\ref{square1} we get that 
\[ \square_{1} f = g^{\alpha\beta} \left(f_{,\alpha\beta}+f_{,\gamma} \widehat{\Gamma}^{\gamma}{}_{\alpha\beta}\right) \]
is also undeformed on the underlying vector space. We used that $\nabla_1$ is a left connection. We can also calculate Riemann tensor using (\ref{Rnabla1}) from which we see that corrections come from $\wedge_{1}$.
\begin{eqnarray*}
\text{Riem}_{1}(\extd x^{i}) &=& (\extd\tens\id)\nabla_1\extd x^i- \widehat\Gamma^i{}_{\mu\gamma}\extd x^\mu \wedge_1\widehat\Gamma^\gamma{}_{\nu\alpha}\extd x^\nu\tens_1\extd x^\alpha\\
& = &-\frac{1}{2}\widehat{R}^i{}_{\alpha\mu\nu} \extd x^\mu \wedge \extd x^\nu \tens_1 \extd x^\alpha - \frac{\lambda \dot{a}^{2}}{2r^3} \epsilon^i{}_{nk} x^k x_j x_m \extd x^m \wedge \extd x^n  \tens_1 \extd x^{j}
\\
&&- \frac{\lambda \dot{a}^{2}}{2r} \left(\epsilon^i{}_{nm} x_j + \epsilon^i{}_{jm} x_n + \epsilon^i{}_{nk} \delta_{jm} x^{k} \right) \extd x^m \wedge \extd x^n  \tens_1 \extd x^{j}
\\
\text{Riem}_{1}(\extd t) &=& -\frac{1}{2}\widehat{R}^0{}_{\alpha\mu\nu} \extd x^\mu \wedge \extd x^\nu \tens_1 \extd x^\alpha
\end{eqnarray*} 
Next step is to calculate $\text{Ricci}_{1}$ which comes out as
\[ \text{Ricci}_1=-\frac{1}{2}\widehat{R}_{\alpha\beta}\extd x^\beta\tens_1\extd x^\alpha\]
with no corrections to the coefficients in this form. The classical Ricci tensor for the Levi-Civita connection in our conventions is
\begin{equation*}
\label{RicciCl}
\widehat{\text{Ricci}} = -\frac{1}{2} \left(\left(2\dot{a}^{2} + a\ddot{a}\right) \delta_{ij} \extd x^{i} \otimes \extd x^{j} - 3 \frac{\ddot{a}}{a} \extd t \otimes \extd t \right)
\end{equation*}
and ${\rm Ricci}_1$ has the same form just with $\tens_1$. The components again depend only on time, hence are central, which means that $\rho=0$ as well. It remains to verify that $\wedge_1({\rm Ricci}_1)=0$ as it should have the same quantum symmetry as $g_1$. So using \eqref{wedge1com1}, we first see that $\extd t \wedge_1 \extd t = 0$ leaving (since the coefficients are time dependent they can be neglected here) 
\[ \delta_{ij} \extd x^{i} \wedge_1 \extd x^{j} = \frac{\lambda}{2r^3} \delta_{ij} \left(r^2 \epsilon^i{}_{nm} x^j + r^2 \epsilon^{ij}{}_m x_n + r^2 \epsilon^i{}_{nk} \delta^j{}_m x^{k} + \epsilon^i{}_{nk} x^k x^j x_m \right) \extd x^m \wedge \extd x^n = 0 \]
From \eqref{S1} we calculate the scalar curvature. Since neither the quantum metric or Ricci tensor have any semiclassical correction, it is straightforward to see that the same is true of the Ricci scalar, i.e.
\begin{equation}\label{FLRWScalar} 
S_1  = -\frac{1}{2}\widehat{S},\quad \widehat{S}=\widehat R_{\mu\nu}g^{\mu\nu}= {6\over a^2}\left({\ddot a}a+  {\dot a^2} \right)
\end{equation}
From Proposition~\ref{dim1}, the quantum dimension of this model comes out as
\[ \dim(M)_1 = \dim(M)-\lambda\omega^{\alpha\beta}g^{\mu\nu}{}_{,\alpha}\Gamma_{\nu\beta\mu} = 4 \]
since the metric depends only on $t$ the $\mathcal{O}(\lambda)$ term vanishes.

\section{Semiquantisation of spherically symmetric metrics}

\subsection{General analysis for the spherical case}

In the previous section we saw that, for a spherically symmetric Poisson tensor, demanding a compatible connection that also satisfied \eqref{nabla1g1} results in a unique quantum Levi-Civita connection for the FLRW metric. Something similar was observed to the Schwarzschild black hole in \cite{BegMa6} which suggests a general phenomenon for  the spherically symmetric case. We prove in the present section that this is generically true. For the metric we choose a  diagonal form
\begin{equation*}\label{SphMetric} \cg = a^2(r,t) \extd t \tens \extd t + b^2(r,t) \extd r \tens \extd r + c^2(r,t) (\extd \theta \tens \extd \theta + \sin^2(\theta) \extd \phi \tens \extd \phi) \end{equation*}
where $a,b,c$ are arbitrary functional parameters. The Poisson tensor is taken to be the same as in Section~3, once again parameterized by
\[ \omega^{23}={f(t,r)\over \sin \theta}=-\omega^{32},\quad \omega^{01}=g(t,r)=-\omega^{10}\]
The Christoffel symbols for the above metric are
\begin{equation}
\label{SphChr}
\begin{array}{lllll}
\displaystyle \widehat{\Gamma}^{0}_{00} = \frac{\partial_t a}{a} , & \displaystyle \widehat{\Gamma}^{0}_{01} = \frac{\partial_r a}{a} , & \displaystyle \widehat{\Gamma}^{0}_{33} = -\frac{b \partial_t b \sin^2(\theta)}{a^2} ,  &  \displaystyle \widehat{\Gamma}^{0}_{11} = -\frac{b \partial_t b}{a^2}    ,  & \displaystyle \widehat{\Gamma}^{0}_{22} = -\frac{c \partial_t c}{a^2} \\

\displaystyle \widehat{\Gamma}^{1}_{00} = -\frac{a \partial_r a}{b^2} , & \displaystyle \widehat{\Gamma}^{1}_{11} = \frac{\partial_r b}{b}, &  \displaystyle \widehat{\Gamma}^{1}_{33} = -\frac{b \partial_r b \sin^2(\theta)}{b^2} , & \displaystyle \widehat{\Gamma}^{1}_{01} = \frac{\partial_t b}{b}  , & \displaystyle \widehat{\Gamma}^{1}_{22} = -\frac{c \partial_r c}{b^2} \\

\displaystyle \widehat{\Gamma}^{2}_{02} = \frac{\partial_t c}{c} , & \displaystyle \widehat{\Gamma}^{2}_{21} = \frac{\partial_r c}{c}, & \displaystyle \widehat{\Gamma}^{2}_{33} = -\sin(\theta)\cos(\theta) \\

\displaystyle \widehat{\Gamma}^{3}_{03} = \frac{\partial_t c}{c} ,\quad & \displaystyle \widehat{\Gamma}^{3}_{31} = \frac{\partial_r c}{c}, & \displaystyle \widehat{\Gamma}^{3}_{23} = \cot(\theta)
\end{array}
\end{equation}
Now for the quantum Levi-Civita connection. 

\begin{theorem}\label{SphLC} For a generic spherically symmetric metric with functional parameters $a,b,c$ and spherically symmetric Poisson tensor, the Poisson-compatibility \eqref{nablaS} and the quantum Levi-Civita condition \eqref{nabla1g1}  require up to normalisation that $g(r,t) = 0$ and $f(r,t) = 1$ and the contorsion tensor components
\[S_{022} = c \del_t c ,\quad S_{122} = c \del_r c ,\quad S_{033} = c \del_t c \sin^2(\theta) ,\quad S_{133} = c \del_r c \sin^2(\theta) \]
\[ S_{120} = S_{123} = S_{223} = S_{320} = S_{130} = S_{132} = S_{230} = S_{233}=0\]
up to the outer antisymmetry of $S_{\mu\nu\gamma}$. The remaining components $S_{\mu 0 \nu}$, $S_{\mu 1 \nu}$ remain undetermined but do not affect $\omega^{\alpha\beta}\nabla_\beta$, which is unique. The relations of the quantum algebra are uniquely determined to $O(\lambda)$ as those of the fuzzy sphere 
\[ [z^i,z^j]=\lambda \eps^{ij}{}_k z^k,\quad [z^i,\extd z^j]=\lambda z^j\eps^i{}_{mn}z^m\extd z^n\]
as in Section~\ref{fuzzyS} and 
\[ [t,x^\mu] = [r,x^\mu] = 0 ,\quad  [x^\mu ,\extd t] = [x^\mu,\extd r] = 0\]
so that $t,r,dt,dr$ are central at order $\lambda$. 
\end{theorem}
\proof The first part is very similar to the proof of Proposition~\ref{FLRWPoisson} but with more complicated expressions. We once again require that either $f=k$ or $g=0$ for $\omega$ to be Poisson. Taking first $f=k$ and leaving $g$ arbitrary gives the Poisson compatibility condition \eqref{nablaS} as
\[ S^1{}_{02} = 0 ,\quad S^3{}_{01} = 0 ,\quad S^3{}_{22} = 0 ,\quad S^3{}_{10} = 0 ,\quad S^0{}_{12} = 0 , \quad S^3{}_{32} = 0,\]
\[ g ab S^0{}_{01} + ab \del_r g + g a\del_r b + gb \del_r a = 0,\quad ab^2 \del_t g + ab g \del_t b + b^2 g \del_t a + a^3 g S^0{}_{11} = 0 \]
\[ c^2 S^3{}_{31} - b^2 S^1{}_{22} + 2c \del_r c = 0, \quad   k c^2  S^0{}_{31} + gb^2 \sin(\theta)  S^1{}_{12} = 0, \quad g \del_t c \sin(\theta) - c S^1{}_{32} = 0 ,\]
\[a^2 g \sin(\theta) S^3{}_{12} - k a^2 S^0{}_{22} - k c \del_t c = 0, \quad g b^2 S^3{}_{02} \sin(\theta) + k b^2 S^1{}_{22} \sin(\theta) - k c\del_r c = 0,\]
\[ S^1{}_{12} + S^0{}_{02} = 0 ,\quad k c S^0{}_{32} + g \del_r c \sin(\theta) = 0,\quad g \sin(\theta) S^3{}_{11} + k S^0{}_{21} = 0,  \]
\[ k c \del_r c + gb^2 \sin(\theta) S^3{}_{02} + kc^2 S^3{}_{31} = 0 ,\quad k d^3 \sin(\theta) S^3{}_{21} - g b^2 \del_t c = 0, \]
\[ c^2 S^3{}_{30} + a^2 S^0{}_{22} a^2 + 2 c \del_r c = 0,\quad g S^3{}_{00} b^2 \sin(\theta) + k a^2 S^0{}_{21} = 0 ,\quad k c^2 S^0{}_{31} - g b^2 S^0{}_{02} = 0 \]
\[ k c^3 \sin(\theta) S^3{}_{20} - a^2 g\del_r c = 0,\quad k c\del_t c - ga^2 S^3{}_{12}\sin(\theta) + k c^2 S^3{}_{30},\quad a^2 S^3{}_{11} - b^2 S^3{}_{00} = 0\]
Once again, the above is enough to determine $\CR$ and can then be solved for $g$ simultaneously with \eqref{nabla1g1} using computer algebra (details omitted) assuming that $a,b,c$ are generic in the sense of invertible and not enjoying any particular relations. The only solution is $g(r,t) = 0$ as in the FLRW case. Now, starting over with $g=0$ and $f$ arbitrary, Poisson compatibility (\ref{nablaS}) gives a number of constraints including
\[ \del_r f = 0  ,\quad \del_t f =0 \]
which again forces us back to $g=0$, $f=k$ (which we set to be 1). Our above reduction of \eqref{nablaS} setting $g=0$ and $k=1$ then gives the contorsion tensor is as stated and by construction we also solved \eqref{nabla1g1}. 

Now we now check \eqref{nabla1g1} for this solution directly and independently of the computer algebra (which then does not require $a,b,c$ generic). For this, the generalised Ricci two-form comes out as
\begin{equation*}
\CR = - \frac{1}{2}  c^2 \sin(\theta) \extd \theta \wedge \extd \phi
\end{equation*} 
giving us  $\widehat\nabla_0 \CR_{\mu\nu} = \widehat\nabla_1 \CR_{\mu\nu}=0$ as well as 
\[ \widehat\nabla_2 \CR_{\mu\nu} = c \sin(\theta) \begin{pmatrix} 
0 & 0 & -\del_t c & 0 
\\ 
0 & 0 & -\del_r c & 0
\\
\del_t c & \del_r c & 0 & 0
\\
0 & 0 & 0 & 0
\end{pmatrix},\quad 
\widehat\nabla_3 \CR_{\mu\nu} = c\sin(\theta) \begin{pmatrix} 
0 & 0 & -\del_t c & 0 
\\ 
0 & 0 & -\del_r c & 0
\\
\del_t c & \del_r c & 0 & 0
\\
0 & 0 & 0 & 0
\end{pmatrix}
\]
Further calculation yields $Q_{0\mu\nu} =Q_{1\mu\nu}=0$ and 
\[ Q_{2\mu\nu} = c \sin(\theta) \begin{pmatrix} 
0 & 0 & -\del_t c & 0 
\\ 
0 & 0 & -\del_r c & 0
\\
\del_t c & \del_r c & 0 & 0
\\
0 & 0 & 0 & 0
\end{pmatrix},\quad 
Q_{3\mu\nu} =c \sin(\theta) \begin{pmatrix} 
0 & 0 & -\del_t c & 0 
\\ 
0 & 0 & -\del_r c & 0
\\
\del_t c & \del_r c & 0 & 0
\\
0 & 0 & 0 & 0
\end{pmatrix}
\]
Substituting, see see that \eqref{nabla1g1} holds in the form $\frac{1}{2}Q_{\gamma\alpha\beta} - \frac{1}{2}\nabla_\gamma \CR_{\mu\nu} = 0$, where $Q$ is the expression \eqref{Q}. This we have solved for the contorsion tensor obeying \eqref{nablaS} and \eqref{nabla1g1} for any $a,b,c$ and this gives us $\omega^{\alpha\beta}\nabla_\beta$ uniquely if these are generic.

Next we take local coordinates $z^1$ and $z^2$ while identifying $(z^3)^2 = 1 - (z^2)^2 - (z^1)^2$.  Then the Poisson tensor becomes 
\[ \omega = \epsilon_{ijk} z^k \extd z^i \wedge \extd z^j = \frac{1}{z^3} \extd z^1 \wedge \extd z^2 \] 
giving the coordinate algebra as stated. Since only $\omega_{23} = -\omega_{32}$ is nonzero, we also have $\{ t,x^\mu \} = \{ r,x^\mu \} = 0$. The `quantising' connection is 
\[\displaystyle \nabla (\extd t) = -\frac{\partial_t a}{a} \extd t \tens \extd t - \frac{b\partial_t b}{a^2} \extd r \tens \extd r - \frac{\partial_r a}{a} (\extd r \tens \extd t + \extd t \tens \extd r) - S^0{}_{0\mu} \extd t \tens \extd x^\mu - S^0{}_{1\mu} \extd r \tens \extd x^\mu
\]
\[
\displaystyle \nabla (\extd r) = - \frac{a \partial_r a}{b^2} \extd t \tens \extd t - \frac{\partial_t b}{b} \extd r \tens \extd r - \frac{\partial_t b}{b} (\extd r \tens \extd t + \extd t \tens \extd r) - S^1{}_{0\mu} \extd t \tens \extd x^\mu - S^1{}_{1\mu} \extd r \tens \extd x^\mu
\]
\[
\displaystyle \nabla (\extd z^i) = -\frac{\partial_r c}{c} \extd r \tens \extd z^i - \frac{\partial_t c}{c} \extd t \tens \extd z^i - \delta_{ab} z^i \extd z^a \tens \extd z^b - S^i{}_{0\mu} \extd t \tens \extd x^\mu - S^i{}_{1\mu} \extd r \tens \extd x^\mu
\]
due to the Christoffel symbols 
\begin{equation*}
\begin{array}{cc}
\displaystyle {\Gamma}^{0}_{\mu\nu} = \begin{pmatrix} 
\frac{\partial_t a}{a} & \frac{\partial_r a}{a} S^0{}_{01} & S^0{}_{02} & S^0{}_{03}
\\ 
\frac{\partial_r a}{a} & \frac{b \partial_t b}{a^2} + S^0{}_{11} & S^0{}_{12} & S^0{}_{13}
\\
0 & 0 & 0 & 0
\\
0 & 0 & 0 & 0
\end{pmatrix}
,& 
\displaystyle {\Gamma}^{1}_{\mu\nu} = \begin{pmatrix} 
S^1{}_{00} + \frac{a \partial_r a}{b^2} & \frac{\partial_t b}{b} & S^1{}_{02} & S^1{}_{03} 
\\ 
\frac{\partial_t b}{b} + S^1{}_{10}  & \frac{\partial_r b}{b} & S^1{}_{12} & S^1{}_{13}
\\
0 & 0 & 0 & 0
\\
0 & 0 & 0 & 0
\end{pmatrix}
\\
\displaystyle {\Gamma}^{2}_{\mu\nu} = \begin{pmatrix} 
S^2{}_{00} & S^2{}_{01} & \frac{\partial_t c}{c} & S^2{}_{03}  
\\ 
S^2{}_{10} & S^2{}_{11} & \frac{\partial_r c}{c} & S^2{}_{12}
\\
0 & 0 & \frac{z^1(1-(z^2)^2)}{(z^3)^2} & \frac{(z^1)^2 z^2}{(z^3)^2}
\\
0 & 0 & \frac{(z^1)^2 z^2}{(z^3)^2} & \frac{z^1(1-(z^1)^2)}{(z^3)^2}
\end{pmatrix}
,&
\displaystyle {\Gamma}^{3}_{\mu\nu} = \begin{pmatrix} 
S^3{}_{00} & S^3{}_{01}  & S^3{}_{03} & \frac{\partial_t c}{c}
\\ 
S^3{}_{10} & S^3{}_{11}  & S^3{}_{12} & \frac{\partial_r c}{c}
\\
0 & 0 & \frac{z^2(1-(z^2)^2)}{(z^3)^2} & \frac{z^1 (z^2)^2}{(z^3)^2}
\\
0 & 0 & \frac{z^1 (z^2)^2}{(z^3)^2} & \frac{z^2(1-(z^1)^2)}{(z^3)^2}
\end{pmatrix}
\end{array}
\end{equation*}
From \eqref{fetacomm} we immediately see that $[t,\extd x^\mu] = [r,\extd x^\mu] = 0$. Furthermore,
\[ [z^1,\extd x^\mu] = - z^{3} \Gamma^\mu{}_{3\beta} \extd x^\beta ,\quad [z^2,\extd x^\mu] = z^{3} \Gamma^\mu{}_{2\beta} \extd x^\beta \]
so we can read of from the Christoffel symbols that $[x^\mu,\extd t] = [x^\mu,\extd r] = 0$. Evaluating the nonzero terms gives
\[ [z^1,\extd z^1] = -\frac{z^1}{z^3} (z^1 z^2 \extd z^1 - ((z^1)^2-1) \extd z^2) ,\quad [z^2,\extd z^2] = \frac{z^2}{z^3} (z^1 z^2 \extd z^2 - ((z^2)^2-1) \extd z^1) \]
\[ [z^2,\extd z^1] = -\frac{z^2}{z^3} (z^1 z^2 \extd z^1 - ((z^1)^2-1) \extd z^2),\quad [z^1,\extd z^2] = \frac{z^1}{z^3} (z^1 z^2 \extd z^2 - ((z^2)^2-1) \extd z^1) \]
which upon using $ \sum_i z^i \extd z^i = 0$ becomes $[z^i,\extd z^j]=\lambda z^j\eps^i{}_{mn}z^m\extd z^n$. 
\endproof
So we see that in generalizing the analysis, we recover the same bimodule structure as in the FLRW case and by extension, that of the fuzzy sphere in Section~\ref{fuzzyS}. The noncommutativity is purely spacial and confined to spatial `spherical shells'; the surfaces of fuzzy spheres at each time and each classical radius $r$. We have checked directly in the proof of the theorem that this is a solution for all $a,b,c$ while for generic $a,b,c$ we showed that it is the only solution, i.e. we are forced into this form from our assumptions and spherical symmetry.  There do in fact exist particular combinations of these metric functional  parameters which permit alternative solutions for $f$ and $g$. In fact we already saw an example in Section~\ref{BRsec} with the Bertotti-Robinson metric which had $f=0$ and $g=-r$. To see why this was allowed, we take a brief look at the Poisson compatibility condition (\ref{nablaS}) again, now with arbitrary $f$ and $g$ and note the particular constraint
\[ S^0{}_{32} fc + g \del_r c \sin(\theta) = 0 ,\quad -S^1{}_{32} fc + g \del_r c \sin(\theta) = 0 \] 
It is clear that with $c$ arbitrary, we cannot have $f=0$ without also having $g=0$. However, allowing $c=$constant means we can also take $f=0$ and $g$ nonzero, as is the case with the Bertotti-Robinson metric. Indeed, this leads to a different contorsion tensor, namely with a flat $\nabla$. In fact this exceptional model was fully solved using algebraic methods in \cite{MaTao} including the quantum Levi-Civita connection to all orders in $\lambda$. 

Proceeding with our generic spherically symmetric metric, for brevity we define the parameters
\[ 
F_1 = {1 \over a^3b} \left( a^2 b \partial^2_r a - a b^2 \partial^2_t b + b^2 \partial_t b \partial_t a - a^2 \partial_r b \partial_r a   \right)
,\quad
F_2 = {a^2 \over b^2} F_1
\]
\[F_3 = \frac{c}{a^3 b} \left( a \partial_t b \partial_r c - ba \partial_t \partial_r c + b \partial_t c \partial_r a \right) ,\quad F_4 = \frac{c}{a^3 b^2} \left(b^2 \partial_t a \partial_t  c + a^2  \partial_r c \partial_r a - b^2 a \partial_{t}^2 c   \right)\]
\[F_5 = \frac{c}{b^3 a^2} \left(a^2 \partial_r b \partial_r  c + b^2  \partial_t c \partial_t b - a^2 b \partial_{r}^2 c \right) ,\quad F_6 = \frac{1}{b^2 a^2} \left( b^2 a^2 + b^2 (\del_t c)^2 - a^2 (\del_r c)^2 \right)\]
\[ F_7 = \frac{1}{a^2 b c} \left( a^2b \del_r^2 - a^2 \del_r b \del_r c - b^2 \del_t b \del_t c \right) ,\quad F_8 = \frac{1}{a b^2 c} \left( -b^2a \del_t^2 + a^2 \del_r a \del_r c + b^2 \del_t a \del_t c \right) \]
\[ F_9 = \frac{1}{a b c} \left( b \del_r a \del_t c + a \del_t b \del_r c - a b \del_r \del_t c \right) \]
in which terms the Riemann tensor for the Poisson-compatible `quantising' connection comes out as
\[ 
\text{Riem}(\extd t) = -F_1 \extd r \wedge \extd t \tens \extd r + C(\extd t), \quad \text{Riem}(dr) =  F_2 \extd r \wedge \extd t \tens \extd t + C(\extd r), 
\]
\[
\text{Riem}(dz^i) =  \delta_{ab} \extd z^i \wedge \extd z^a \tens \extd z^b + C(\extd z^i),
\]
where we have collected in the tensor $C$ all contributions coming from the undetermined components of the contorison tensor, namely
\begin{eqnarray*}
C(\extd x^\iota) &=& + \nabla_\mu S^\iota{}_{0\alpha}  \extd x^\mu \wedge \extd t \tens \extd x^\alpha + \nabla_\mu S^\iota{}_{1\alpha}\extd x^\mu \wedge \extd r \tens \extd x^\alpha
\\
&& + \left( S^\iota{}_{0\kappa} S^\kappa{}_{\nu\alpha} + S^\iota{}_{0\alpha} S^0{}_{0\nu} + S^\iota{}_{1\alpha} S^1{}_{0\nu}\right) \extd t \wedge \extd x^\nu \tens \extd x^\alpha
\\
&&+ \left(S^\iota{}_{1\kappa} S^\kappa{}_{\nu\alpha} + S^\iota{}_{0\alpha} S^0{}_{1\nu} + S^\iota{}_{1\alpha} S^1{}_{1\nu}\right) \extd r \wedge \extd x^\nu \tens \extd x^\alpha
\\
&& +\left(S^\iota{}_{0\alpha} S^0{}_{i\nu} + S^\iota{}_{1\alpha} S^1{}_{i\nu} \right) \extd z^i \wedge \extd x^\nu \tens \extd x^\alpha 
\end{eqnarray*}
We also have the classical Ricci tensor for the Levi-Civita connection
\begin{eqnarray*} 
\widehat{\rm Ricci} &=& -\frac{1}{2} \left( (F_2 + 2F_8) \extd t \tens \extd t - (F_1 + 2F_7) \extd r \tens \extd r + F_9 (\extd t \tens \extd r + \extd r \tens \extd t) \right.
\\
&& \left. - \frac{1}{(z^3)^2}(F_6 + F_5 - F_4) \delta_{ij} \extd z^i \tens \extd z^j  \right) 
\end{eqnarray*}
and, for later reference, the Einstein tensor  
\begin{eqnarray*} 
\widehat{\rm G} &=& -\frac{1}{2} \left( \frac{a^2}{c^2}(F_5 + 2F_6) \extd t \tens \extd t - \frac{a^2}{c^2}(F_6 - 2F_4) \extd r \tens \extd r + F_9 (\extd t \tens \extd r + \extd r \tens \extd t) \right.
\\
&& \left. - \frac{1}{(z^3)^2} ( F_5 - F_4 - \frac{c^2}{b^2} F_1) \delta_{ij} \extd z^i \tens \extd z^j  \right) 
\end{eqnarray*}
Before continuing, we turn briefly to the quantity $\omega^{\beta \alpha} (R^\iota{}_{\nu\mu\alpha}+S^\iota{}_{\mu\nu;\alpha})$ which appears in several formulas in Section~2, most importantly, the quantum Levi-Civita connection condition (\ref{nabla1g1}). In particular, we note that it is surprising simple with the only nonzero components  
\[ \omega^{2 \alpha} (R^2{}_{22\alpha}+S^2{}_{22;\alpha}) =  \frac{z^1 z^2}{z^3} ,\quad \omega^{2 \alpha} (R^3{}_{22\alpha}+S^3{}_{22;\alpha}) =  \frac{(z^2)^2 - 1}{z^3} ,\]
\[ \omega^{3 \alpha} (R^3{}_{23\alpha}+S^3{}_{32;\alpha}) =  \frac{z^2 z^1}{z^3}, \quad \omega^{3 \alpha} (R^3{}_{23\alpha}+S^3{}_{32;\alpha}) =  \frac{(z^2)^2 - 1}{z^3} , \]
\[  \omega^{2 \alpha} (R^2{}_{32\alpha}+S^2{}_{32;\alpha}) =  \frac{1 - (z^1)^2}{z^3} ,\quad \omega^{2 \alpha} (R^3{}_{32\alpha}+S^3{}_{32;\alpha}) =  -\frac{z^1 z^2}{z^3}, \]
\[ \omega^{3 \alpha} (R^2{}_{33\alpha}+S^2{}_{33;\alpha}) =  \frac{1 - (z^1)^2}{z^3} ,\quad \omega^{3 \alpha} (R^3{}_{33\alpha}+S^3{}_{33;\alpha}) =  -\frac{z^1 z^2}{z^3} \]
The undetermined components of $S$ do not contribute. In general the `quantising' connection always enters in combination with the Poisson tensor e.g. $\omega^{\alpha\beta} \Gamma^\iota{}_{\beta\gamma}$ so the same argument as in the proof of Theorem~\ref{SphLC} applies and we the undetermined components $S_{\mu 1\nu}$ or $S_{\mu 2\nu}$ in geometrically relevant expressions, as demonstrated by the generalized Ricci 2-form which now is     
\[ \mathcal{R} = -{1\over2} c^2 \epsilon_{mnk} z^{k} \extd z^m \wedge \extd z^n = -\frac{c^2}{z^3} \extd z^1 \wedge \extd z^2  \]
From this we have the quantum wedge product 
\begin{equation*}
\extd t \wedge_1 \extd x^\mu = \extd t \wedge \extd x^\mu ,\quad \extd x^\mu  \wedge_1 \extd t = \extd x^\mu  \wedge \extd t , \quad \extd r \wedge_1 \extd x^\mu = \extd r \wedge \extd x^\mu ,\quad \extd x^\mu  \wedge_1 \extd r = \extd x^\mu  \wedge \extd r
\end{equation*}
\begin{equation*}
\label{WedgeSph} 
\extd z^i \wedge_1 \extd z^j =\extd z^i \wedge \extd z^j + {\lambda \over 2} \left( 3z^i z^j - \delta^{ij} \right) \extd z^1 \wedge \extd z^2
\end{equation*} 
and also 
\[ \{\extd z^i, \extd z^j \}_1 = \lambda \left( 3z^i z^j - \delta^{ij} \right) \extd z^1 \wedge \extd z^2  \]
Our next step is to calculate the quantum metric. 
\begin{equation*}
\label{g1exGen}
g_{1} = g_{\mu\nu} \extd x^{\mu} \otimes_{1} \extd x^{\nu} + \frac{\lambda c^2}{2(z^3)^2}  \epsilon_{3ij}  \left( z^3  \extd z^i \tens_1 \extd z^j -  z^i \extd z^3 \tens_1 \extd z^j  \right)
\end{equation*}
Working with metric components $\tilde g_{ij}$ (in the middle) we get
\[ h = \frac{c^2 (2-(z^3)^2)}{(z^3)^3}\epsilon_{3ij} \extd z^i \tens_1 \extd z^j \]
Meanwhile, for the inverse metric with components $\tilde g^{ij}$ we get 
\[ (\extd z^1,\extd z^2)_{1} = g^{23} + \frac{\lambda}{2} \frac{z^3}{c^2},\quad (\extd z^2,\extd z^1)_{1} = g^{32} - \frac{\lambda}{2} \frac{z^3}{c^2}  \]
\[ (\extd t,\extd t)_{1} = g^{00},\quad (\extd r,\extd r)_{1} = g^{11} ,\quad (\extd z^1,\extd z^1)_{1} = g^{22} ,\quad (\extd z^2,\extd z^2)_{1} = g^{33}   \]
Now, Lemma~\ref{Gamma1} gives the quantum connection as
\[
\nabla_1 (\extd t) =  -\widehat{\Gamma}^{0}{}_{\mu\nu} \extd x^\mu \tens_1 \extd x^\nu + \frac{\lambda}{2 (z^3)^2} \frac{c \del_t c}{a^2} \left(  \epsilon_{ijk} z^k \extd z^i \tens_1 \extd z^j - \epsilon_{3ij} z^i \extd z^j \tens_1 \extd z^3 \right) 
\]
\begin{equation}
\label{SphCon}
\nabla_1 (\extd r) = -\widehat{\Gamma}^{1}{}_{\mu\nu} \extd x^\mu \tens_1 \extd x^\nu - \frac{\lambda}{2 (z^3)^2} \frac{c \del_r c}{b^2} \left(  \epsilon_{ijk} z^k \extd z^i \tens_1 \extd z^j - \epsilon_{3ij} z^i \extd z^j \tens_1 \extd z^3 \right) 
\end{equation}
\[
\nabla_1 (\extd z^a) = - \widehat{\Gamma}^{i}{}_{\mu\nu} \extd x^\mu \tens_1 \extd x^\nu + \frac{\lambda}{2} \left( \epsilon_{ijk} z^k z^a \extd z^i \tens_1 \extd z^j - \frac{1}{(z^3)^2} \epsilon^a{}_{i3}  \extd z^3 \tens_1 \extd z^i \right)
\]
Lastly, we calculate the associated braiding. Its nonzero contributions at first order in $\lambda$ are
\[ \sigma_1 (\extd z^i \tens_1 \extd z^j ) = \extd z^j \tens_1 \extd z^i + \lambda \left( \epsilon_{abc} z^c z^i z^j + \epsilon^j{}_{ac} z^c \delta^i{}_b + \epsilon^{ij}{}_{c} z^c \delta_{ab} \right) \extd z^a \tens_1 \extd z^b\]
when calculated using \eqref{sigma1}.Once again, a useful, if rather nontrivial check of the above formulae would be to explicitly calculate $\nabla_1 g_1 = 0$. Meanwhile, from Proposition~\ref{i1} quantum antisymmetric lift is 
\begin{equation}
\label{i1GSph1}
i_1 (\extd x^\mu \wedge \extd x^\nu) = {1 \over 2} \left(\extd x^\mu \tens_1 \extd x^\nu - \extd x^\nu \tens_1 \extd x^\mu \right) + \lambda I (\extd x^\mu \wedge \extd x^\nu) 
\end{equation}
The functorial choice gives $I (\extd x^\mu \wedge \extd x^\nu) = 0$, but we leave this general. 

\subsection{Laplace operator and Curvature Tensor}

Following from the previous section, we first calculate the Laplace operator. From Theorem~\ref{square1} we get that 
\[ \square_{1} f = g^{\alpha\beta} \left(f_{,\alpha\beta}+f_{,\gamma} \widehat{\Gamma}^{\gamma}{}_{\alpha\beta}\right) \]
as with the flat FLRW metric, is undeformed in the underlying algebra. Then, \eqref{Riem1} gives the quantum Riemann tensor as 
\begin{eqnarray*}
{\rm Riem}_1 (\extd t) &=& -\frac{1}{2}\widehat{R}^0{}_{\alpha\mu\nu} \extd x^\mu \wedge \extd x^\nu \tens_1 \extd x^\alpha - \frac{\lambda}{2 (z^3)^2} \left(\epsilon_{3ij} z^3  ( F_3  \extd t - F_4 \extd r ) \wedge \extd z^i \tens_1 \extd z^j    \right.  
\\
&& \left. + \epsilon_{3ij} z^i ( F_3  \extd t - F_4 \extd r )  \wedge \extd z^j \tens_1 \extd z^3  \right)
\end{eqnarray*}
\begin{eqnarray*}
{\rm Riem}_1 (\extd r) &=& -\frac{1}{2}\widehat{R}^1{}_{\alpha\mu\nu} \extd x^\mu \wedge \extd x^\nu \tens_1 \extd x^\alpha + \frac{\lambda}{2 (z^3)^2} \left(\epsilon_{3ij} z^3  ( F_5  \extd t - F_3 \extd r ) \wedge \extd z^i \tens_1 \extd z^j    \right.  
\\
&& \left. + \epsilon_{3ij} z^i ( F_5  \extd t - F_3 \extd r )  \wedge \extd z^j \tens_1 \extd z^3  \right)
\end{eqnarray*}
\begin{equation*} 
{\rm Riem}_1 (\extd z^\mu ) = -\frac{1}{2}\widehat{R}^\mu{}_{\alpha\mu\nu} \extd x^\mu \wedge \extd x^\nu \tens_1 \extd x^\alpha + \frac{\lambda F_6 }{2 (z^3)^2} (1+(z^3)^3) \extd z^1 \wedge \extd z^2  \tens_1 \extd z^\mu  
\end{equation*}
Using the lift map \eqref{i1GSph1} and the tensor formula \eqref{Ricci1} we get the quantum Ricci tensor as
\begin{eqnarray*}
{\rm Ricci_1} &=& -\frac{1}{2}\widehat{R}_{\mu\nu} \extd x^\mu \tens_1 \extd x^\nu - \frac{\lambda}{4 (z^3)^2} (F_6 + F_5 - F_4)\epsilon_{3ij} \left( z^3   \extd z^i \tens_1 \extd z^j -  z^i \extd z^3 \tens_1 \extd z^j  \right)
\\
&&  - \frac{3\lambda}{4 z^3} F_6 \epsilon_{3ij} \extd z^i \tens_1 \extd z^j - {\lambda\over 2}\widehat{R}^\alpha{}_{\gamma\eta\zeta}I^{\eta\zeta}{}_{\alpha\nu}\extd x^\nu\tens_1 \extd x^\gamma
\end{eqnarray*}
Lastly, we fix $I$ so that we have both $\wedge_1 ({\rm Ricci}_1) = 0$ and ${\rm flip}(*\tens *){\rm Ricci}_1={\rm Ricci_1}$. For the latter, it is easiest to consider the quantum Ricci tensor with components in the middle so that from Section~\ref{QRiemRicci} we have
\begin{eqnarray*}
\rho &=& -\frac{1}{4 (z^3)^3} \left((F_5 - F_4) ( 2-(z^3)^2 ) + 2F_6 ( 1+(z^3)^2 ) \right) \epsilon_{3ij} \extd z^i \tens_1 \extd z^j 
\\
&& - \frac{1}{2}\widehat{R}^\alpha{}_{\gamma\eta\zeta}I^{\eta\zeta}{}_{\alpha\nu}\extd x^\nu\tens_1 \extd x^\gamma
\end{eqnarray*}
where, for comparison, $\rho=-\frac{1}{2}\rho_{\mu\nu} \extd x^\mu \tens_1 \extd x^\nu$. Now for the reality condition to hold, since the coefficients are real and $\lambda$ imaginary, we require this to be antisymmetric. We also have
\begin{eqnarray*}
\wedge_1 ({\rm Ricci}_1) &=& -\frac{3\lambda}{2 z^3} F_6 \extd z^1 \wedge \extd z^2 - {\lambda\over 2}\widehat{R}^\alpha{}_{\gamma\eta\zeta}I^{\eta\zeta}{}_{\alpha\nu}\extd x^\nu \wedge \extd x^\gamma 
\end{eqnarray*}
Putting this together results in 
\[ \widehat{R}^\alpha{}_{\gamma\eta\zeta}I^{\eta\zeta}{}_{\alpha\nu}\extd x^\nu\tens_1 \extd x^\gamma = -\frac{3}{2 z^3} F_6 \epsilon_{3ij} \extd z^i \tens_1 \extd z^j \] 
This answer for the contraction of the lift map with the Riemann tensor is unique, but the same is not true of the lift map itself and we are left with a large moduli of possible solutions with most components of $I$ undetermined. It is however constructive to examine the simplest possible form by setting these to zero, leaving us with 
\[ i_1(\extd z^1 \wedge \extd z^2) = \frac{1}{2} \left( \extd z^1 \tens_1 \extd z^2 - \extd z^2 \tens_1 \extd z^1 \right) - \frac{3\lambda}{4 z^3} \delta_{ij} \extd z^i \tens_1 \extd z^j \]
as the only part with an $\mathcal{O}(\lambda)$ contribution and which is the same as for the fuzzy sphere seen previously. This results in 
\[ \rho = -\frac{1}{4 (z^3)^3} (F_6 + F_5 - F_4) ( 2-(z^3)^2 ) \epsilon_{3ij} \extd z^i \tens_1 \extd z^j \]
which we note has the same structure as $h$ for the quantum metric, but with different coefficients. The quantum Ricci tensor (with components on the left) is now
\begin{eqnarray*}
{\rm Ricci_1} &=& -\frac{1}{2}\widehat{R}_{\mu\nu} \extd x^\mu \tens_1 \extd x^\nu - \frac{\lambda}{4 (z^3)^2} (F_6 + F_5 - F_4)\epsilon_{3ij} \left( z^3   \extd z^i \tens_1 \extd z^j - z^i \extd z^3 \tens_1 \extd z^j  \right)
\end{eqnarray*}
The scalar curvature, using \eqref{S1}, has no corrections and comes out as
\[ {S}_1 = -{1\over 2}\widehat{S}, \quad \widehat{S}=\widehat R_{\mu\nu}g^{\mu\nu} = \frac{2}{c^2}(F_6 + 2F_5 - 2F_4) - \frac{2}{b^2} F_1\]
Note that it depends only on $t$ and $r$ and is therefore central in the algebra. From Proposition~\ref{dim1}, the quantum dimension comes out as 
\[ \dim(M)_1 = \dim(M)-\lambda\omega^{\alpha\beta}g^{\mu\nu}{}_{,\alpha}\Gamma_{\nu\beta\mu} = 4 \]
It might also be of interest to think about a quantum Einstein tensor. While a general theorem has not been established, we could consider a `naive' construction by analogy to the classical expression. Since the quantum and classical dimensions are the same, we could take for example 
\[ G_1 = \rm{Ricci}_1 - \frac{1}{2} S_1 g_1  \]
which has the same form as the classical case. This can be written as
\[ G_1 =-\frac{1}{2} G_{1\mu\nu} \extd x^\mu \tens_1 \extd x^\nu = -\frac{1}{2} \extd x^\mu \bullet \widetilde{G}_{1\mu\nu} \tens_1 \extd x^\nu \]
where $\widetilde{G}_{1\mu\nu} = G_{1\mu\nu} - \lambda \omega^{\alpha\beta} \widehat{G}_{\gamma\nu,\alpha} \Gamma^\gamma{}_{\beta\mu}$ and as previously the hat denotes that this is for the Levi-Civita connection. Now since in our case $S_1$ is purely classical and central, this can be expressed in component form as
\[ \widetilde G_{1\mu\nu} = {\widetilde{\widetilde{R_1}}}_{\mu\nu} - \frac{1}{2} \widehat S\, \widetilde g_{\mu\nu} 
\]
following the same pattern as how the components of ${\rm Ricci}_1$ and $g_1$ are written. If we write $\widetilde{G}_{1\mu\nu} = \widehat{G}_{\mu\nu} + \lambda \Sigma_{\mu\nu}$ then
\begin{eqnarray*} 
\Sigma &=& \rho - \frac{1}{4} \widehat{S}\, h = \frac{1}{4 (z^3)^3} ( F_5 - F_4 - \frac{c^2}{b^2} F_1) ( 2-(z^3)^2 ) \epsilon_{3ij} \extd z^i \tens_1 \extd z^j
\end{eqnarray*}
where $\Sigma=-\frac{1}{2}\Sigma_{\mu\nu} \extd x^\mu \tens_1 \extd x^\nu$ and is manifestly antisymmetric corresponding to ${\rm flip}(*\tens *){G}_1={G}_1$. Indeed, $G_1$ is both quantum symmetric and obeys the reality condition since ${\rm Ricci}_1$, $g_1$ and $\hat S$ is real and central. However, we are not claiming that the above $G_1$ is canonical. 

With the results of this section, we can calculate the quantum Levi-Civita connection and related quantities for all metrics of the form \eqref{SphMetric} simply by choosing appropriate parameters for $a$,$b$ and $c$. We look at some examples doing just this.

\subsection{FLRW metric case} Comparing the above results with those of the FLRW metric in Section~\ref{flrw}, there is a disparity that previously the metric appeared undeformed while now it has a quantum correction. We resolve this here. We first specialise the general theory above to the FLRW metric
\[ \cg = - \extd t \tens \extd t + a^2(t)( \extd r \tens \extd r + r^2 \delta_{ij} \extd z^i \tens \extd z^j) \]
where we identify the parameters
\[ a(r,t) = 1 , \quad b(r,t) = a(t)  ,\quad c(r,t) = r a(t). \]
This gives us the `quantising' connection up to undetermined but irrelevant contorsion tensor components (which are set to zero for simplicity)
\[
\displaystyle \nabla (\extd t) = - a \dot a \extd r \tens \extd r,\quad \displaystyle \nabla (\extd r) = - a \dot a ( \extd r \tens \extd t + \extd t \tens \extd r )
\]
\[
\displaystyle \nabla (\extd z^i) = - \frac{1}{r} \extd r \tens \extd z^i - \frac{\dot a}{a} \extd t \tens \extd z^i - \delta_{ab} z^i \extd z^a \tens \extd z^b 
\]
Meanwhile, for the classical Ricci tensor of the Levi-Civita connection we have  
\[ \widehat{\rm Ricci} = -\frac{1}{2} \left(-3 \frac{\ddot{a}}{a} \extd t \tens \extd t + \left(2\dot{a}^{2} + a\ddot{a}\right) ( \extd r \tens \extd r + r^2 \delta_{ij} \extd z^i \tens \extd z^j) \right) \]
and the curvature scalar is as in \eqref{FLRWScalar}. Now, the quantum metric comes out as 
\begin{equation}\label{FLRWg1}
g_{1} = g_{\mu\nu} \extd x^{\mu} \otimes_{1} \extd x^{\nu} + \frac{\lambda r^2}{2(z^3)^2} \eps_{3ij} \left( z^3 \extd z^i \tens_1 \extd z^j -  \extd z^3 \tens_1 z^i\extd z^j  \right)
\end{equation}
where $x^\mu$ refers to coordinates $t,r,z^1,z^2$ as we used polar coordinates. Equivalently, $\tilde g_{ij}$ (where the components are in the middle) has quantum correction
\begin{equation}\label{FLRWh} h = \frac{a(t)^2r^2 (2-(z^3)^2)}{(z^3)^3}\epsilon_{3ij} \extd z^i \tens_1 \extd z^j \end{equation}
For the inverse metric with components $\tilde g^{ij}$ we obtain
\[ (\extd z^1,\extd z^2)_{1} = g^{23} + \frac{\lambda}{2} \frac{z^3}{r^2},\quad (\extd z^2,\extd z^1)_{1} = g^{32} - \frac{\lambda}{2} \frac{z^3}{r^2}  \]
\[ (\extd t,\extd t)_{1} = g^{00},\quad (\extd r,\extd r)_{1} = g^{11} ,\quad (\extd z^1,\extd z^1)_{1} = g^{22} ,\quad (\extd z^2,\extd z^2)_{1} = g^{33}   \]
The quantum connection is
\[
\nabla_1 (\extd t) =  -\widehat{\Gamma}^{0}{}_{\mu\nu} \extd x^\mu \tens_1 \extd x^\nu + \frac{\lambda}{2 (z^3)^2} a\dot{a} r^2 \left(  \epsilon_{ijk} z^k \extd z^i \tens_1 \extd z^j - \epsilon_{3ij} z^i \extd z^j \tens_1 \extd z^3 \right) 
\]
\[
\nabla_1 (\extd r) = -\widehat{\Gamma}^{1}{}_{\mu\nu} \extd x^\mu \tens_1 \extd x^\nu - \frac{\lambda}{2 (z^3)^2} \left(  \epsilon_{ijk} z^k \extd z^i \tens_1 \extd z^j - \epsilon_{3ij} z^i \extd z^j \tens_1 \extd z^3 \right) 
\]
\[
\nabla_1 (\extd z^a) = - \widehat{\Gamma}^{i}{}_{\mu\nu} \extd x^\mu \tens_1 \extd x^\nu + \frac{\lambda}{2} \left( \epsilon_{ijk} z^k z^a \extd z^i \tens_1 \extd z^j - \frac{1}{(z^3)^2} \epsilon^a{}_{i3}  \extd z^3 \tens_1 \extd z^i \right)
\]
Then, by computing $(F_6 + F_5 - F_4) = r^2 (2\dot{a}^{2} + a\ddot{a})$ the quantum Ricci tensor is
\[
{\rm Ricci}_1  =  \widehat{R}_{\mu\nu} \extd x^\mu \tens_1 \extd x^\nu  - \frac{\lambda r^2}{4 (z^3)^2} (2\dot{a}^{2} + a\ddot{a})\epsilon_{3ij} \left( z^3   \extd z^i \tens_1 \extd z^j -  z^i \extd z^3 \tens_1 \extd z^j  \right)
\]
With components in the middle, this comes out as 
\[ \rho = -\frac{1}{4 (z^3)^3} r^2 (2\dot{a}^{2} + a\ddot{a}) ( 2-(z^3)^2 ) \epsilon_{3ij} \extd z^i \tens_1 \extd z^j \]
and in either case $S_1 = -{1\over 2}\widehat{S}$ in our conventions.

Now these results appear at first sight to be at odds with Section~\ref{flrw} since there the quantum metric  from  \eqref{g1ex} looks the same as classical when written in Cartesian coordinates. We first write it terms of $\extd z^i$ by   writing $\extd x^i = r\extd z^i-z^i\extd r$ and note that since $z^i\extd r=z^i\bullet\extd r$, we can take such $z^i$ terms to the other side of $\tens_1$. Since also $\extd z^i\bullet z^i=O(\lambda^2)$ (sum over $i$), we find
\[ \cg_1 = - \extd t \tens_1 \extd t + a^2(t)( \extd r \tens_1 \extd r + r^2 \delta_{ij} \extd z^i \tens_1 \extd z^j) \] 
This begins to look like (\ref{FLRWg1}) but note that only $z^1,z^2$ (say) are coordinates with $z^3$ a function of them. In particular, 
\[ \extd z^3 = -(z^3)^{-1} \bullet ( z^1 \bullet \extd z^1 +  z^2 \bullet \extd z^2) = -(z^3)^{-1}  ( z^1  \extd z^1 +  z^2  \extd z^2) - \frac{\lambda}{2} \epsilon_{3ij} z^i \extd z^j\]
would be needed to reduce to the form of (\ref{FLRWg1}) where the first term has only $\extd z^a\tens_1\extd z^b$ for $a,b=1,2$. Equivalently, we show that we have the same $\tilde g_{\mu\nu}$. Considering only the angular part $\delta_{ij} \extd z^i \tens_1 \extd z^j=\extd z^1 \tens_1 \extd z^1 + \extd z^2 \tens_1 \extd z^2 +\extd z^3 \tens_1 \extd z^3$ and examine the last term more closely (sum over repeated indices understood)
\begin{eqnarray*}
\extd z^3 \tens_1 \extd z^3 &=& (z^3)^{-1} \bullet z^{ a} \bullet \extd z^{ a} \tens_1 (z^3)^{-1} \bullet z^{ b} \bullet \extd z^{ b}
\\
&=& \left(\frac{z^{ a}}{z^3} + \frac{\lambda}{2} \epsilon_{ a  c} z^{ c} \right) \bullet \extd z^{ a} \tens_1 \left(\frac{z^{ b}}{z^3} + \frac{\lambda}{2} \epsilon_{ b  d} z^{ d}\right) \bullet \extd z^{ b}
\\
&=& \extd z^{ a} \bullet \left(\frac{z^{ a}}{z^3} + \frac{\lambda}{2} \epsilon_{ a  c} z^{ c} \right) \bullet \left(\frac{z^{ b}}{z^3} + \frac{\lambda}{2} \epsilon_{ b  d} z^{ d}\right) \tens_1 \extd z^{ b} +  [\frac{z^{ a}}{z^3},\extd z^{ a}] \bullet \frac{z^{ b}}{z^3} \tens_1 \extd z^{ b}
\\
&=& \extd z^{ a} \bullet \frac{z^{ a}z^{ b}}{(z^3)^2} \tens_1 \extd z^{ b} + \frac{\lambda}{2} \left( \extd z^{ a} \frac{1}{(z^3)^3} ( \epsilon_{ a  b} + z^3 \epsilon_{ b  d} z^{ a} z^{ d} + z^3 \epsilon_{ a  c} z^{ b} z^{ c} )  \tens_1  \extd z^{ b} \right.
\\
&& \left. - \extd z^{ a} \frac{1}{(z^3)^3} \epsilon_{ c  b} z^{c} z^{a}  \tens_1 \extd z^{ b} \right)
\\
&=&\extd z^{ a} \bullet \frac{z^{ a}z^{ b}}{(z^3)^2} \tens_1 \extd z^{ b} + \frac{\lambda}{2} \extd z^{ a} \frac{(2-(z^3)^2)}{(z^3)^3}\epsilon_{ a  b} \tens_1 \extd z^{ b}
\end{eqnarray*}
The $\bullet$ in the first term is left unevaluated so as to obtain $\tilde g_{ij}$ and we clearly see that we now have the  same semiclassical correction $h$ as in (\ref{FLRWh}).  We can perform a similar calculation for the quantum Ricci tensor in Section~\eqref{RicciCl}, making the same coordinate transformation as for the metric 
\[ {\rm Ricci}_1= -\frac{1}{2} \left(-3 \frac{\ddot{a}}{a} \extd t \otimes_1 \extd t + \left(2\dot{a}^{2} + a\ddot{a}\right) ( \extd r \tens_1 \extd r + r^2 \delta_{ij} \extd z^i \tens_1 \extd z^j) \right) \]
Indeed, since $t$ and $r$ are central in the algebra, the procedure is simply a repeat of that for the metric and clearly results in the same $\rho$ as above. Thus we obtain the same results as in Section~\ref{flrw} but only after allowing for the change of variables in the noncommutative algebra and $\tens_1$.

\subsection{Schwarzschild metric} We now look at some examples of well known metrics that fit the above analysis. For the first, we reexamine the Schwarzschild metric case in  \cite[Sec. 7.2]{BegMa6}. There it was found that (as we would now expect), the quantum Levi-Civita condition is satisfied for a spherically symmetric Poisson tensor. A difference however, is that in \cite{BegMa6} the torsion tensor was restricted to being rotationally invariant. By contrast, no such assumption is made here yet we are still led to a unique (con)torsion from Theorem~\ref{SphLC} up to some  undetermined components which we show do not enter into the quantum metric, quantum connection etc. Here the metric is
\[ \cg = -\left( 1 - \frac{r_S}{r} \right) \extd t \tens \extd t + \left( 1 - \frac{r_S}{r} \right)^{-1} \extd r \tens \extd r + r^2 \delta_{ij} \extd z^i \tens \extd z^j \]
so our three functional parameters are 
\[ a(r,t) = \left( 1 - \frac{r_S}{r} \right)^{\frac{1}{2}}, \quad b(r,t) = \left( 1 - \frac{r_S}{r} \right)^{-\frac{1}{2}} ,\quad c(r,t) = r \]
giving the `quantising' connection up to undetermined but irrelevant contorsion tensor components (which are set to zero for simplicity)
\[
\displaystyle \nabla (\extd t) = - \left(1-\frac{r_S}{r}\right)^{-\frac{1}{2}} \frac{r_S}{r^2} ( \extd r \tens \extd t + \extd t \tens \extd r )
\]
\[
\displaystyle \nabla (\extd r) = - \left(1-\frac{r_S}{r}\right)^{\frac{3}{2}} \frac{r_S}{r^2} \extd t \tens \extd t + \left(1-\frac{r_S}{r}\right)^{-\frac{1}{2}} \frac{r_S}{r^2} \extd r \tens \extd r 
\]
\[
\displaystyle \nabla (\extd z^i) = - \frac{1}{r} \extd r \tens \extd z^i - \delta_{ab} z^i \extd z^a \tens \extd z^b 
\]
As a check, by transforming into spherical polars and likewise neglecting the irrelevant components, we can recover the `quantising' connection in \cite{BegMa6}. In particular
\[
\displaystyle \nabla (\extd \theta) = - \frac{1}{r} \extd r \tens \extd \theta + \cos(\theta) \sin(\theta) \extd \phi \tens \extd \phi
\]
\[
\displaystyle \nabla (\extd \phi) = - \frac{1}{r} \extd r \tens \extd \phi - \cot(\theta) (\extd \theta \tens \extd \phi + \extd \phi \tens \extd \theta)
\]  
which agrees with \cite{BegMa6}. Obviously, the classical Ricci tensor for the Levi-Civita connection vanishes for the Schwarzschild metric, likewise for the curvature scalar. 

Now, the quantum metric comes out as 
\begin{equation*}
\label{g1exGen2}
g_{1} = g_{\mu\nu} \extd x^{\mu} \otimes_{1} \extd x^{\nu} + \frac{\lambda r^2}{2(z^3)^2} \eps_{3ij} \left( z^3 \extd z^i \tens_1 \extd z^j -  \extd z^3 \tens_1 z^i\extd z^j  \right)
\end{equation*}
While $\tilde g_{ij}$ (components in the middle) is
\[ h = \frac{r^2 (2-(z^3)^2)}{(z^3)^3}\epsilon_{3ij} \extd z^i \tens_1 \extd z^j \]
For the inverse metric with components $\tilde g^{ij}$ we get 
\[ (\extd z^1,\extd z^2)_{1} = g^{23} + \frac{\lambda}{2} \frac{z^3}{r^2},\quad (\extd z^2,\extd z^1)_{1} = g^{32} - \frac{\lambda}{2} \frac{z^3}{r^2}  \]
\[ (\extd t,\extd t)_{1} = g^{00},\quad (\extd r,\extd r)_{1} = g^{11} ,\quad (\extd z^1,\extd z^1)_{1} = g^{22} ,\quad (\extd z^2,\extd z^2)_{1} = g^{33}   \]
The quantum connection is
\[
\nabla_1 (\extd t) =  -\widehat{\Gamma}^{0}{}_{\mu\nu} \extd x^\mu \tens_1 \extd x^\nu  
\]
\[
\nabla_1 (\extd r) = -\widehat{\Gamma}^{1}{}_{\mu\nu} \extd x^\mu \tens_1 \extd x^\nu - \frac{\lambda r}{2 (z^3)^3} \left( 1 - \frac{r_S}{r} \right) \left(  \epsilon_{3ij} \extd z^i \tens_1 \extd z^j - \epsilon_{3ij} z^i z^3 \extd z^j \tens_1 \extd z^3 \right) 
\]
\[
\nabla_1 (\extd z^a) = - \widehat{\Gamma}^{i}{}_{\mu\nu} \extd x^\mu \tens_1 \extd x^\nu + \frac{\lambda}{2} \left( \epsilon_{ijk} z^k z^a \extd z^i \tens_1 \extd z^j - \frac{1}{(z^3)^2} \epsilon^a{}_{i3}  \extd z^3 \tens_1 \extd z^i \right)
\]
Meanwhile, from calculating the parameter $F_6 + F_5 - F_4=0$, we see that analogous to the classical case, the quantum Ricci tensor also vanishes 
\[
{\rm Ricci}_1  =  0 ,\quad \rho = 0
\]
as does $S_1$.

\subsection{Bertotti-Robinson metric with fuzzy spheres} Another interesting example is the Bertotti-Robinson metric, discussed in the context of a different differential algebra in Section~\ref{BRsec}. In order to draw a comparison between this case and the previous one, we define our metric as
\[ \cg= -a^2 r^{2\alpha}\extd t\tens\extd t + b^2 r^{-2} \extd r\tens\extd r + c^2 \delta_{ij} \extd z^i \tens \extd z^j \]
To chime with the conventions in this section, we relabel the constant terms and compared to the metric in Section~\ref{BRsec}, the off diagonal component is zero (either by diagonalising or setting the corresponding coefficient to zero). So our three functional parameters are 
\[ a(r,t) = a r^{\alpha} , \quad b(r,t) = b r^{-1}  ,\quad c(r,t) = c \]
As explained after Theorem~\ref{SphLC}, the theorem in this case does not give a unique quantum geometry but does give one. Dropping the undetermined and irrelevant contorsion components, Poisson-connection come out as
\[
\displaystyle \nabla (\extd t) = - \frac{\alpha}{r} (\extd t \tens \extd r + \extd r \tens \extd t) ,\quad 
\displaystyle \nabla (\extd r) = - \alpha r^{3\alpha} \frac{a^2}{c^2} \extd t \tens \extd t + \frac{1}{r} \extd r \tens \extd r
\]
\[
\displaystyle \nabla (\extd z^i) = - \delta_{ab} z^i \extd z^a \tens \extd z^b 
\]
This is markedly different from that in Section~\ref{BRsec} (apart from the different choice of coordinates), in particular with regard to the bimodule relations since previously $t$ was not central. We also have the Ricci tensor for the Levi-Civita connection
\[\widehat{\rm Ricci} = -\frac{1}{2} \left(\alpha^2 r^{2\alpha} \frac{a^2}{b^2} \extd t \tens \extd t - \frac{\alpha^2}{r^2} \extd r \tens \extd r + \delta_{ij} \extd z^i \tens \extd z^j \right)\]
with the corresponding scalar curvature
\[ \widehat{S} =\widehat{R}_{\mu\nu}g^{\mu\nu}=  \frac{2}{c^2} - \frac{2\alpha^2}{b^2} \]
The quantum metric is 
\begin{equation*}
\label{g1exGen3}
g_{1} = g_{\mu\nu} \extd x^{\mu} \otimes_{1} \extd x^{\nu} + \frac{\lambda c^2 }{2(z^3)^2} \epsilon_{3ij}\left( z^3   \extd z^i \tens_1 \extd z^j - z^i \extd z^3 \tens_1 \extd z^j  \right)
\end{equation*}
While $\tilde g_{ij}$ (components in the middle) has the deformation term
\[ h = \frac{c^2 (2-(z^3)^2)}{(z^3)^3}\epsilon_{3ij} \extd z^i \tens_1 \extd z^j \]
For the inverse metric with components $\tilde g^{ij}$ we get 
\[ (\extd z^1,\extd z^2)_{1} = g^{23} + \frac{\lambda}{2} \frac{z^3}{c^2},\quad (\extd z^2,\extd z^1)_{1} = g^{32} - \frac{\lambda}{2} \frac{z^3}{c^2}  \]
\[ (\extd t,\extd t)_{1} = g^{00},\quad (\extd r,\extd r)_{1} = g^{11} ,\quad (\extd z^1,\extd z^1)_{1} = g^{22} ,\quad (\extd z^2,\extd z^2)_{1} = g^{33}   \]
Now, the quantum connection is
\[
\nabla_1 (\extd t) =  -\widehat{\Gamma}^{0}{}_{\mu\nu} \extd x^\mu \tens_1 \extd x^\nu ,\quad \nabla_1 (\extd r) = -\widehat{\Gamma}^{1}{}_{\mu\nu} \extd x^\mu \tens_1 \extd x^\nu 
\]
\[
\nabla_1 (\extd z^a) = - \widehat{\Gamma}^{i}{}_{\mu\nu} \extd x^\mu \tens_1 \extd x^\nu + \frac{\lambda}{2} \left( \epsilon_{ijk} z^k z^a \extd z^i \tens_1 \extd z^j - \frac{1}{(z^3)^2} \epsilon^a{}_{i3}  \extd z^3 \tens_1 \extd z^i \right)
\]
Again, calculating the parameter $F_6 + F_5 - F_4=1$, the quantum Ricci tensor is
\[
{\rm Ricci}_1  =  -\frac{1}{2}\widehat{R}_{\mu\nu} \extd x^\mu \tens_1 \extd x^\nu  - \frac{\lambda}{4 (z^3)^2} \epsilon_{3ij}\left( z^3   \extd z^i \tens_1 \extd z^j - z^i \extd z^3 \tens_1 \extd z^j  \right)
\]
With components in the middle, this comes out as 
\[ \rho = -\frac{1}{4 (z^3)^3} ( 2-(z^3)^2 ) \epsilon_{3ij} \extd z^i \tens_1 \extd z^j \]
and in either case $S_1 = -{1\over 2}\widehat{S}$ in our conventions.

\section{Semiquantisation of PP-Wave Spacetimes}

The examples so far have concerned only spherically symmetric spacetimes. We used the symmetry as a basis to construct the Poisson tensor and `quantising' connection. However, as the choice of Poisson tensor (or any two-form) is not canonical, this is evidently not the only route to such a construction, but begs the question of how one is to find some appropriate structure. An interesting case is that of pp-wave spacetimes. A spacetime with nonvanishing curvature is referred to as a pp-wave spacetime if and only if it admits a covariantly constant, null bivector \cite{Grif}. These are exact solutions of Einstein's field equations which model massless radiation, such as gravitational waves. What makes them interesting here is the necessary existence of a covariantly constant bivector, indeed the definition is contingent only on this and not the form of the metric.  In this case we can canonically take the `quantising' connection to be the Levi-Civita as we can see from (\ref{nablaT}) with $T=0$.
\\
\\
We work in null coordinates $\{ v,u,z,\bar{z}\}$ where the bar denotes the complex conjugate and which can be related to standard Euclidean coordinates by 
\[ v=t+x^1,\quad u=t-x^1,\quad z=x^2+ix^3 \]
The metric is 
\[ \cg=\mathcal{H}(u,z,\bar{z})\extd u \tens \extd u + \extd u \tens \extd v + \extd v \tens \extd u - \extd z \tens \extd \bar{z} - \extd \bar{z} \tens \extd z  \]
where $H(u,z,\bar{z})$ is some arbitrary function. To construct the bivector associated with this metric, we consider the complex null tetrad given by
\[ l = \extd v,\quad n=\extd u - {\mathcal{H} \over 2}\extd v,\quad m=\extd \bar{z},\quad \bar{m}=\extd z\]
Here, indices are denoted by lower case Latin letters and range from $1$ to $4$. As null vectors these satisfy $l^a l_a=n^a n_a=m^a m_a=\bar{m}^a \bar{m}_a = 0$ and $l^a n_a = - m^a \bar{m}_a = 1$.  From this a two-form $U$ and its conjugate can be constructed as 
\[ U = l \wedge \bar{m} , \quad \bar{U} = l \wedge  m\]
which is covariantly constant 
\[ \widehat{\nabla} U = \widehat{\nabla} \bar{U} = 0\]
and null $U_{ab}U^{ac}=\bar{U}_{ab}\bar{U}^{ac}=0$. This tensor is unique and its existence is guarantied by the choice of metric, indeed the construction could be reversed and the metric derived by assuming this two form. For our purposes, we define
\[ \omega=a_1 U + a_2\bar{U}=a_1 \extd u \wedge \extd \bar{z} + a_2 \extd u \wedge \extd z\]
Where $a_1$ and $a_2$ are two arbitrary constants and $\omega^{ab} \omega_{bc} = 0$. Obviously we have
\[ \widehat{\nabla} \omega_{ab} = 0\]
so we can take $\nabla=\widehat{\nabla}$. The nonzero Christoffel symbols are
\begin{equation}
\label{PPChr}
\begin{array}{lll}
\displaystyle \widehat{\Gamma}^{1}_{11} = {1\over2} \mathcal{H}_{,u},\quad  & \displaystyle \widehat{\Gamma}^{1}_{12} = {1\over2} \mathcal{H}_{,z} ,\quad & \displaystyle \widehat{\Gamma}^{1}_{13} = {1\over2} \mathcal{H}_{,\bar{z}} \\
\displaystyle \widehat{\Gamma}^{3}_{11} = -{1\over2} \mathcal{H}_{,\bar{z}},\quad  & \widehat{\Gamma}^{0}_{11} = -{1\over2} \mathcal{H}_{,z}  & 
\end{array}
\end{equation}
The nonzero components of the Riemann tensor, taking into account the symmetries $R_{abcd}=-R_{bacd}$ and $R_{abcd}=-R_{abdc}$, are 
\begin{equation}
\label{PPRiemann}
\begin{array}{ccc}
R_{2323} = -\partial_{z}^{2}  \mathcal{H} \quad & R_{2324} = R_{2423} = -\partial_{z} \partial_{\bar{z}} \mathcal{H} \quad  &  R_{2424} = -\partial_{\bar{z}}^{2}  \mathcal{H}
\end{array}
\end{equation}
And Ricci tensor
\begin{equation}
\label{PPRicci}
\text{Ricci}_{22} = -\partial_{z} \partial_{\bar{z}} \mathcal{H}
\end{equation}
Regarding the physical interpretation, a wave propagating in Minkowski space along the $x^1$-axis satisfies  
\[ (\partial_{t}^{2} - \partial_{x^1}^{2}) h = \partial_{u} \partial_{v} h = 0\]
and evidently applies to the function $\mathcal{H}$ (which is independent of $v$). Hence, the metric corresponds to massless radiation with a wave profile given by $\mathcal{H}$ traveling in the $x^1$ direction. For the case of a zero stress-energy-momentum tensor, the vacuum solution $\text{Ric}_{ab} = 0$ corresponds to purely gravitational radiation or gravitation waves. The 2-surface with $u=v=$ const. are referred to as wave surfaces (hence plane waves) and there are no longitudinal modes. 

The nonzero bimodule relations are now
\[ [ v, z ] = a_1, \quad [v, \bar{z}] = a_2 ,\quad [v, \extd v]= -\frac{1}{2}(a_1 \mathcal{H}_{,z} + a_2 \mathcal{H}_{,\bar{z}}) du\]
Note that this calculus is associative. When cast in Euclidean form these are 
\[ [t,x^1]=0, \quad [t,x^2]=\frac{1}{2}(a_1 + a_2), \quad [t,x^3]=-\frac{i}{2}(a_1 - a_2),\] 
\[ \quad [x^1,x^2]= \frac{1}{2}(a_1 + a_2), \quad [x^1,x^3]= \frac{i}{2} (-a_1 + a_2), \quad [x^2,x^3]=0 \] 
\[ [t,\extd t] = [t,\extd x^1] = [x^1,\extd t] = [x^1,\extd x^1] = \left(\frac{a_1}{4}  \left( \mathcal{H}_{,x^2} - i \mathcal{H}_{,x^3} \right) + \frac{a_2}{4}  \left( \mathcal{H}_{,x^2} + i \mathcal{H}_{,x^3} \right) \right) \extd x^1 \]  
\[ - \left(\frac{a_1}{4} \left( \mathcal{H}_{,x^2} - i \mathcal{H}_{,x^3} \right) + \frac{a_2}{4}  \left( \mathcal{H}_{,x^2} + i \mathcal{H}_{,x^3} \right) \right) \extd t \]

\subsection{Quantum Metric}

As before, we continue to calculate the exterior algebra and quantum metric. The only nonzero component of $H$ is
\[ H^{11} = (a_1 \mathcal{H}_{,zz} + a_2 \mathcal{H}_{,z\bar{z}}) \extd u \wedge \extd z + (a_1 \mathcal{H}_{,\bar{z}z} + a_2 \mathcal{H}_{,\bar{z}\bar{z}}) \extd u \wedge \extd \bar{z} \]
Furthermore, it turns out that the only correction to the quantum wedge product is in the term 
\[\extd v \wedge_1 \extd v = (a_1 \mathcal{H}_{,zz} + a_2 \mathcal{H}_{,z\bar{z}}) \extd u \wedge \extd z + (a_1 \mathcal{H}_{,\bar{z}z} + a_2 \mathcal{H}_{,\bar{z}\bar{z}}) \extd u \wedge \extd \bar{z} \]
with all other having no order $\lambda$ contribution. Now, checking against (\ref{omegapoissonT}) and (\ref{omegapoisson}) shows $\omega$ to be Poisson. With a `quantising' connection, we can now look for a quantisation of pp-wave spacetimes by looking at (\ref{nabla1g1}). We calculate the Ricci two-form and find
\[ \mathcal{R} = 0 \]
so we immediately see that (\ref{nabla1g1}) holds since the contorsion is also zero. So the there exists a quantum Levi-Civita connection for the above. As it turns out, the quantum metric is undeformed at order $\lambda$
\[ g_1 = g_{\mu\nu} \extd x^\mu \otimes_1 \extd x^\nu \]
and similarly in the other form, so $h=0$. The same result holds for the inverse metric which is simply $(\ ,\ )_1 = (\ ,\ )$. It also turns out that all semi-classical corrections vanish in Lemma~\ref{Gamma1} and we are left with 
\[ \nabla_1 (\extd x^\iota)= -\widehat{\Gamma}^\iota_{\mu\nu}\extd x^\mu\tens_1 \extd x^\nu  \]
i.e., the same coefficients as classically. Lastly, for the Laplace operator and $f=f(v,u,z,\bar{z})$ from Theorem~\ref{square1} we obtain  
\[ \square_{1} f = g^{\alpha\beta} \left(f_{,\alpha\beta}+f_{,\gamma} \widehat{\Gamma}^{\gamma}{}_{\alpha\beta}\right) \]
So that this too remains classical. Indeed, looking at the bimodule relations, we see that the only terms that might experience any contribution at $\mathcal{O}(\lambda)$ must be functions of $v$. Since our metric is not, the same is true of the Christoffel symbols and curvature tensor (from \eqref{PPChr} and below). Also, looking at the expressions for the Riemann tensor, it is evident that it does not contain any terms involving $\extd v \wedge \extd v$ so the modified wedge product also fails to contribute. Hence we do not see any corrections to the non-commutative Riemann tensor and Ricci tensors either,
\[ \text{Riem}_1(\extd x^{\mu}) =  -\frac{1}{2}\widehat{R}^\mu{}_{\nu\alpha\beta} \extd x^\alpha \wedge \extd x^\beta \tens_1 \extd x^\nu \]
\[ \text{Ricci}_1 =  -\frac{1}{2}\widehat{R}_{\mu\nu} \extd x^{\mu} \tens_1 \extd x^{\nu} \]
with $\rho=0$. It should be remembered that  the pp-wave is only a very small part of a far larger class of solutions.  More interesting analysis may be possible within the Newman-Penrose formalism by relinquishing this requirement.

\section{Conclusions}

In this paper we extended the study of Poisson-Riemannian geometry introduced in \cite{BegMa6} to include a formula for the quantum Laplace-Beltrami operator at semiclassical order (Theorem~\ref{square1}) and we also looked at the lifting map needed to define a reasonable Ricci tensor in a constructive approach to that.  Our second main piece of analysis was Theorem~\ref{SphLC} for spherically symmetric Poisson tensors on spherically symmetric spacetimes. We found that when the metric components are sufficiently generic (in particular the coefficient of the angular part of the metric is not constant), any quantisation has to have $t,\extd t,r,\extd r$ central (so they remain classical) and nonassociative fuzzy spheres\cite{BegMa4} at each value of time and radius. This a startling degree of rigidity in which the quantum geometry is uniquely determined and in which the Laplace-Beltrami operator has no corrections at order $\lambda$. Key to the theorem was condition (\ref{nabla1g1}) from \cite{BegMa6} needed for the existence of a quantum torsion free and fully quantum metric compatible (i.e. quantum Levi-Civita) connection $\nabla_1$. If one wanted to avoid this conclusion then \cite{BegMa6} says that we {\em can} drop this condition (\ref{nabla1g1}), which would then open the door to a larger range of spherically symmetric models, but with a new  physical effect of $\nabla_1 g_1=O(\lambda)$ in its antisymmetric part. One can also drop our other assumption in the analysis that $\omega$ obeys (\ref{omegapoisson}) for the Jacobi identity. In that generic (nonassociative algebra) context we noted that spherical symmetry and Poisson-compatibility leads to a unique contorsion tensor, while imposing the Jacobi identity leads to half the modes of $S$ being undetermined but in such a way that the contravariant connection $\omega^{\alpha\beta}\nabla_\beta$ more relevant to the quantum geometry is still unique. This suggests an interesting direction for the general theory. 

The paper also included detailed calculations of the quantum metric, quantum Levi-Civita connection and quantum  Laplacian for a number of models, some of them, such as the FLRW, Schwarzschild and the time-central Bertotti-Robinson model being covered by Theorem~\ref{SphLC}. The important case of the FLRW model was first solved directly in Cartesian coordinates both as a warm up and as an independent check of the main theorem (the needed quantum change of coordinates was provided in Section~4.3). Two models not covered by our analysis of spherical symmetry are the 2D bicrossproduct model for which most of the algebraic side of the quantum geometry but not the quantum Laplacian was already found in \cite{BegMa5}, and the non-time central but spherically symmetric Bertotti-Robinson model for which the full quantum geometry was already found in \cite{MaTao} (this case is not excluded by Theorem~\ref{SphLC} since the coefficient of the angular metric is constant). In both cases the quantum spacetime algebra is the much-studied Majid-Ruegg spacetime $[x_i,t]=\lambda x_i$ in \cite{MaRue}. The non-time central Bertotti-Robinson model quantises $S^{n-1}\times dS_2$ and the quantum Laplacian in Section~\ref{BRsec} is quite similar to the old `Minkowski spacetime' Laplacian for this spacetime algebra which has previously led to variable speed of light\cite{AmeMa} in that provided wave functions are normal ordered, one of the double-differentials becomes a finite-difference (the main difference from\cite{AmeMa} is that this time there is an actual quantum geometry forcing the classical metric not to be flat\cite{MaTao}). However, when we analysed this within Poisson-Riemannian geometry we found no order $\lambda$ correction to the quantum Laplacian. We traced this to the formula for the bullet product in Poisson-Riemannian geometry in \cite{BegMa6} being realised on the classical space by an antisymmetric deformation, which is analogous to Weyl-ordered rather than left or right normal ordered functions in the noncommutative algebra being identified with classical ones. Our conclusion then is that order $\lambda$ predictions from such models\cite{AmeMa} were an artefact of the hypothesised normal ordering assumption and that Theorem~\ref{square1}  is a more stringent test within the  paradigm of Poisson-Riemannian geometry or semi-quantum gravity. We should not then be too surprised that order $\lambda$ corrections are more rare than one might naively have expected from the formula in Theorem~\ref{square1}. We see this again for the the pp-wave model where there is no order $\lambda$ correction.
The 2D bicrossproduct model in Section~\ref{bicross} {\em does} however have an order $\lambda$ deformation to the Laplacian even within Poisson-Riemannian geometry and we were able to solve the deformed massless wave equation at order $\lambda$ using Kummer functions (i.e. it is effectively the Kummer equation). This behaviour is reminiscent of the minimally coupled black hole in the wave operator approach of \cite{Ma:alm} (which was again without the actual quantum geometry) and its frequency dependent gravitational redshift and speed of light. 

 It is less clear at the present time how to draw immediate physical conclusions from our formulae for the quantum metric $g_1$ and quantum Ricci tensor ${\rm Ricci}_1$. In the FLRW model for example we found that $g_1$ looks identical to the classical metric but of course as an element of the quantum tensor product $\Omega^1\tens_1\Omega^1$. The physical  understanding of how quantum tensors relate to classical ones is suggested here as a topic of further work. Another topic on which we made only a tentative comment at the end of Section~4.2, is what should be the quantum Einstein tensor. Its deformation could perhaps be reinterpreted as an effective change to the stress energy tensor. This is another direction for further work.

\end{document}